\documentclass[preprint,aps,12pt,showpacs,nofootinbib,tightenlines,amsmath,amssymb]{revtex4}

\usepackage{amsmath}
\usepackage{graphicx}
\usepackage{amssymb}
\begin{document}
\def\intdk{\int\frac{d^4k}{(2\pi)^4}}
\def\sla{\hspace{-0.17cm}\slash}
\newcommand{\intp}[1]{\int\frac{d^4{#1}}{(2\pi)^4}}
\hfill

\title{Quantum Structure of Field Theory and Standard Model \\
Based on Infinity-free Loop Regularization/Renormalization\footnote{ The article in honor of Freeman Dyson's 90th birthday. } }
\author{Yue-Liang Wu}
\email{ylwu@itp.ac.cn}
\affiliation{ State Key Laboratory of Theoretical Physics (SKLTP)\\
Kavli Institute for Theoretical Physics China (KITPC) \\
Institute of Theoretical Physics, Chinese Academy of Science, Beijing,100190 \\
University of Chinese Academy of Sciences,  P.R.China}


\begin{abstract}
To understand better the quantum structure of field theory and standard model in particle physics, it is necessary to investigate carefully the divergence structure in quantum field theories (QFTs) and work out a consistent framework to avoid infinities. The divergence has got us into trouble since developing quantum electrodynamics in 1930s. Its treatment via the renormalization scheme is satisfied not by all physicists, like Dirac and Feynman who have made serious criticisms. The renormalization group analysis reveals that QFTs can in general be defined fundamentally with the meaningful energy scale that has some physical significance, which motivates us to develop a new symmetry-preserving and infinity-free regularization scheme called loop regularization (LORE). A simple regularization prescription in LORE is realized based on a manifest postulation that a loop divergence with a power counting dimension larger than or equal to the space-time dimension must vanish. The LORE method is achieved without modifying original theory and leads the divergent Feynman loop integrals well-defined to maintain the divergence structure and meanwhile preserve basic symmetries of original theory. The crucial point in LORE is the presence of two intrinsic energy scales which play the roles of ultraviolet cut-off $M_c$ and infrared cut-off $\mu_s$ to avoid infinities. As $M_c$ can be made finite when taking appropriately both the primary regulator mass and number to be infinity to recover the original integrals, the two energy scales $M_c$ and $\mu_s$ in LORE become physically meaningful  as the characteristic energy scale and sliding energy scale respectively. The key concept in LORE is the introduction of irreducible loop integrals (ILIs) on which the regularization prescription acts, which leads to a set of gauge invariance consistency conditions between the regularized tensor-type and scalar-type ILIs. An  interesting observation in LORE is that the evaluation of ILIs with ultraviolet-divergence-preserving (UVDP) parametrization naturally leads to Bjorken-Drell's analogy between Feynman diagrams and electric circuits, which enables us to treat systematically the divergences of Feynman diagrams and understand better the divergence structure of QFTs. The LORE method has been shown to be applicable to both underlying and effective QFTs. Its consistency and advantages have been demonstrated in a series of applications, which includes the Slavnov-Taylor-Ward-Takahaski identities of gauge theories and supersymmetric theories, quantum chiral anomaly, renormalization of scalar interaction and power-law running of scalar mass, quantum gravitational effects and asymptotic free power-law running of gauge couplings.
\end{abstract}


\maketitle

\section{ Introduction}

Quantum Field Theory (QFT) is the greatest successful framework established based on the special relativity and quantum mechanics. QFT with appropriate symmetries has successfully been applied to describe the microscopic world in elementary particle physics, nuclear physics, condensed matter physics and statistical physics. In the standard model of particle physics, the basic forces of nature are governed by the gauge symmetries $SU(3)_c\times SU(2)_L\times U(1)_Y$ which are well characterized by the QFT. In the framework of perturbation treatment of QFT, there is the well-known ultraviolet (UV) divergence problem due to the infinite Feynman integrals with closed loops of virtual particles, which may destroy the symmetries of original theory. In fact, the divergences appeared when developing quantum electrodynamics (QED) in1930s and 1940s by many physicists including M. Born, W. Heisenberg, P. Jordan, P. Dirac, E. Fermi, F.Bloch, A. Nordsieck, V.F. Weisskopf, R. Oppenheimer, H. Bethe, S. Tomonaga, J. Schwinger, R.P. Feynman, F. Dyson\cite{Dirac,Fermi,Bloch,Weiss,Oppen,Lamb,Bethe,Foley, Tomonaga, Schwinger,Feynman,Dyson}.  It was Dyson who made a systematic analysis to demonstrate the equivalence among various frameworks of QED and provided a sensible treatment to remove the divergences by introducing the renormalization scheme\cite{Dyson}. Mathematically, the divergence arises from the integral region where all particles in the loop have large energies and momenta. Physically, it may be understood due to very short proper-time between particle emission and absorption with the loop being thought of as a sum over particle paths, or from the very short wavelength or high frequency fluctuations of the fields in the path integral. The renormalization scheme treats the divergences by absorbing them into the fields and couplings through the redefinitions of fields and coupling constants, while the renormalization scheme is satisfied not by all physicists, like Dirac and Feynman who made their criticisms:

 ``Most physicists are very satisfied with the situation. They say: `Quantum electrodynamics(QED) is a good theory and we do not have to worry about it any more.' I must say that I am very dissatisfied with the situation, because this so-called `good theory' does involve neglecting infinities which appear in its equations, neglecting them in an arbitrary way. This is just not sensible mathematics. Sensible mathematics involves neglecting a quantity when it is small, not neglecting it just because it is infinitely great and you do not want it." by Dirac\cite{PDirac,Helge}, and 

``The shell game that we play$\dots$ is technically called `renormalization'. But no matter how clever the word, it is still what I would call a dippy process! Having to resort to such hocus-pocus has prevented us from proving that the theory of quantum electrodynamics(QED) is mathematically self-consistent. It's surprising that the theory still hasn't been proved self-consistent one way or the other by now; I suspect that renormalization is not mathematically legitimate." by Feynman\cite{RFeynman}.

In general, QFT becomes well-defined only when it can be regularized properly and the infinities can be avoided in the regularized theory, so that the renormalization involves only finite quantities. In fact, as it was emphasized by David Gross\cite{DG} that there is in principle no infinities in QCD as it can be described by a single finite gauge coupling with a behavior of asymptotic freedom\cite{QCDAF}.  In particular, the development of renormalization group technique enables one to define QFTs fundamentally with the meaningful energy scale that has some physical significance. However, the usual regularization schemes are known all to bear some limitations for providing a satisfied description on QFTs.  So that we have developed a symmetry-preserving and infinity-free regularization scheme called loop regularization(LORE) method\cite{YLWU1,YLWU2}. Such a LORE method has been demonstrated to be a finite regularization scheme that introduces intrinsically two meaningful energy scales to avoid infinities without spoiling symmetries of original theory. The gauge symmetry is preserved by a set of gauge invariance consistency conditions in LORE. In the practical calculations, the LORE method is manifestly simple at one loop level and becomes very sensible at high loop level to understand better the overlapping divergence structure of Feynman diagrams. 

Let us first make a brief comment on several regularization schemes often used in literature, such as: cut-off regularization\cite{COR}, Pauli-Villars regularization\cite{PV}, Schwinger's proper time regularization\cite{PR}, dimensional regularization\cite{DR}, BPHZ regularization\cite{BPHZ}, lattice regularization\cite{LR}, differential regularization\cite{DFR}. All the regularizations have their advantages and disadvantages. The naive cut-off regularization sets an upper bound to the integrating loop momentum, which in general destroys Lorentz invariance, translation invariance and gauge invariance for gauge theories. Such a scheme may be adopted to treat QFTs in statistical mechanics and in certain low energy dynamical systems, where the divergent behavior in the theories plays a key role, and Lorentz invariance becomes unimportant and gauge symmetry is not concerned at all. Thus the cut-off scheme is obviously unsuitable to be used for QFTs of elementary particles, where the Lorentz invariance and Yang-Mills gauge invariance play an important role. The Pauli-Vallars regularization is only suitable for the calculation of QED, but not applicable to the non-Abelian gauge theories as it destroys the non-Abelian gauge invariance due to the introduction of superheavy particles. A higher covariant derivative Pauli-Villars regularization was proposed to maintain the gauge symmetry in non-Abelian gauge theories\cite{AAS}, while it was shown that such a regularization violates unitarity\cite{LMR} and also leads to an inconsistent discription on QCD \cite{MR}.  The widely-used dimensional regularization is defined by making an analytical extension for the space-time dimensions of original theories, it can preserve gauge symmetries and be suitable to carry out computations for gauge theories, such as QED and QCD. Despite its great success, it remains questionable with some important properties in original theories\cite{Bonneau:1989xv, DR}, which includes that the spinor matrix $\gamma_5$ and chirality cannot be well defined in the extended dimension, the dimensional regularization cannot be applied directly to supersymmetric theories which require exact dimension of space-time, like supersymmetric theories. Thus the so-called dimensional reduction regularization was introduced\cite{DRED1} as a variant of dimensional regularization, in which the continuation from dimension d=4 to d=n  is made by a compactification, where the number of field components remains unchanged and  the momentum integrals are n-dimensional, which may cause ambiguities in the finite parts of the amplitudes and also in the divergent parts of high order corrections.In the practical application, it seems to hold only at one loop level and becomes inconsistent with analytical continuation concerning $\gamma_5$ \cite{DRED2}. The BPHZ prescription is actually a regularization-independent subtraction scheme up to the finite part, while to proceed the practical calculation for extracting the finite part by means of the forest formula, one still needs to use a concrete regularization scheme. As the subtraction process is based on expanding around an external momentum, it modifies the structure of Feynman integral amplitudes, thus the gauge invariance is potentially destroyed if applying the BPHZ subtraction scheme to non-Abelian gauge theories. In addition, the unitarity, locality and causality may not rigorously hold in such a subtraction scheme. In the lattice regularization, both space and time are made to be the discrete variables, the method can preserve gauge symmetries, but Lorentz invariance is not manifest. The lattice regularization may have a great advantage for the nonperturbative calculations by a numerical method, but it may lead to a very complicated perturbative calculation. 

We now turn to the LORE method. The development of LORE method is not for working out a much simpler regularization scheme, but for finding out whether there exists in principle a symmetry-preserving and infinity-free regularization method without modifying the original theory, so as to overcome the shortages and limitations in the widely-used regularization schemes and answer to the criticisms raised by Dirac and Feynman on the treatment of infinities via a renormalization scheme to remove the divergences. The consistency and applicability of the LORE method have been demonstrated in a series of works\cite{Cui:2008uv, Cui:2008bk, Ma:2005md, Ma:2006yc, cui:2011, DW,HW, Tang:2008ah, Tang:2010cr,Tang2011}. It has explicitly been shown at one loop level that the LORE method can preserve non-Abelian gauge symmetry~\cite{Cui:2008uv} and supersymmetry~\cite{Cui:2008bk}. It can lead to a consistent result for the chiral anomaly\cite{Ma:2005md} and radiatively induced Lorentz and CPT-violating Chern-Simons term in an extended QED\cite{Ma:2006yc} as well as QED trace anomaly\cite{cui:2011}. It has also been applied to derive the dynamically generated spontaneous chiral symmetry breaking of low energy QCD, and obtain consistent dynamical quark masses and mass spectra of light scalar and pseudoscalar mesons in a chiral effective field theory\cite{DW}, as well as understand the chiral symmetry restoration in chiral thermodynamic model\cite{HW}. The LORE method enables us to carry out consistently the quantum gravitational contributions to gauge theories with asymptotic free power-law running\cite{Tang:2008ah,Tang:2010cr,Tang2011}. The LORE method has also been applied to clarify the issue raised in\cite{HGG} for the process $H\to \gamma\gamma$ through a W-boson loop in the unitary gauge, and show that even for the finite amplitudes of Feynman diagrams, it still needs to adopt a consistent regularization for ensuring the cancellation between tensor-type and scalar-type divergent integrals\cite{HTW} in order to obtain the finite amplitudes. Recently, it has been shown that the LORE method enables us to demonstrate consistently the general divergence structure of QFTs by high-loop order calculations\cite{HW1,HLW}.  In particular, it is interesting to observe in \cite{HW1} that the evaluation of ILIs from Feynman integrals by adopting the so-called ultraviolet divergence-preserving (UVDP) parametrization naturally leads to the Bjorken-Drell's circuit analogy between Feynman diagrams and electric circuits\cite{BD}, which allows us to show the one-to-one correspondence between the divergences of the UVDP parameters and the subdiagrams of Feynman diagrams.

In this article, I mainly review and summarize the recent progresses on the development of the LORE method and some works done with J.W. Cui, D. Huang, Y.L.Ma, Y. Tang.

\section{Concept of Irreducible Loop Integrals (ILIs) }

The key concept in LORE is the introduction of irreducible loop integrals (ILIs) which are evaluated from the Feynman integrals by using the Feynman parametrization and ultraviolet divergence-preserving (UVDP) parametrization. In fact, the ``loop'' regularization is named as the simple regularization prescription of LORE acts on the ILIs.

To illustrate conceptually the ILIs, let us first consider one loop calculations of Feynman diagrams. It can be demonstrated that all Feynamn integrals of the one-particle irreducible graphs can be evaluated into the following sets of loop integrals by using Feynman parametrization
    \begin{eqnarray}
     I_{-2\alpha} = \int \frac{d^4 k}{(2\pi)^4}\ \frac{1}{(k^2 - {\cal M}^2)^{2+\alpha}}\ ,
     \qquad \alpha = -1, 0, 1,  \cdots
    \end{eqnarray}
for scalar type integrals and
    \begin{eqnarray}
  & & I_{-2\alpha\ \mu\nu} = \int \frac{d^4 k}{(2\pi)^4}\
    \frac{k_{\mu}k_{\nu}}{(k^2 - {\cal M}^2)^{3 + \alpha} }\ , \nonumber \\
  & & I_{-2\alpha\ \mu\nu\rho\sigma} = \int \frac{d^4 k}{(2\pi)^4}\
    \frac{k_{\mu}k_{\nu} k_{\rho}k_{\sigma} }{(k^2 - {\cal M}^2)^{4+ \alpha} }\ , \qquad \alpha =-1, 0, 1,
      \cdots
    \end{eqnarray}
for tensor type integrals. The subscript ($-2\alpha$) labels the power counting dimension of
energy momentum in the integrals. Two special cases with $\alpha = -1$ and $\alpha = 0$
correspond to the quadratic divergent integrals ($I_2$, $I_{2 \mu\nu \cdots}$) and
the logarithmic divergent integrals ($I_0$, $I_{0 \mu\nu \cdots}$) respectively. The mass factor
${\cal M}^2$ is in general a function of Feynman parameters and external momenta $p_i$,
${\cal M}^2 = {\cal M}^2 (m_1^2, p_1^2, \cdots)$.

The above loop integrals define the irreducible loop integrals (ILIs) at one-loop order\cite{YLWU1}. For high loop overlapping Feynman integrals, the corresponding ILIs are defined to be the integrals in which there exist no longer in the denominator the overlapping momentum factors $(k_i-k_j + p_{ij})^2$ $(i\ne j)$ which appear in the original overlapping Feynman integrals of loop momenta $k_i$ ($i=1,2,\cdots $), and there have no scalar momentum factors $k^2$ in the numerator. In evaluating any loop integrals into the corresponding ILIs, the algebraic computing for multi $\gamma$ matrices involving loop momentum $k\sla$ such as $k\sla\gamma_{\mu}k\sla$ should be carried out first and be expressed in terms of the independent components of $\gamma$ matrices: $\gamma_\mu$, $\sigma_{\mu\nu}$, $\gamma_5\gamma_{\mu}$, $\gamma_5$. 

To demonstrate the necessity of introducing the ILIs and yielding a consistent regularization scheme for regularizing divergent integrals, let us examine the following tensor-type and scalar-type divergent integrals.
\begin{eqnarray}
I_{2\mu\nu} &=& \int d^4 k \frac{k_\mu k_\nu}{(k^2-{\cal M}^2)^2} ,\quad I_{2} = \int d^4 k \frac{1}{(k^2-{\cal M}^2)^2} ,\quad  I^{\prime}_{2} = \int d^4 k \frac{k^2}{(k^2-{\cal M}^2)^3} \nonumber \\
I_{0\mu\nu} &=& \int d^4 k \frac{k_\mu
k_\nu}{(k^2-{\cal M}^2)^3} ,\quad I_{0} = \int d^4 k \frac{1}{(k^2-{\cal M}^2)^2} , \quad I^\prime_0 = \int d^4 k
\frac{k^2}{(k^2-{\cal M}^2)^3} \nonumber \\
I_{-2\mu\nu} &=& \int d^4 k \frac{k_\mu k_\nu}{(k^2-{\cal M}^2)^4} ,\quad I_{-2} = \int d^4 k \frac{1}{(k^2-{\cal M}^2)^3} , 
\end{eqnarray}
where  $I_{2 \mu\nu }$, $I_{0\mu\nu }$, $I_2$, $I_0$ are the corresponding quadratic and logarithmic divergent tensor-type and scalar-type ILIs, and  $I_{-2 \mu\nu }$, $I_{-2}$ are the corresponding convergent ILIs. Note that $I^{\prime}_2$ and $I^{\prime}_0$ should not be regarded as the ILIs according to the definitions of ILIs, they are actually related to the ILIs as follows
\begin{equation}
I^{\prime}_2 = I_2 + {\cal M}^2 I_{0},\qquad  I^{\prime}_0  = I_0 + {\cal M}^2 I_{-2}
\end{equation}

From general Lorentz decompositions, we may express the tensor-type ILIs in terms of the scalar-type ILIs with the following relations
\begin{eqnarray}\label{relations}
   & & I_{2\mu\nu } = \frac{1}{4} a_2\  I_2\ g_{\mu\nu}, \qquad I_{0\mu\nu} = \frac{1}{4} a_0\ I_0\ g_{\mu\nu}, \qquad I_{-2\mu\nu} = \frac{1}{4} a_{-2}\ I_{-2}\ g_{\mu\nu}
\end{eqnarray}
Here the definition for $a_2$ differs from the early one in\cite{YLWU1} and also our other papers by a factor of 4. As the ILIs $I_{-2}$ and $I_{-2 \mu\nu}$ are convergent integrals, we can safely carry out the integrals and the value of $a_{-2}$ is determined to be $a_{-2} = 2/3$. Here we have directly adopted the tensor manipulation by multiplying $g^{\mu\nu}$ on both sides of  $I_{-2\mu\nu} = \frac{1}{4} a_{-2}\ I_{-2}\ g_{\mu\nu}$ or one can simply replace $k_{\mu}k_{\nu}$ with $k^2 g_{\mu\nu}/4$.  For other relations concerning divergent ILIs, if applying for the naive analysis of Lorentz decomposition and tensor manipulation by multiplying $g^{\mu\nu}$ on both sides, one has
\begin{eqnarray}
& & \mbox{left-side} = g^{\mu\nu} I_{2\mu\nu} = I_2 + {\cal M}^2 I_0 = I^{\prime}_2 ,\quad 
\mbox{right-side} =  a_2 I_2,  \nonumber \\
& &  \mbox{left-side} = g^{\mu\nu} I_{0\mu\nu} = I_0 + {\cal M}^2 I_{-2} = I^{\prime}_0,\quad \mbox{right-side} = a_0\ I_0,  \nonumber
\end{eqnarray}
which leads to the results
\begin{equation}
a_2 = I^{\prime}_2 /I_2 = 1 + {\cal M}^2 I_0/I_2 , \quad a_0 = I^{\prime}_0/I_0 = 1 + {\cal M}^2 I_{-2}/I_0  \nonumber 
\end{equation}
or the following relations
\begin{eqnarray}\label{NR}
& & I_{2\mu\nu} = \frac{1}{4} g_{\mu\nu}\ I^{\prime}_2 = \frac{1}{4} g_{\mu\nu}\ I_2 +  \frac{1}{4} g_{\mu\nu} {\cal M}^2\ I_0 ,  \nonumber \\
& & I_{0\mu\nu} = \frac{1}{4} g_{\mu\nu} \ I^{\prime}_0 = \frac{1}{4} g_{\mu\nu} \ I_0 + g_{\mu\nu} {\cal M}^2 I_{-2} =  \frac{1}{4} g_{\mu\nu} \ I_0 - \frac{i}{32\pi^2} g_{\mu\nu} \nonumber
\end{eqnarray}
where we have performed the integration for the convergent integral $I_{-2}$.  For the divergent integrals, the tensor manipulation and integration do not in general commute with each other as they are not well defined without adopting a proper regularization scheme. As a consequence, the resulting divergent integrations become inconsistent. 

To obtain consistent relations, let us first examine the time component of the tensors on both sides of Eq.(\ref{relations})
\begin{equation}
I_{2\ 00 } = \frac{1}{4} a_2\  I_2\ g_{00},\qquad I_{0\ 00 } = \frac{1}{4} a_0\  I_0\ g_{00}\  ,
\end{equation}
by rotating the four-dimensional energy momentum into Euclidean space via the Wick rotation, and integrating safely over the zero component of energy momentum $k_0$ on both sides as such an integration is convergent. For the quadratic divergent ILIs, we have
\begin{eqnarray}
 I_{2} & = & -i\int \frac{d^4 k}{(2\pi)^4}\ \frac{1}{k^2 + {\cal M}^2} = -i\int \frac{d^3 k}{(2\pi)^4}\ \int dk_0 \frac{1}{k^2_0 + \bf{k}^2 + {\cal M}^2}  \nonumber \\
 & = & -i\int \frac{d^3 k}{(2\pi)^4}\ 2\frac{1}{\sqrt{\bf{k}^2 + {\cal M}^2}}
 \tan^{-1} \left( k_0/\sqrt{\bf{k}^2 + {\cal M}^2}\right) |_{k_0=0}^{k_0=\infty} \nonumber \\
 & = & -i\int \frac{d^3 k}{(2\pi)^3}\ \frac{1}{2\sqrt{\bf{k}^2 + {\cal M}^2}}
\end{eqnarray}
on the right-hand side, and
\begin{eqnarray}
 I_{2\ 00} & = & -i\int \frac{d^4 k}{(2\pi)^4}\ \frac{k_0^2}{(k^2 + {\cal M}^2)^2} = -i\int \frac{d^3 k}{(2\pi)^4}\ \int dk_0 \frac{k_0^2}{(k^2_0 + \bf{k}^2 + {\cal M}^2)^2} \nonumber \\
 & = & -i\int \frac{d^3 k}{(2\pi)^4}\ \int dk_0 \left( \frac{1}{k^2_0 + \bf{k}^2 + {\cal M}^2} -  \frac{\bf{k}^2 + {\cal M}^2}{(k^2_0 + \bf{k}^2 + {\cal M}^2)^2} \right) \nonumber \\
 & = & -i\int \frac{d^3 k}{(2\pi)^4}\ \int dk_0 \left( \frac{1}{k^2_0 + \bf{k}^2 + {\cal M}^2} -  \frac{1}{2}  \frac{1}{k^2_0 + \bf{k}^2 + {\cal M}^2}\right) \nonumber \\
 & & -\frac{k_0}{k^2_0 + \bf{k}^2 + {\cal M}^2}|_{k_0=0}^{k_0=\infty}  \nonumber \\
 & = & \frac{-i}{2}\int \frac{d^3 k}{(2\pi)^4}\ 2\frac{1}{\sqrt{\bf{k}^2 + {\cal M}^2}}
 \tan^{-1} \left( k_0/\sqrt{\bf{k}^2 + {\cal M}^2}\right) |_{k_0=0}^{k_0=\infty} \nonumber \\
 & = & \frac{-i}{2}\int \frac{d^3 k}{(2\pi)^3}\ \frac{1}{2\sqrt{\bf{k}^2 + {\cal M}^2}} 
\end{eqnarray}
on the left-hand side. Similarly, for the logarithmic divergent ILIs, we yield
\begin{eqnarray}
 I_{0} & = & i\int \frac{d^4 k}{(2\pi)^4}\ \frac{1}{(k^2 + {\cal M}^2)^2} = i\int \frac{d^3 k}{(2\pi)^4}\ \int dk_0 \frac{1}{(k^2_0 + \bf{k}^2 + {\cal M}^2)^2}  \nonumber \\
 & = & i\int \frac{d^3 k}{(2\pi)^4}\   \frac{1}{ 2(\bf{k}^2 + {\cal M}^2) }  \int dk_0 \frac{1}{k^2_0 + \bf{k}^2 + {\cal M}^2} \nonumber \\ 
& &  -\frac{k_0}{(\bf{k}^2 + {\cal M}^2 ) (k^2_0 + \bf{k}^2 + {\cal M}^2) }|_{k_0=0}^{k_0=\infty}  \nonumber \\
 & = & i\int \frac{d^3 k}{(2\pi)^4}\    \frac{1}{2 (\bf{k}^2 + {\cal M}^2 ) }   \frac{2}{\sqrt{\bf{k}^2 + {\cal M}^2} }
  \tan^{-1} \left( k_0/\sqrt{\bf{k}^2 + {\cal M}^2}\right) |_{k_0=0}^{k_0=\infty} \nonumber \\
 & = & i\int \frac{d^3 k}{(2\pi)^3}\ \frac{1}{ 4 (\bf{k}^2 + {\cal M}^2 )  \sqrt{\bf{k}^2 + {\cal M}^2} }
\end{eqnarray}
on the right-hand side, and
\begin{eqnarray}
 I_{0\ 00} & = & i\int \frac{d^4 k}{(2\pi)^4}\ \frac{k_0^2}{(k^2 + {\cal M}^2)^3} = i\int \frac{d^3 k}{(2\pi)^4}\ \int dk_0 \frac{k_0^2}{(k^2_0 + \bf{k}^2 + {\cal M}^2)^3} \nonumber \\
 & = & i\int \frac{d^3 k}{(2\pi)^4}\ \int dk_0 \left( \frac{1}{(k^2_0 + \bf{k}^2 + {\cal M}^2)^2} -  \frac{\bf{k}^2 + {\cal M}^2}{(k^2_0 + \bf{k}^2 + {\cal M}^2)^3} \right) \nonumber \\
 & = & i\int \frac{d^3 k}{(2\pi)^4}\ \int dk_0 [ \frac{1}{(k^2_0 + \bf{k}^2 + {\cal M}^2)^2} -  \frac{1}{4(\bf{k}^2 + {\cal M}^2)}  \frac{1}{k^2_0 + \bf{k}^2 + {\cal M}^2}   \nonumber \\
 & & -  \frac{1}{4}  \frac{1}{(k^2_0 + \bf{k}^2 + {\cal M}^2)^2} ]   -\frac{k_0^3+ 2k_0(\bf{k}^2 + {\cal M}^2)}{2(\bf{k}^2 + {\cal M}^2) (k^2_0 + \bf{k}^2 + {\cal M}^2)^2 }|_{k_0=0}^{k_0=\infty}  \nonumber \\
 & = & i \int \frac{d^3 k}{(2\pi)^4}\  \frac{3}{4}  \frac{1}{(\bf{k}^2 + {\cal M}^2) \sqrt{\bf{k}^2 + {\cal M}^2}}
 \tan^{-1} \left( k_0/\sqrt{\bf{k}^2 + {\cal M}^2}\right) |_{k_0=0}^{k_0=\infty} \nonumber \\
 & &  - \frac{1}{4}  \frac{2}{(\bf{k}^2 + {\cal M}^2) \sqrt{\bf{k}^2 + {\cal M}^2}} \tan^{-1} \left( k_0/\sqrt{\bf{k}^2 + {\cal M}^2}\right) |_{k_0=0}^{k_0=\infty}  \nonumber \\
 & = & i \int \frac{d^3 k}{(2\pi)^3}\ \frac{1}{16 (\bf{k}^2 + {\cal M}^2) \sqrt{\bf{k}^2 + {\cal M}^2} } 
\end{eqnarray}
on the left-hand side. 

By comparing the results on the left-hand and right-hand sides, which are obtained by integrating over the convergent part of quadratic and logarithmic divergent ILIs, we can safely determine the coefficients  
\begin{equation}
a_2 = 2 ; \qquad a_0 = 1
\end{equation}
which is completely different from the results yielded from the naive analysis of Lorentz decomposition and tensor manipulation by simply multiplying $g^{\mu\nu}$ on both sides.  

It demonstrates the manifestation and significance of introducing ILIs. As the integration over the zero component of momentum $k_0$ is convergent, all algebraic manipulation made in the calculation must be safe and valid. We would like  to address that the above demonstration for obtaining the consistent relations between the divergent ILIs is nothing to do with any regularization schemes. Nevertheless, it is valid only for one of the Lorentz components rather than for all components of Lorentz tensor in a covariant way. Thus to obtain the consistent relation in a covariant way between the tensor-type and scalar-type divergent ILIs, it is necessary to look for a proper regularization scheme which can make the divergent ILIs be well redefined or regularized.

\section{Infinity-Free Loop Regularization } 

A crucial point in LORE is the presence of two intrinsic energy scales which are introduced from the string-mode regulators in the regularization prescription acting on ILIs, they play the roles of the ultraviolet (UV) cut-off and infrared (IR) cut-off to avoid infinities without spoiling symmetries in original theories. The two energy scales are actually shown to become physically meaningful as the characteristic energy scale and sliding energy scale. Such a feature of LORE method may be understood better from the so-called folk's theorem emphasized by Weinberg\cite{SW3,SW1} that: any quantum theory that at sufficiently low energy and large distances looks Lorentz invariant and satisfies the cluster decomposition principle will also at sufficiently low energy look like a quantum field theory. It indicates that in any case there should exist a characteristic energy scale (CES) $M_c$ to make the statement of sufficiently low energy become meaningful. In general,  the CES $M_c$ can be either a fundamental-like energy scale (such as the string scale $M_s$ in string theory) or a dynamically generated energy scale of effective theories (like the chiral symmetry breaking scale $\Lambda_{\chi}$ in chiral perturbation theory and the critical temperature in superconductivity). Furthermore, the idea and analysis of renormalization group developed by Wilson\cite{KGW} and Gell-Mann-Low\cite{GML} allow one to deal with physical phenomena at any interesting energy scale by integrating out the physics at higher energy scales, which implies that one can define the renormalized theory at any interesting renormalization scale. It further indicates the existence of both characteristic energy scale (CES) $M_c$ and sliding energy scale(SES) $\mu_s$ which is not related to masses of particles or fields and can be chosen to be at any scale of interest.  Actually, the physical effects above the CES $M_c$ are integrated in the renormalized couplings and fields. 

To realize the above ideas, it needs to work out a consistent regularization prescription. A consistent regularization prescription without modifying the original theories is achieved by operating on the ILIs, so that the concept of ILIs becomes crucial in LORE. The regularization prescription is simple: firstly rotating the momentum to the Euclidean space by a Wick rotation, then replacing the loop integrating variable $k^2$ and the loop integrating measure $\int{d^4k}$ of the ILIs by the corresponding regularized ones $[k^2]_l$ and $\int[d^4k]_l$:
\begin{eqnarray}
 k^2 & \rightarrow & [k^2]_l \equiv k^2+M^2_l\ ,  \\
 \int{d^4k}\  {\cal F}(k^2)  & \rightarrow & \int[d^4k]_l \   {\cal F}(k^2 +M^2_l )  
\equiv  \int{d^4k}  \lim_{N, M_i^2}\sum_{l=0}^{N}c_l^N \  {\cal F}(k^2 +M^2_l )  \nonumber \\
&  = & \lim_{N, M_i^2}\sum_{l=0}^{N}c_l^N \  \int{d^4k}  \  {\cal F}(k^2 +M^2_l )
\end{eqnarray}
where $M_l^2$ ($ l= 0,1,\ \cdots $) are regarded as regulator masses and ${\cal F}(k^2)$ represents any integration function. The notation $\lim_{N, M_i^2}$ denotes the limiting case $\lim_{N\to \infty, M_i^2\rightarrow \infty}$ ($i=1,2,\cdots, N$). Note that the order of the integration and limiting operations is exchanged in the last step as it is supposed that the regularized integral has been well-defined with such a regularization scheme.

The coefficients $c_l^N$ are chosen with a postulation that a loop divergence with the power counting dimension larger than or equal to the space-time dimension vanishes, which is manifest based on the fact that any loop divergence has a power counting dimension less than the space-time dimension. Such a postulation means that
\begin{eqnarray}
 \int d^4 k \  \lim_{N, M_i^2}\sum_{l=0}^{N}c_l^N  (k^2 + M_l^2)^n = 0, \quad  (n= 0, 1, \cdots)\label{cl conditions}
\end{eqnarray}
which  leads to the following conditions for regulators
\begin{eqnarray}
\sum_{l=0}^{N}c_l^N  (M_l^2)^n = 0, \quad  (n= 0, 1, \cdots)\label{cl conditions}
\end{eqnarray}
with the initial conditions $M_0^2 \equiv \mu_s^2 = 0$ and $c_0^N = 1$, which are required to recover the original integrals in the limits $M_i^2 \to \infty$ ($i=1,2,\cdots,N$ ) and $N\to \infty$.

To yield the simplest solution of conditions given in Eq. (\ref{cl conditions}), so that the coefficients $c_l^N$ can completely be determined and independent of the regulator masses,  it is natural to take the string-mode regulators
\begin{equation}
M_l^2=\mu_s^2+lM_R^2, \qquad l=0,1,2,\cdots 
\end{equation}
which leads the coefficients $c_l^N$ to be uniquely determined to be 
\begin{equation}
 c_l^N=(-1)^l\frac{N!}{(N-l)!l!}\label{mus}
\end{equation}
which is a sign-associated Combinations. Actually, $(-1)^l c_l^N$ is the number of combinations of $N$ regulators taken $l$ at a time.

When applying the above regularization prescription and solution of regulators to ILIs, we can regularize divergent ILIs to be the well-defined ILIs in the Euclidean space-time:
\begin{eqnarray}
I_{-2\alpha}^R&=& i (-1)^{\alpha} \lim_{N,
M_i^2}\sum_{l=0}^{N}c_l^N\intdk\frac{1}{(k^2 + M^2 + M_l^2)^{2+\alpha}},\nonumber\\
I_{-2\alpha\ \mu\nu}^R&=& -i (-1)^{\alpha} \lim_{N,
M_i^2}\sum_{l=0}^{N}c_l^N\intdk\frac{k_{\mu}k_\nu}{(k^2+M^2+M_l^2)^{3+\alpha}},  \\
I_{-2\alpha\ \mu\nu\rho\sigma}^R&=& i (-1)^{\alpha} \lim_{N,
M_i^2}\sum_{l=0}^{N}c_l^N\intdk\frac{k_{\mu}k_{\nu}k_{\rho}k_{\sigma}}{(k^2+M^2+M_l^2)^{4+\alpha}} \nonumber 
\end{eqnarray}
with $\alpha=-1,0,1,\cdots $ and $i=1,2,\cdots$. Where  the subscript $``R"$ means well redefined or regularized divergent ILIs. To be more explicit, for the regularized quadratically and logarithmically divergent ILIs $I_2^R$ and $I_0^R$,  we can safely carry out the integration and obtain the following finite results\cite{YLWU1}:
\begin{eqnarray}
I_2^R&=&\frac{-i}{16\pi^2}\{M_c^2-\mu_M^2 - \mu_M^2[\ln\frac{M_c^2}{\mu_M^2}-\gamma_E+ \varepsilon (\frac{\mu_M^2}{M_c^2}) + \varepsilon^{\prime}(\frac{\mu_M^2}{M_c^2}) -1 ]\}  \\
I_0^R&=&\frac{i}{16\pi^2}[\ln\frac{M_c^2}{\mu_M^2}-\gamma_E+  \varepsilon (\frac{\mu_M^2}{M_c^2})],\qquad \mu_M^2=\mu_s^2+M^2
\end{eqnarray}
where $M_c$ is an intrinsic mass scale which is defined as
\begin{equation}
M_c^2\equiv \lim_{N,M_R} M_R^2 \sum_{l=1}^{N}c_l^N(l \ln l) =\lim_{N,M_R\to \infty } \left( \frac{M_R^2}{\ln N} \right) 
\end{equation}
and $\gamma_E$ is the Euler constant and defined here as
\begin{equation}\label{y}
 \gamma_E \equiv \lim_N \gamma_W(N) = \lim_{N}\{ \ \sum_{l=1}^{N} c_l^N \ln l +
     \ln [\ \sum_{l=1}^{N} c_l^N\ l \ln l \ ] \} =0.577215 \cdots \ .
 \end{equation}
The special function $\varepsilon(x)$ with $x=\mu_M^2/M_c^2$ and $\varepsilon^{\prime}(x) \equiv \partial_x \varepsilon (x)$ is introduced via the following definition
\begin{eqnarray}
 \varepsilon (x) & = & - \lim_{N,M_R}\sum_{l=1}^{N} c_l^N\
     \ln (1 + \frac{\mu_M^2}{lM_R^2} ) = \lim_{N\to \infty} \sum_{n=1}^{\infty} \frac{(-)^{n-1} L_n(N) }{n\ n!} x ^n \nonumber \\
& = & \sum_{n=1}^{\infty} \frac{(-)^{n-1} }{n\ n!} x ^n = \int_0^x d\sigma \frac{1-e^{-\sigma}}{\sigma}
\end{eqnarray}
which is an incomplete gamma function with the property $\varepsilon(x) \to x\ $ at $x\to 0$. In obtaining these results, we have used the following interesting functional limits introduced in\cite{YLWU1}
      \begin{eqnarray}
      W_N & \equiv &  \sum_{l=1}^{N} c_l^N \ ( l \ln l )  =  \sum_{l=1}^{N} (-1)^l \frac{N!}{(N-l)!\ l!}
       \ ( l \ln l )  \stackrel{\rm N\rightarrow \infty}{=} 1/ \ln N  \\
      \Gamma_N & \equiv & \sum_{l=1}^{N} c_l^N \  \ln l = \sum_{l=1}^{N} (-1)^l \frac{N!}{(N-l)!\ l!}\  \ln l
     \stackrel{\rm N\rightarrow \infty}{=} \ln \ln N + \gamma_E \\
     E_N^{(n)} & \equiv & -\sum_{l=1}^{N} c_l^N \  \frac{1}{l^n}  =  \sum_{l=1}^{N} (-1)^{l-1} \frac{N!}{(N-l)!\ l!}\  \frac{1}{l^n}\stackrel{\rm N\rightarrow \infty}{=} \frac{(\ln N)^n }{n!}
     \end{eqnarray}
which are applied to yield the functional limit
\begin{equation}
L_n = \lim_{N}L_n(N) \equiv - \lim_{N}\sum_{l=1}^{N} c_l^N\frac{n!}{l^n}\
    [\ \sum_{l'=1}^{N} c_{l'}^N\ l' \ln l' \ ]^n = 1 
\end{equation}
with $n = 1, 2 \cdots$.

It is interesting to note that $M_c$ provides an UV `cutoff', and $\mu_s $ sets an IR `cutoff' when $M^2 =0$.  In a theory without infrared divergence, $\mu_s$ can safely run to $\mu_s=0$.  In general, the mass scale $M_c$ can be made finite when taking appropriately both the primary regulator mass $M_R$ and regulator number $N$ to approach infinity in such a way that the ratio $M_R^2/\ln N$ is kept to be a finite quantity. Namely, one can reasonably require the primary regulator mass square $M^2_R$ going to be infinity logarithmically via $\ln N$ as $N\to \infty$. In fact, taking the primary regulator mass $M_R$ to be infinity is necessary to recover the original integrals, and setting the regulator number $N$ to be infinity is needed to make the regularized theory independent of the regularization prescription.

So far we arrive at a truly infinity-free LORE method. With the simple regularization prescription in LORE, one can easily prove by an explicit calculation that the regularized divergent ILIs get the consistent relations in a covariant form\cite{YLWU1}
\begin{eqnarray}\label{NR}
& & I^R_{2\mu\nu} = \frac{1}{2} g_{\mu\nu}\ I^R_2; \qquad I^R_{0\mu\nu} = \frac{1}{4} g_{\mu\nu} \ I^R_0 
\end{eqnarray}
which becomes manifest that the LORE method also maintains the original divergent structure of integrals when taking $M_c\to \infty$ and $\mu_s\to 0$.

 In general, we may come to the conclusion that any consistent regularization scheme which can provide the well redefined or regularized divergent tensor-type and scalar-type ILIs must  result in the consistent relations 

 In comparison with the dimensional regularization, there is a correspondence: $\ln \frac{M_c^2}{\mu^2} \to \frac{2}{\epsilon}$ with $M_c \to \infty$ and $\epsilon=4-d \to 0$, which indicates that the function $\varepsilon(x)$ approaches to zero much faster than the polynomial of $\epsilon$ in the dimensional regularization. This can be seen explicitly from the expression: $\varepsilon(x) \simeq x \sim e^{-\frac{2}{\epsilon}} \to 0 $ in the limit $M_c \to \infty$ and $\epsilon \to 0$. 
 
 On the other hand,  there are two distinguishing features between the LORE method and dimensional regularization: one is that LORE method is in principle an infinity-free regularization as the intrinsic UV cut-off mass scale $M_c$ can be made finite, whereas in dimensional regularization the extended dimension parameter $\epsilon = 4-d$ must be taken to be zero $\epsilon =0$ in order to recover the original theory at 4-dimensional space-time $d=4, \epsilon=0$, thus the divergences of Feynman integrals cannot in general be avoided by using dimensional regularization without modifying the original theory; the other is that LORE method maintains the divergent structure of original integrals, whereas dimensional regularization cannot keep the  divergent structure of original integrals as it suppresses all divergences to the logarithmic divergence due to the mathematical identity $\int d^dk (k^{2})^n =0$ ($n\geq -1$). It shows the advantage of LORE method as the quadratic structure is involved for scalar or Higgs interactions\cite{HW1} and plays an important role for effective field theories with dynamically generated spontaneous symmetry breaking\cite{DW}.

Before proceeding, we would like to address an important issue in all regularization schemes that for a divergent integral it is in general not appropriate to shift the integration variables. While in evaluating the ILIs in LORE,  it often needs to shift the integration variables before making the regularization prescription, which is justified as the LORE method is translational invariance. In fact, one can take the regularization prescription in LORE before shifting the integration variables, and the resulting consequence is the same as the one when shifting the integration variables first, which may be examined by the following logarithmic divergent Feynman integral:
\begin{eqnarray}
L={\intdk}\frac{1}{k^2-m_1^2}\frac{1}{(k-p)^2-m_2^2}
\end{eqnarray}
Following the LORE method, we shall first evaluate the Feynman integral into an ILI. By using the Feynman parametrization method, the Feynman integral can be written as follows 
\begin{eqnarray}
L&=&{\intdk}\int_0^1dx\frac{1}{\{(1-x)(k^2-m_1^2)+x[(k-p)^2-m_2^2]\}^2}\nonumber\\
&=&\int_0^1dx{\intdk}\frac{1}{( (k-xp)^2 -M^2)^2}
\end{eqnarray}
with 
\[ M^2=(1-x)m_1^2+xm_2^2-x(1-x)p^2\] \  .
When shifting the integration variable, we yield the standard scalar type ILI
\begin{eqnarray}
L&=& \int_0^1dx{\intdk}\frac{1}{( k^2 -M^2)^2} = \int_0^1dx\  I_0  \  .
\end{eqnarray}
After making a Wick rotation and applying the regularization
prescription in LORE, one can obtain the regularized Feynman integral as follows
\begin{eqnarray}
L^R= i \int_0^1dx\lim_{N, M_i^2}\sum_{l=0}^{N}c_l^N\intdk\frac{1}{(k^2+M^2+M_l^2)^2}  \   .
\end{eqnarray}

On the other hand, one may first apply for the regularization
prescription of LORE before shifting the integration variable, 
$(k-xp)^2 \to (k-xp)^2 + M_l^2$, one then has
\begin{eqnarray}
L^{\prime R}= i \lim_{N, M_i^2}\sum_{l=0}^{N}c_l^N\intdk\frac{1}{[(k-xp)^2+M^2+M_l^2]^2}
\end{eqnarray}
which is considered to be a well-defined integral, so that one can safely shift
the integration variable:
\begin{eqnarray}
L^{\prime R}=\int_0^1dx\lim_{N,
M_l^2}\sum_{l=0}^{N}c_l^N\intdk\frac{1}{(k^2+M^2+M_l^2)^2} \equiv
L^R
\end{eqnarray}
which shows that one can safely shift the integration variables and evaluate any Feynman integrals into ILIs before applying for the regularization prescription of LORE.  In fact, it was shown in the calculation of triangle anomaly that even for the linear divergent integral, one should first make a shift of integral variable in order to avoid the ambiguities and obtain a consistent result\cite{Ma:2005md}.

\section{Gauge Invariance Consistency Conditions }

The most important features required for a consistent regularization scheme are that the regularization method should preserve the basic symmetry principle of original theory, such as gauge invariance, Lorentz invariance and translational invariance, and meanwhile maintain the initial but well-defined divergent structure of original theory. 

The LORE method shows its advantages as it can lead to a set  of gauge invariance consistency conditions with maintaining the divergent structure of original theories. Such consistency conditions are presented by the consistent relations between the regularized tensor-type and scalar-type ILIs\cite{YLWU1}:
\begin{eqnarray}\label{CC}
& & I_{2\mu\nu}^R = \frac{1}{2} g_{\mu\nu}\ I_2^R, \quad
I_{2\mu\nu\rho\sigma }^R = \frac{1}{8} (g_{\mu\nu}g_{\rho\sigma} +
g_{\mu\rho}g_{\nu\sigma} +
g_{\mu\sigma}g_{\rho\nu})\ I_2^R  , \nonumber \\
& & I_{0\mu\nu}^R = \frac{1}{4} g_{\mu\nu} \ I_0^R, \quad
I_{0\mu\nu\rho\sigma }^R = \frac{1}{24} (g_{\mu\nu}g_{\rho\sigma} +
g_{\mu\rho}g_{\nu\sigma} + g_{\mu\sigma}g_{\rho\nu})\ I_0^R .
\end{eqnarray}
which become the necessary and sufficient conditions to ensure the gauge symmetry in QFTs. 

Note that the dimensional regularization scheme also leads to the same conditions as it is known to preserve gauge invariance, while the resulting $I^{R}_2$ in dimensional regularization is suppressed to be a logarithmic divergence multiplying by the mass scale ${\cal M}^2$, or vanishes $I^{R}_2=0$ when ${\cal M}^2 =0$. 

To see explicitly how the above consistency conditions are necessary and sufficient for preserving gauge invariance, let us first examine the calculations for QED and QCD vacuum polarization diagrams\cite{YLWU1}.  In general, we may consider the gauge theory with Dirac spinor fields $\psi_n$ ($n=1,\cdots, N_f$) interacting with Yang-Mill gauge fields $A_{\mu}^a$ ($a=1, \cdots, d_G$). Here $d_G =1$ is U(1) group for QED and $d_G=8$ is SU(3) group for QCD. The Lagrangian is given by
   \begin{eqnarray} \label{GT}
  {\cal L}  =  \bar{\psi}_n (i\gamma^{\mu}D_{\mu} - m) \psi_n
   - \frac{1}{4} F^a_{\mu\nu}F_a^{\mu\nu}
   \end{eqnarray}
 where 
 \begin{eqnarray}
  & & F_{\mu\nu}^a  =  \partial_{\mu} A_{\nu}^a - \partial_{\nu} A_{\mu}^a
  -g f_{abc}A_{\mu}^b A_{\nu}^b \nonumber  \\
  & & D_{\mu}\psi_n  = (\partial_{\mu} + ig T^a A_{\mu}^a)\psi_n
   \end{eqnarray}
with $T^a$ the generators of gauge group and $f_{abc}$ the structure function of the gauge group $ [T^a, \ T^b ] = i f_{abc} T^c $. To quantize the gauge theory, by adding the gauge fixing term and introducing the corresponding Faddeev-Popov ghost term with the ghost fields $c^a$ to fix the gauge. In the covariant gauge, a Lagrangian ${\cal L'}$ with the following form has to be added
 \begin{eqnarray}
{\cal L'} = -\frac{1}{2\xi} (\partial^{\mu} A_{\mu}^a )^2 + \partial^{\mu}\bar{c}^a (
\partial_{\mu} c^a + g f_{abc}c^b A_{\mu}^c )
 \end{eqnarray}
 where $\xi$ is an arbitrary parameter. Thus the completed Lagrangian is given by
 \begin{eqnarray}\label{GTQCD}
 \hat{{\cal L}} & = & {\cal L}+{\cal L'} = \bar{\psi}_n (i\gamma^{\mu}D_{\mu} - m) \psi_n
   - \frac{1}{4} F^a_{\mu\nu}F_a^{\mu\nu} \nonumber \\
   & - & \frac{1}{2\xi} (\partial^{\mu} A_{\mu}^a )^2 + \partial^{\mu}\bar{c}^a (
\partial_{\mu} c^a + g f_{abc}c^b A_{\mu}^c ) \  .
  \end{eqnarray}
 Based on this whole Lagrangian,  one can derive the Feynman
 rules for propagators and vertex interactions.
 
Let us evaluate the vacuum polarization diagrams of gauge fields at one-loop order. There are four non-vanishing one-loop diagrams (see Figures (1a)-(1d))

\begin{figure}[ht]
\begin{center}
\includegraphics[scale=0.5]{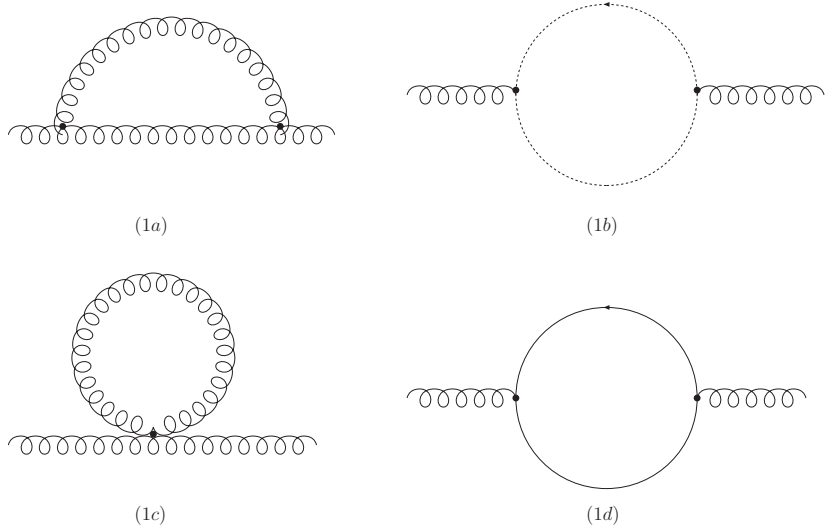}
 \caption{One loop vacuum polarization diagrams}\label{01}
\end{center}
\end{figure}

We may denote their contributions to the vacuum polarization function as $\Pi_{\mu\nu}^{(i)ab}$ ($i=1,2,3,4$) respectively. The first three diagrams (Fig. (1a)-(1c)) arise from pure Yang-Mills gauge interactions, their contributions to the vacuum polarization function are labeled as  $\Pi_{\mu\nu}^{(g) an} \equiv \Pi_{\mu\nu}^{(1) ab} + \Pi_{\mu\nu}^{(2) ab} + \Pi_{\mu\nu}^{(3) ab} $. The diagram in Fig.(1d) is from the fermionic loop and its contribution to the vacuum polarization function is denoted by $\Pi_{\mu\nu}^{(f) ab} = \Pi_{\mu\nu}^{(4)ab}$, which is the case in QED. The total contributions to the vacuum polarization function are given by summing over all the four diagrams
  \begin{eqnarray}
  \Pi_{\mu\nu}^{ab} = \sum_{i=1}^4 \Pi_{\mu\nu}^{(i) ab}
  \equiv \Pi_{\mu\nu}^{(g) ab} + \Pi_{\mu\nu}^{(f) ab}  \  .
  \end{eqnarray}
Gauge invariance means $k^{\mu} \Pi_{\mu\nu}^{ab} = \Pi_{\mu\nu}^{ab} k^{\nu} = 0$ which holds for any gauge theories with arbitrary fermion number $N_f$, which indicates that both parts $\Pi_{\mu\nu}^{(f) ab}$ (like QED) and $\Pi_{\mu\nu}^{(g) ab}$ (like QCD) should satisfy the generalized Ward identities
  \begin{eqnarray}
 & &  k^{\mu} \Pi_{\mu\nu}^{(f) ab} = \Pi_{\mu\nu}^{(f) ab} k^{\nu} = 0\ , \qquad 
 k^{\mu} \Pi_{\mu\nu}^{(g) ab} = \Pi_{\mu\nu}^{(g) ab} k^{\nu} = 0  \  .
   \end{eqnarray}
   
In terms of ILIs, the vacuum polarization function  $\Pi_{\mu\nu}^{(f) ab}$ from fermionic loop has the following simple form\cite{YLWU1}
  \begin{eqnarray}
    \Pi_{\mu\nu}^{(f) ab}
     & = & -g^2 4N_f C_2  \delta_{ab} \  \int_{0}^{1} dx\ [\ 2 I_{2\mu\nu} (m)
     - I_2(m) g_{\mu\nu} \nonumber \\
     & & + 2x(1-x) (p^2 g_{\mu\nu} - p_{\mu}p_{\nu} ) I_0(m) 
    \end{eqnarray}
where the gauge invariance is spoiled by the quadratic divergent ILIs and can be preserved only when the regularized ILIs satisfy the consistency condition
    \begin{equation}
  I_{2\mu\nu}^R(m) = \frac{1}{2} I_{2}^R(m) \ g_{\mu\nu}  \  . \nonumber   
  \end{equation}
Under this condition the regularized vacuum polarization function $\Pi_{R \mu\nu}^{(f) ab}$ becomes gauge invariant and takes the simple form
  \begin{eqnarray}
    \Pi_{R \mu\nu}^{(f) ab} = -g^2 4N_f C_2  \delta_{ab} \
    (p^2 g_{\mu\nu} - p_{\mu}p_{\nu} ) \ \int_{0}^{1} dx\  2x(1-x) I_0^R(m)  \  .
  \end{eqnarray}

The vacuum polarization function $\Pi_{\mu\nu}^{(g) ab}$ for the Yang-Mills gauge fields receives contributions from three diagrams. By summing over all contributions and expressing the tensor type ILIs in terms of the scalar type ones with parameters $a_0$ and $a_{-2}$, the gauge field vacuum polarization function $\Pi_{\mu\nu}^{(g) ab} $ can be written as follows\cite{YLWU1} 
  \begin{eqnarray}
 & & \Pi_{\mu\nu}^{(g) ab}  =  g^2 C_1 \delta_{ab}(p^2g_{\mu\nu} - p_{\mu}p_{\nu})
  \ \int_{0}^{1} dx\ \{ \  [1 + 4x(1-x)]\ I_0\  \nonumber \\
  & & + \frac{1}{4}\lambda \Gamma(3)\ [\ \left(\ 1 + 6x(1-x)(a_0 + 2) - 3a_0 \right)
  I_{0}\   \\
  & & - 2x(1-x) \left(\ 1 + 12x(1-x)\ \right) p^2\ I_{-2} \ ]  +  \frac{1}{8}\lambda^2 \Gamma(4)\ a_{-2}\ x(1-x)\ p^2\ I_{-2}\ \} \nonumber  \\
   & & + g^2 C_1 \delta_{ab}\ \int_{0}^{1} dx\ \{\ 2(\ 2I_{2\mu\nu} - I_{2}g_{\mu\nu}\ )
  +  \lambda \Gamma(3)\frac{a_0 -1}{2}\ p_{\mu}p_{\nu}\ x(1-x)\ p^2 \ I_{-2}\ \} \nonumber 
   \end{eqnarray}
which shows that both quadratically divergent integrals and logarithmically divergent term can in general destroy the gauge invariance. 
It is manifest that the gauge invariance can be preserved only when the regularized divergent ILIs satisfy the consistency conditions 
  \begin{eqnarray}
   & & I_{2\mu\nu}^R = \frac{1}{2} I_{2}^R\ g_{\mu\nu}; \qquad I_{0\mu\nu}^R = \frac{1}{4} I_{0}^R\ g_{\mu\nu}   \  .  \nonumber
   \end{eqnarray}

After adopting the consistency conditions, the regularized gauge field vacuum polarization function $\Pi_{\mu\nu}^{ab} $ gets the gauge invariant form
 \begin{eqnarray}
  \Pi_{R \mu\nu}^{ab} & = & \Pi_{R \mu\nu}^{(g) ab} + \Pi_{R \mu\nu}^{(f) ab} \nonumber \\
  & = & g^2\delta_{ab}\ (p^2g_{\mu\nu} - p_{\mu}p_{\nu}) \ \int_{0}^{1} dx\ \{ \
  C_1\ [\ 1 + 4x(1-x) + \lambda/2 \ ] \ I_0^R\  \nonumber \\
   & - & N_f C_2\ 8x(1-x)\  I_0^R(m)  - 4C_1 \lambda\ [\
   1 - \lambda /8\  ] \ x(1-x)\ p^2\ I_{-2}^R\  \  \}  \  .
 \end{eqnarray}
 
From such an example, it is seen that the quadratically divergences may not necessarily be a harmful source for the gauge invariance as they eventually cancel each other as long as they satisfy the consistency conditions. In contrast to dimensional regularization in which the quadratically divergent tadpole graphs vanish due to the analytical extension of space-time dimension, whereas the tadpole graph of gauge fields in LORE plays an essential role for maintaining the gauge invariance. Actually, it is the tadpole graph that leads to the manifest gauge invariant form of the vacuum polarization function when keeping the divergence structure of original integrals. 
 
In general, once the consistency conditions for the regularized ILIs hold, the divergent structure of the theories can be well characterized by two regularized scalar type ILIs $I_0^R$ and $I_2^R$. The quadratic term $I_2^R$ for gauge interactions cancel each other due to gauge invariance, only the logarithmic term $I_0^R$ is left and the theory can be regularized with the redefinitions of coupling constants and quantum fields. In this case, one can in principle arrive at a regularization-independent scheme by just introducing the ILIs and applying for the consistency conditions. Mathematically, one only needs to prove the existence of a consistent regularization which can result in the consistency conditions between the regularized ILIs.

\section{Overlapping Divergence Structure and UVDP Parametrization}

To deal with consistently and systematically the divergences in QFTs, a more careful treatment has to be paid for Feynman diagrams beyond one-loop order as it concerns a new feature occurring in {\em overlapping} structure of high loop Feynman diagram, which happens when two divergent loops share a common propagator. For that, it turns out to be very useful to introduce the so-called UVDP parametrization in evaluating the ILIs from Feynman diagrams\cite{YLWU1,HW1}. 

To illustrate the new feature arising from overlapping divergent structure, we may consider one particular contribution to the photon vacuum polarization diagrams at two-loop order in QED (see Fig. 2) 
\begin{figure}[ht]
\begin{center}
  \includegraphics[scale=0.9]{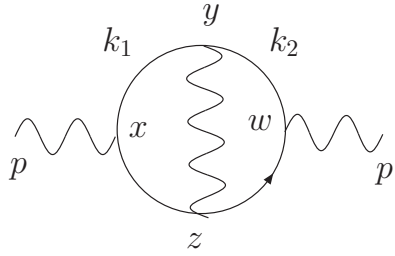}
  \caption{Two-Loop Vacumm Polarization }\label{qedB1}
\end{center}
\end{figure}

\begin{figure}[ht]
\begin{center}
  \includegraphics[scale=0.9]{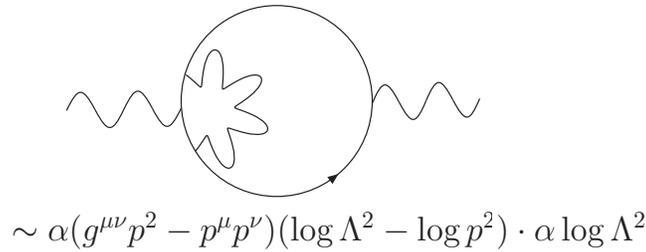}\\
  \caption{Overlapping Divergence Structure }\label{vertex insertion}
\end{center}
\end{figure}

\begin{figure}[ht]
\begin{center}
  \includegraphics{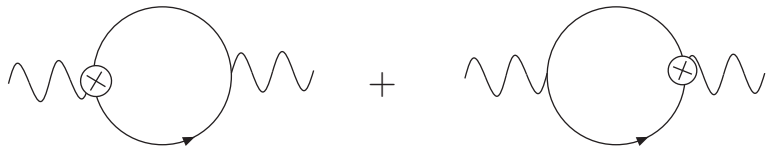}\\
  \caption{Conterterm Diagrams}\label{counterinsertion}
\end{center}
\end{figure}

\begin{figure}[ht]
\begin{center}
  \includegraphics[scale=0.8]{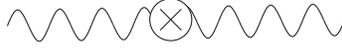}
  \caption{Overall or Local Counterterm}\label{counterlocal}
\end{center}
\end{figure}

As described in the usual textbooks of QFTs\cite{Peskin:1995ev}, the divergences in the two-loop photon vacuum polarization diagram shown in Fig. 2 can arise from three regions of momentum spaces. One of divergent contributions to the diagram in Fig.2 comes from the region where there is a large momentum passing through the left subdiagram, which indicates that the three points $x,y,$ and $z$ in position space are very close together, while the point $w$ must be farther away. In this region, the virtual photon gives large corrections to the vertex $x$. Inserting the divergent part of one-loop vertex corrections into the rest of diagram and integrating over the momentum $k_1$, which will give the expression identical to the one-loop photon vacuum polarization correction multiplied by the additional logarithmic divergence, as it is shown in Fig.\ref{vertex insertion}. A similar divergent contribution to the diagram in Fig.1 comes from the region with a large momentum passing through the right subdiagram as shown in Fig. \ref{vertex insertion}.
It then brings the double logarithmic term $\log^2\Lambda^2$ in the region where both $k_1$ and $k_2$ become large. While the $\log p^2\log \Lambda^2$ term is resulted from the region where $k_2$ is large but $k_1$ is small. The same term can arise from the region where $k_1$ is large but $k_2$ is small. Such terms like $\log p^2\log \Lambda^2$ are called {\em nonlocal} or {\em harmful} divergences as they cannot be canceled by the ordinary substraction scheme by introducing the corresponding two loop counterterms in the Lagrangian. These divergences must be canceled by two types of counterterm diagrams. Thus one can build diagrams of order $\alpha^2$ by inserting the order-$\alpha$ counterterm vertex into the one-loop vacuum polarization diagram
(see Fig. \ref{counterinsertion}).
Such two diagrams are expected to cancel the harmful divergences shown in Fig.\ref{vertex insertion}. Once these counterterm diagrams are added, the remaining divergence becomes exactly local and can be canceled by the two-loop overall counterterm. It can diagrammatically be shown in Fig. \ref{counterlocal}.

The above description is general and standard in the textbooks and has no any question in principle. Nevertheless, to carry out the practical calculations for those diagrams, it raises some conceptional problems. One needs to integrate over two loop momentums $k_1$ and $k_2$ one by one. Suppose that one first integrates over the loop momentum $k_1$,  which corresponds to integrate over the left subdiagram with the left vertex insertion. One then integrates over the loop momentum $k_2$, which is actually the overall divergence of the whole diagram as indicated from its divergent behavior. It then comes to the question which loop momentum integral represents the right subdiagram and the corresponding correction to the right vertex. It seems  that there is nothing to do with it as one has already integrated over both loop momenta in the diagram. While it is noticed that when carrying out the calculations by using the Feynman parametrization and UVDP parametrization to combine the momenta in the denominator, the integrals for the UVDP parameters are actually logarithmic divergent, which is exactly equal to that of the vertex correction at one-loop order. It is then expected that the divergence of right subdiagram is actually converted into the parameter space. Thus the question becomes whether we can figure out, for a given divergence in the UVDP parameter space, the origin of such a divergence in the original Feynman diagrams. It has been shown in ref. \cite{HW1} that there does exist an exact correspondence between the UVDP parameter integrals and those from the original loop momenta. 

To demonstrate the correspondence of divergent structure between the UVDP parameter integrals and loop momentum integrals, it is very useful to consider the scalar-type ILIs. As discussed by 't Hooft and Veltman\cite{DR}, a general two-loop Feynman diagram can be reduced to the general $\alpha\beta\gamma$ integrals:
\begin{equation}\label{alpha-beta-gamma exp}
I_{\alpha\beta\gamma}=\int\frac{d^4k_1}{(2\pi)^4}\int\frac{d^4k_2}{(2\pi)^4}
\frac{1}{(k_1^2-m_1^2)^\alpha(k_2^2-m_2^2)^\beta[(-k_1-k_2+p)^2-m_3^2]^\gamma} \  ,
\end{equation}
where $m_i^2$ are in general the functions of the external momenta $p$ and Feynman parameters. Notice that the scalar-type overlapping divergence integrals in QED at two-loop level can be reduced to the following two types of integrals by using the Feynman parametrization:
\begin{eqnarray}
I_{111} &=& \int\frac{d^4k_1}{(2\pi)^4}\int\frac{d^4k_2}{(2\pi)^4}
\frac{1}{(k_1^2-m_1^2)(k_2^2-m_2^2)[(k_1-k_2+p)^2-m_3^2]} \  , \\
I_{121} &=& \int\frac{d^4k_1}{(2\pi)^4}\int\frac{d^4k_2}{(2\pi)^4}
\frac{1}{(k_1^2-m_1^2)(k_2^2-m_2^2)^2[(k_1-k_2+p)^2-m_3^2]}  \  ,
\end{eqnarray}
which are the two special cases of the general $\alpha\beta\gamma$ integrals with $\alpha =\beta=\gamma=1$ and $\alpha=\gamma = 1,\,  \beta=2$. 

Before making a general discussion and analysis on the regularization and renormalization for the general $\alpha\beta\gamma$ integrals, we may briefly describe the UVDP parametrization. Such a method is introduced for combining the denominator propagating factors similar to Feynman parameterization. The UVDP parametrization enables us to convert the divergences in the momentum space into the ones in the UVDP parameter space, which can well be regularized by the LORE method. For the simplest case with only two factors in the denominator, it can be combined by using the UVDP parametrization:
\begin{eqnarray}
\frac{1}{AB} &=& \int^\infty_0 \frac{du}{(1+u)^2}\frac{dv}{(1+v)^2}\delta(1-\frac{1}{1+u}-\frac{1}{1+v}) \frac{1}{[\frac{A}{1+u}+\frac{B}{1+v}]^2}
\end{eqnarray}
For a more general case, it can be expressed by the UVDP parametrization as following form:
\begin{equation}\label{UVDP gen}
\frac{1}{A_1^{m_1}A_2^{m_2}\cdots A_n^{m_n}} = \int^\infty_0 \prod^n_{i=1}\frac{dv_i}{(1+v_i)^2}\delta(\sum^n_{i=1}\frac{1}{1+v_i}-1)\frac{\prod^n_{i=1}\frac{1}{(1+v_i)^{m_i-1}}} {[\sum^n_{i=1}\frac{A_i}{1+v_i}]^{\sum^n_{i=1}m_i}}
\end{equation}

\begin{figure}[ht]
\begin{center}
  \includegraphics{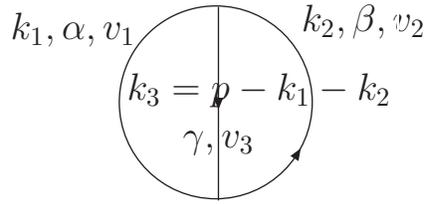}\\
  \caption{General $\alpha\beta\gamma$ diagram}\label{abcoverall}
\end{center}
\end{figure}

\begin{figure}[ht]
\begin{center}
  \includegraphics{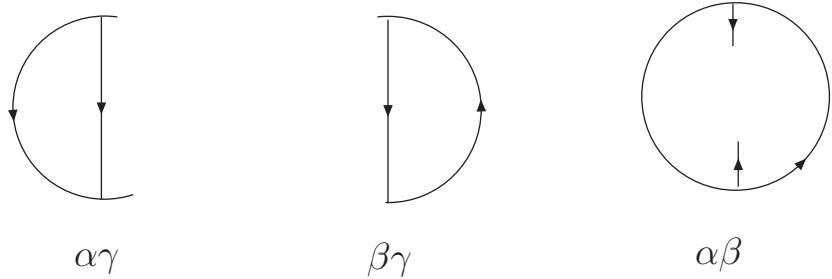}\\
  \caption{Diagrams for subdivergences}\label{subdiag}
\end{center}
\end{figure}

\begin{figure}[ht]
\begin{center}
  \includegraphics{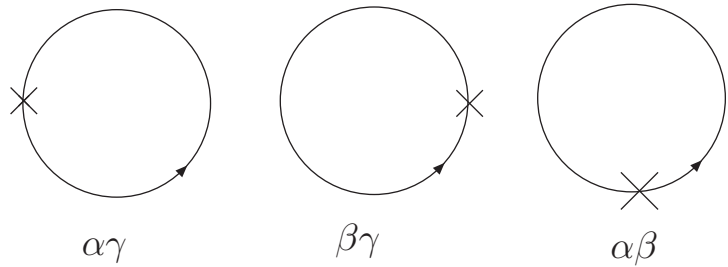}\\
  \caption{Diagrams for counterterms }\label{stdiag}
\end{center}
\end{figure}

Let us now pay attention how to disentangle the overlapping divergences and deal with consistently the divergences contained in the UVDP parameter space due to the overlapping structure. It is seen from the general form of Eq.(\ref{alpha-beta-gamma exp}) that there are generally one overall integral $\alpha\beta\gamma$ and three subintegrals ($\alpha\beta$, $\beta\gamma$ and $\gamma\alpha$) as represented diagrammatically with three corresponding subdiagrams (Fig.(\ref{abcoverall})and Fig.(\ref{subdiag})). Their corresponding counterterm diagrams shown in Fig.(\ref{stdiag}) are usually taken to cancel the harmful divergences.

The power counting to the general $\alpha\beta\gamma$ integral  indicates that the overlapping divergences occur with two cases: (i) $\alpha+\beta+\gamma=4$, and (ii) $\alpha+\beta+\gamma=3$.  To adopt the LORE method, it needs firstly to evaluate the general $\alpha\beta\gamma$ integral into ILIs. By applying for the UVDP parametrization and getting rid of the cross terms of momenta in the
denominator, the resulting ILIs are given as follows\cite{HW1}
\begin{eqnarray}
I_{\alpha\beta\gamma} &=& \int\frac{d^4k_1}{(2\pi)^4}\int\frac{d^4k_2}{(2\pi)^4} \frac{\Gamma(\alpha+\beta+\gamma)}{\Gamma(\alpha)\Gamma(\beta)\Gamma(\gamma)}\int^\infty_0\prod^3_{i=1}\frac{dv_i}{(1+v_i)^2} \delta(1-\sum^3_{i=1}\frac{1}{1+v_i}) \nonumber\\
&& \frac{\frac{1}{(1+v_1)^{\alpha-1}}\frac{1}{(1+v_2)^{\beta-1}}\frac{1}{(1+v_3)^{\gamma-1}}} {\{\frac{1}{1+v_1}(k_1^2-m_1^2)+\frac{1}{1+v_2}(k_2^2-m_2^2)+\frac{1}{1+v_3}[(-k_1-k_2+p)^2-m_3^2]\}^{\alpha+\beta+\gamma}} \nonumber\\
&=&
\frac{\Gamma(\alpha+\beta+\gamma)}{\Gamma(\alpha)\Gamma(\beta)\Gamma(\gamma)}
\int^\infty_0\prod^3_{i=1}\frac{dv_i}{(1+v_i)^{1+\alpha_i}}
\delta(1-\sum^3_{j=1}\frac{1}{1+v_j})\int\frac{d^4k_1}{(2\pi)^4}\int\frac{d^4k_2}{(2\pi)^4}
\nonumber\\
&& \frac{1}{[(\frac{1}{1+v_1}+\frac{1}{1+v_3}){k}_1^2 +\frac{3+v_1+v_2+v_3}{(2+v_1+v_2)(1+v_3)}{k}_2^2+\frac{1}{3+v_1+v_2+v_3}p^2-\sum^3_{j=1}\frac{m_j^2}{1+v_j}]^ {\alpha+\beta+\gamma}}  \nonumber \\
\end{eqnarray}
with $\alpha_i$ (i=1,2,3) denoting $\alpha$, $\beta$, $\gamma$ respectively. To obtain the result in the second equality, the following momentum replacement has been made as a consequence of momentum translation
\begin{eqnarray}
{k}_1 &\to & k_1+\frac{1+v_1}{2+v_1+v_3} k_2 - \frac{1+v_1}{3+v_1+v_2+v_3}p  \  ,  \nonumber \\
{k}_2 &\to & k_2-\frac{1+v_2}{3+v_1+v_2+v_3}p  \  ,
\end{eqnarray} 
which is convergent with respect to one of momentum integrations $k_i$s due to $\alpha+\beta+\gamma \geq 3$. After integrating over $k_1$ without losing generality, and making a scaling transformation for the momentum
\begin{equation} \label{scaling}
k_2^2 = \frac{(2+v_1+v_3)(1+v_2)}{3+v_1+v_2+v_3} l_+^2
\end{equation}
we can arrive at the ILIs with a more symmetric form
\begin{eqnarray}\label{abc int_ed0}
I_{\alpha\beta\gamma}  &=& \frac{i}{16\pi^2}
\frac{\Gamma(\alpha+\beta+\gamma-2)}{\Gamma(\alpha)\Gamma(\beta)\Gamma(\gamma)}
\int^\infty_0\prod^3_{i=1}\frac{dv_i}{(1+v_i)^2}
\delta(1-\sum^3_{j=1}\frac{1}{1+v_j})F(v_k)\nonumber\\
&& \int\frac{d^4 l_+}{(2\pi)^4} \frac{1}{[l_+^2-{\cal
M}^2(p^2,m_k^2,v_k)]^{\alpha+\beta+\gamma-2}}
\end{eqnarray}
with
\begin{eqnarray}
F(v_k)  &=&
\frac{(1+v_1)^{\alpha+1}(1+v_2)^{\beta+1}(1+v_3)^{\gamma+1}}
{(3+v_1+v_2+v_3)^{2}} \\
{\cal M}^2 &=& \sum^3_{j=1}
\frac{m_j^2}{1+v_j}-\frac{1}{3+v_1+v_2+v_3}p^2
\end{eqnarray}

It is seen that the momentum integral over $l_+$ corresponds to the overall divergence, which can easily be carried out to characterize the overall divergences. The overall divergence occurs in two cases, one for the logarithmic divergence with $\alpha+\beta+\gamma=4$ and the other for the quadratic divergence with $\alpha+\beta+\gamma=3$. To discuss the one-to-one correspondence of the divergences in UVDP parameter space and sub diagrams, it has to consider some explicit values of $\alpha, \beta,\gamma$.

\section{Divergence Treatment in the UVDP Parameter Space}

In QFTs, the theorem on the cancelation of harmful divergences caused from the overlapping structure is crucial as only the harmless divergences and finite terms can be absorbed into the overall counterterms. To demonstrate such a theorem, it is important to keep track of treating the overlapping divergences. At two loop level, the harmful divergences have the forms like $M_c^2\cdot\log\frac{M_c^2}{-p^2}$ and $\log \frac{M_c^2}{-p^2}\cdot \log\frac{M_c^2}{-p^2}$. 

To show explicitly how to treat the overlapping divergences in the UVDP parameter space, consider the case with $\alpha=\gamma=1, ~\beta=2$. The corresponding ILIs can simply be read off from the above general ILIs of $\alpha\beta\gamma$ integral
\begin{eqnarray}\label{I121}
I_{121} &=& \frac{i}{16\pi^2}\int^\infty_0\prod^3_{i=1}dv_i \delta(1-\sum^3_{j=1}\frac{1}{1+v_j}) \frac{1}{(3+v_1+v_2+v_3)^2(1+v_2)}\nonumber\\
&& \int\frac{d^4k_2}{(2\pi)^4} \frac{1}{[k_2^2-{\cal M}(p^2,m_k^2,v_k)]^2}\nonumber\\
&\to & -\frac{1}{(16\pi^2)^2}\int^\infty_0\prod^3_{i=1}dv_i \delta(1-\sum^3_{j=1}\frac{1}{1+v_j}) \frac{1}{(3+v_1+v_2+v_3)^2(1+v_2)}\nonumber\\
&& \cdot (\ln\frac{M_c^2}{\mu_M^2}-\gamma_E+\varepsilon(\frac{\mu_M^2}{M_c^2})) \   ,
\end{eqnarray}
where the integral over the loop momentum $k_2$ has a logarithmic divergence, which is an overall divergence and has been regularized by applying for the LORE method. For simplicity, we may only keep the quadratic and logarithmic parts and drop all the terms $\varepsilon(\frac{{\cal M}^2}{M_c^2})$ by taking $\mu_s =0$.

It is seen that the divergence in the region of UVDP parameter space at $v_1, v_3\rightarrow \infty$ reflects the divergence of subdiagram $\alpha\gamma$. To extract the divergence, it is useful to focus on the region $v_1, v_3 > v_o$ with $v_o\gg 1$, and $v_2\ll 1$ or $v_2\rightarrow 0$ which is ensured by the delta function. Thus the integration is well characterized in the domain $\int^\infty_{v_o} dv_1 \int^\infty_{v_o} dv_3 \int^\infty_0 dv_2$. With such a treatment,  some insignificant terms in comparison with $v_1$ and $v_3$ can be neglected and ${\cal M}\to m_2^2$, the above integral $I_{121}$ is simplified into the following form
\begin{eqnarray}
I_{121} &\simeq & -\frac{1}{(16\pi^2)^2}\int^\infty_{v_o} dv_1\int^\infty_{v_o} dv_3 \int^\infty_0 dv_2 \delta(1-\frac{1}{1+v_2})\nonumber\\
&& \frac{1}{(v_1+v_3)^2} (\ln\frac{M_c^2}{m_2^2}-\gamma_E)\nonumber\\
&=  & -\frac{1}{(16\pi^2)^2}
(\ln\frac{M_c^2}{m_2^2}-\gamma_E)\int^\infty_{v_o} dv_1
\frac{1}{v_1+v_o}
\end{eqnarray}
where the convergent integrations over $v_2$ and $v_3$ have been performed. The integration over $v_1$ becomes divergent and has to be regularized appropriately. As such a divergence is a kind of scalar-type divergent ILI in the UVDP parameter space, it is suitable to be regularized by the LORE method. To regularize the UVDP parameter integrals by applying for the LORE method, it is useful to convert them into a manifest form of ILI through multiplying a free mass-squared scale $q_o^2$ to the UVDP parameter $v_1$ and define the momentum-like integration variable $q_1^2\equiv q_o^2v_1$
\begin{eqnarray}\label{treatment}
\int^\infty_{v_o} d(q_o^2v_1)\frac{1}{q_o^2v_1+q_o^2v_o} & = & \int^\infty_{0} dq_1^2 \frac{1}{q_1^2+ \mu_o^2 } \nonumber\\
& \to & \ln\frac{M_c^2}{\mu_o^2 }-\gamma_E  
\end{eqnarray}
with $\mu_o^2 \equiv 2q_o^2v_o $. Where we have made the replacement $q_1^2 \to q_1^2 + q_o^2 v_o$ to shift the integrating region, and applied the LORE method to regularize the divergent ILIs in the UVDP parameter space. The free mass scale $\mu_o^2$ will be determined by a suitable criterion, such as the cancellation of harmful divergences of different diagrams. 

The above treatment can be extended to any divergent UVDP parameter integrals. Thus such a prescription in LORE enables us to treat all divergent ILIs in the UVDP parameter space.  

With the above analysis, the general form of overlapping divergence in the integral $I_{121}$ is given by
\begin{equation}\label{I_121}
I_{121} \simeq
-\frac{1}{(16\pi^2)^2} (\ln\frac{M_c^2}{m_2^2}-\gamma_E) \cdot(\ln\frac{M_c^2}{\mu_o^2}-\gamma_E) \  .
\end{equation}

To demonstrate the exact cancelation of harmful divergence in $I_{121}$, it needs to consider the corresponding counterterm
diagram ($\alpha\gamma$) (Fig. (\ref{stdiag})) which leads to the integral
\begin{equation}\label{I_121 alpha-gamma}
I_{121}^{(c)(\alpha\gamma)} = - \int\frac{d^4k_2}{(2\pi)^4}\frac{1}{(k_2^2-m_2^2)^2}\textsc{DP}
\{\int\frac{d^4k_1}{(2\pi)^4} \frac{1}{(k_1^2-m_1^2)[(k_1-k_2+p)^2-m_3^2]}\} \  ,
\end{equation}
with $\textsc{DP}\{\}$ representing the divergent part. One can easily carry out such a counterterm integral 
\begin{eqnarray}\label{I_121 alpha-gamma fin}
I_{121}^{(c)(\alpha\gamma)} = \frac{1}{(16\pi^2)^2} (\ln\frac{M_c^2}{m_2^2}-\gamma_E) \cdot(\ln\frac{M_c^2}{\mu^2}-\gamma_E) \  ,
\end{eqnarray}
where the first part from the integration of internal loop momentum $k_2$ and the second factor comes from the subintegral $(\alpha\gamma)$ part contained in $\textsc{DP}\{\}$.

As shown explicitly from the above integrated expressions, there does exist the exact correspondence between the UVDP parameter space and subdiagram. Once taking the free mass scale to be $\mu_o^2 = \mu^2 $, two divergent terms cancel each other exactly. It becomes manifest that the divergence in the UVDP parameter space at the region $v_1, v_3\rightarrow\infty$,$v_2\to 0$ reproduces that of subintegral $(\alpha\gamma)$ in the momentum space, i.e., the integration over $k_1$. It is interesting to notice that the divergences of $I_{121}$ can in general be factorized and written as the product of two divergent integrals which correspond to the overall divergence  from the integration over $k_2$  and the sub-divergence from the subintegral $k_1$ $(\alpha\gamma)$. The latter is converted into the divergence from the UVDP parameter integral in the region $v_1, v_3\rightarrow \infty$. 

So far we have shown the general feature when applying the LORE method to disentangle the two-loop overlapping divergences, it can be generalized to high loop diagrams.

\section{Evaluation of ILIs and Circuit Analogy of Feynman Diagrams}

As it is seen that the concept of ILIs and the introduction of UVDP parametrization are very useful in developing LORE method to treat overlapping divergences. To generalize the correspondence between the divergences in the UVDP parameter space and in the subintegrals to more complicated cases, it has been demonstrated in ref.\cite{HW1} that the evaluation of ILIs by adopting the UVDP parametrization method naturally leads to the Bjorken-Drell's analogy between the Feynman diagrams and the electrical circuits. Such an analogy was originally motivated for discussing the analyticity properties of Feynman diagrams from the causality requirement\cite{BD}. Though two motivations are different, it arrives at the same circuit analogy of Feynman diagrams. 

Let us first provide a standard procedure to evaluate systematically the ILIs in LORE and merge it with Bjorken-Drell's circuit analogy of Feynman diagrams. For that, we may follow the definitions and notations by Bjorken and Drell. For a general connected Feynman diagram, the external momenta of the diagram will always be denoted by $p_1,..., p_m$ with the direction of entering the diagram. Thus, the overall momentum conservation leads to the condition
\begin{equation}\label{pcons}
\sum^m_{s=1}p_s=0
\end{equation}
For each internal line, we may assign a momentum $k_j$ with a specified direction and a mass $m_j$. At each vertex, the law of momentum conservation gives the following conditions
\begin{equation}\label{kcons}
\sum^m_{s=1}\bar{\epsilon}_{is}p_s + \sum^n_{j=1}\epsilon_{ij}k_j =0
\end{equation}
where $\epsilon_{ij}$ is chosen to be $+1$ when the internal line $j$ enters vertex $i$, while $-1$ when the internal line $j$ leaves vertex $i$, otherwise $\epsilon_{ij}$ is defined to be 0. $\bar{\epsilon}_{is}$ has the similar definition for the external
lines which, by convention, are always taken to enter vertices.

For a given diagram which has a definite number $k$ of internal loops, one has the freedom to choose the concrete internal loops and assigns each loop a momentum $l_r$ which will be integrated out along the loop. For each internal line $j$, we can always make the following decomposition in terms of the loop momentum $l_r$
\begin{equation}\label{decomp}
k_j=q_j+\sum^k_{r=1}\eta_{jr}l_r  \  ,
\end{equation}
with $q_j$ being another kind of internal momentum introduced for the momentum conservation. Where $\eta_{jr}$ is chosen to be $+1$ if the $j$th internal line lies on the $r$th loop and the momenta $k_j$ and $l_r$ are parallel, and $-1$ if the $j$th line lies on the $r$th loop but $k_j$ and $l_r$ are antiparallel, otherwise $\eta_{jr}$ is 0. The internal momentum $q_j$ will be determined after we adopt the UVDP parametrization to evaluate the ILIs. From the decomposition Eq.(\ref{decomp}) and the conditions 
\begin{equation}
\sum^n_{j=1}\epsilon_{ij}\eta_{jr}=0  \  , 
\end{equation}
which is a consequence of the definitions of $\epsilon_{ij}$ and $\eta_{jr}$ given in Eqs. (\ref{kcons}) and (\ref{decomp}), we can immediately obtain the following momentum conservation laws for each vertex:
\begin{equation}\label{Vertex q}
\sum^m_{s=1}\bar{\epsilon}_{is}p_s + \sum^n_{j=1}\epsilon_{ij}q_j =0  \  .
\end{equation}

Let us begin with the general structure of Feynman integral:
\begin{equation}
I(p_1,...,p_m)=\int d^4l_1...d^4 l_k
\frac{N}{(k_1^2-m_1^2)^{\alpha_1}...(k_n^2-m_n^2)^{\alpha_n}}  \  ,
\end{equation}
where the numerator $N$ represents a general matrix element which can be the products of external momenta, internal momenta, spin matrices, wave functions. After adopting the UVDP parametrization, the above integral has the following form:
\begin{eqnarray}\label{genI}
I(p_1,...,p_m) &=& \int d^4l_1...d^4 l_k
\frac{\Gamma(\sum^{n}_{j=1}\alpha_j)}{\Gamma(\alpha_1)...\Gamma(\alpha_n)}\int^\infty_0
\prod^n_{i=1}\frac{dv_i}{(1+v_i)^{\alpha_i+1}}  \nonumber \\
& & \cdot  \delta(1-\sum^n_{j=1}\frac{1}{1+v_j}) \frac{N}{[\sum^n_{j=1}\frac{k_j^2-m_j^2}{1+v_j}]^{\sum^n_{j=1}\alpha_j}}\nonumber\\
&=& \frac{\Gamma(\sum^{n}_{j=1}\alpha_j)}
{\Gamma(\alpha_1)...\Gamma(\alpha_n)}  \int d^4l_1...d^4 l_k\  \int^\infty_0
\prod^n_{i=1}\frac{dv_i}{(1+v_i)^{\alpha_i+1}}
\delta(1-\sum^n_{j=1}\frac{1}{1+v_j})  \nonumber\\
&& \cdot  \frac{N}{[\sum^n_{j=1}\frac{q_j^2-m_j^2}{1+v_j}+2\sum_{j,r}\frac{q_j\eta_{jr}l_r}{1+v_j}
+\sum_{j,r,r'}\frac{\eta_{jr}\eta_{jr'}l_r
l_{r'}}{1+v_j}]^{\sum^n_{j=1}\alpha_j}}  \  .
\end{eqnarray}
To evaluate the ILIs, the cross terms in the denominator have to be eliminated, which leads to the following conditions: 
\begin{equation}\label{Loop q}
\sum^n_{j=1}\frac{\eta_{jr}q_j}{1+v_j}=0; \quad r=1,\cdots, k
\end{equation}

When combining the above conditions with the ones in Eq. (\ref{Vertex q}), we are able to determine the momenta $q_j$.  Such a procedure is equivalent to the shifting of the loop momenta. It is useful to get an alternative interesting understanding on Eqs.(\ref{Vertex q}) and (\ref{Loop q}) by putting them into a more heuristic form:
\begin{eqnarray}
& & \sum_{q_j,~p_s}~(q_j+p_s)=0,\quad \mbox{ all momenta entering~vertex~i }\label{KCurr} \\
& & \sum_{q_j }\frac{q_j}{1+v_j}=0,\qquad  \quad \mbox{ in~any~given~loop~r} \label{KVolt} 
\end{eqnarray}
\begin{figure}[ht]
  \includegraphics[scale=0.6]{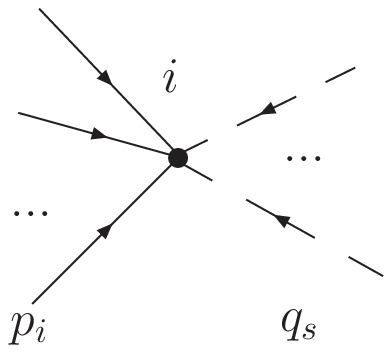}\label{Kvertex}~~~~~~~~~~~
  \includegraphics[scale=0.8]{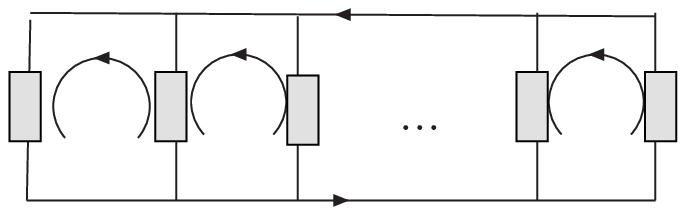}\\
  \caption{Current conservation for each vertex i and conservation of the voltage in any loop r}\label{Kloop}
\end{figure}
which shows the analogy between the Feynman diagrams and electrical circuits. 

In such a circuit analogy of Feynman diagrams, the momenta are associated with the currents, i.e., $q_j$ are the internal currents flowing in the circuit and $p_s$ the external currents entering it, and the UVDP parameters $\frac{1}{1+v_j}$ are associated with the resistance of the $j$th line, or $v_j$ can be regarded as the conductance of the $j$th line. Thus Eqs. (\ref{KVolt}) and (\ref{KCurr}) correspond to the Kirchhoff's laws, i.e.,  Eq. (\ref{KVolt}) means that the sum of ``voltage drop" around any closed loop is zero, and Eq. (\ref{KCurr}) shows that the sum of ``currents" flowing a vertex is zero. The positivity of the UVDP parameter $v_i$ as the ``conductance" is related to the
causality of propagation for the free particles.\\
\begin{figure}[ht]
\begin{center}
  \includegraphics{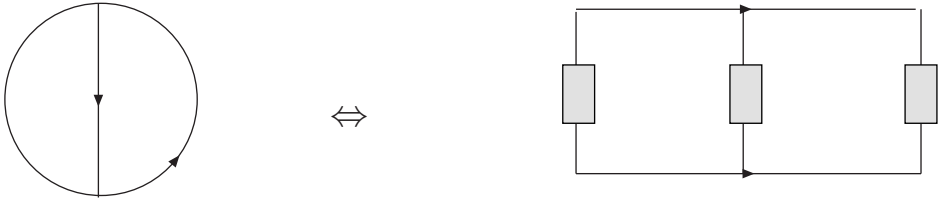}\\
  \caption{Electrical circuit analogy of Feynman diagrams}\label{}
\end{center}
\end{figure}

To yield the standard form of ILIs, it needs further to make the quadratic terms of the momentum $l_r$ be diagonal. This can be reached by an orthogonal transformation $O$ 
\begin{equation}
L = O L^\prime, \qquad O^T M O = diag(\lambda_1, \cdots, \lambda_k) \equiv diag(\lambda_+, \lambda_{- (1)}, \cdots, \lambda_{- (k-1)} )
\end{equation}
where $L^T=(l_1,\cdots, l_k)$ is the transpose of the vector $L$ and $M_{rr'}$ is a symmetric matrix with the definitions
\begin{equation}
\sum_{j,r,r'}\frac{\eta_{jr}\eta_{jr'}l_r
l_{r'}}{1+v_j} = \sum_{r,r'} l_r M_{rr'} l_{r'}\equiv L^TML, \qquad M_{rr'} = \sum_j\frac{\eta_{jr}\eta_{jr'}}{1+v_j}
\end{equation}
The eigenvalues $\lambda_r$ ($r=1,\cdots, k$) or $\lambda_+$, $\lambda_{-(r)}$ $(r=1,\cdots, k-1)$ correspond to the eigenvectors $L'= (l'_1,\cdots, l'_k)^T\equiv (l'_+, l'_{-(1)}, \cdots, l'_{- (k-1)})^T$. As the transformation matrix $O$ is orthogonal, the integration measure is unchanged $d^4l^\prime_1\cdots d^4 l^\prime_k=d^4l_1 \cdots d^4 l_k$, and the integral Eq.(\ref{genI}) can be simplified as:
\begin{eqnarray}\label{genI2}
 I(p_1,...,p_m) 
&=&\frac{\Gamma(\sum^{n}_{j=1}\alpha_j)}{\Gamma(\alpha_1)...\Gamma(\alpha_n)}
\int^\infty_0 \prod^n_{i=1}\frac{dv_i}{(1+v_i)^{\alpha_i+1}}
\delta(1-\sum^n_{j=1}\frac{1}{1+v_j}) \cdot \nonumber\\
&&\int d^4l'_1...d^4
l'_k\frac{N}{[\sum^n_{j=1}\frac{q_j^2-m_j^2}{1+v_j}
+\sum_{r}\lambda_r l_r^{'2}]^{\sum^n_{j=1}\alpha_j}} \  .
\end{eqnarray}
As $\sum^n_{j=1}\alpha_j\geq 2k-1$ for a generic k-loop integrals with $k\geq 2$ and $n > k$, we can safely integrate out the loop momenta for the convergent integrals. When the numerator $N$ contains no $l^{'}_i$ terms, we can integrate out the last $(k-1)$ internal loop momenta $l'_2, l'_3,..., l'_k$ and obtain
the following form of ILIs\cite{HW1}:
\begin{eqnarray} \label{genI3}
I(p_1,...,p_m) &=&
\frac{\Gamma(\sum^{n}_{j=1}\alpha_j-2k+2)}{\Gamma(\alpha_1)...\Gamma(\alpha_n)}
\int^\infty_0 \prod^n_{i=1}\frac{dv_i}{(1+v_i)^{\alpha_i+1}} 
\delta(1-\sum^n_{j=1}\frac{1}{1+v_j}) \cdot \nonumber\\
&&\frac{1}{(\det|M|)^{2}} \int d^4l_+
\frac{1}{[\sum^n_{j=1}\frac{q_j^2-m_j^2}{1+v_j} +
l_+^{2}]^{\sum^n_{j=1}\alpha_j-2(k-1)}}
 \end{eqnarray}
with the definition of the determinant for the matrix $M$
\begin{equation}
\det|M|=\prod^k_{r=1}\lambda_r \equiv \lambda_+\prod^{k-1}_{r=1}\lambda_{-(r)}
\end{equation}
where a rescaling transformation $l_+\rightarrow \sqrt{\lambda_+}l'_+$ has been made. Where the ILIs for the momentum integral on $l_+$ reflect the overall divergence of the Feynman diagram. From the above expression, it is clearly seen that the UV divergences contained in the loop momentum integrals on $l'_{-(r)}$ $(r=1,\cdots, k-1)$ for the original loop subdiagrams are now characterized by the possible zero eigenvalues $\lambda_{-(r)}\to 0 $ $(r=1,\cdots, k-1)$ of the matrix $M$ in the allowed regions of the parameters $v_i$ $(i=1,\cdots, n)$. Namely, each zero eigenvalue $\lambda_{-(r)}\to 0$ resulted from some infinity values of parameters $v_i$ in the UVDP parameter space leads to a singularity for the parameter integrals, which corresponds to the divergence of subdiagram in the relevant loop momentum integral.

By applying the general LORE formulae to the above integration over the momentum $l_+$, we have:
\begin{eqnarray}\label{genI4}
& & I(p_1,...,p_m) =
\frac{\Gamma(\sum^{n}_{j=1}\alpha_j-2k+2)}{\Gamma(\alpha_1)...\Gamma(\alpha_n)}\int^\infty_0
\prod^n_{i=1}\frac{dv_i}{(1+v_i)^{\alpha_i+1}}
\delta(1-\sum^n_{j=1}\frac{1}{1+v_j}) 
\nonumber\\
&& \qquad  \frac{1}{(\det|M|)^{2}} \lim_{N,M_i^2}\sum^N_{l=0}c_l^N \int
d^4 l_+ \frac{i (-1)^{\sum^n_{j=1}\alpha_j} }{[\sum^n_{j=1}\frac{q_j^2-m_j^2}{1+v_j}
+ l_+^{2}+M_l^2 ]^{\sum^n_{j=1}\alpha_j-2(k-1)}}\nonumber\\
&& \qquad = \frac{\Gamma(\sum^{n}_{j=1}\alpha_j-2k+2)}{\Gamma(\alpha_1)...\Gamma(\alpha_n)}\int^\infty_0
\prod^n_{i=1}\frac{dv_i}{(1+v_i)^{\alpha_i+1}}
\delta(1-\sum^n_{j=1}\frac{1}{1+v_j})\nonumber\\
&& \qquad \frac{1}{(\det|M|)^2}\  I^R_{-2\alpha}({\cal M}^2)
\end{eqnarray}
with
\begin{equation}\label{MASS}
\alpha = \sum^n_{j=1}\alpha_j - 2k,\qquad {\cal M}^2 = \sum^n_{j=1}(m_j^2-q_j^2)/(1+v_j)
\end{equation}
where $I^R_{-2\alpha}({\cal M}^2)$ is the regularized ILI for the possible overall divergence of the Feynman diagram.

The above general procedure explicitly realizes the UVDP parametrization and  demonstrates systematically the evaluation of ILIs, which illustrates the advantage when merging the LORE method with the Bjorken-Drell analogy between Feynman diagrams and electrical circuit diagrams.

In order to demonstrate explicitly the correspondence between two kinds of divergences in the UVDP parameter space and in the momentum space, it is useful to apply the above general procedure to the $\alpha\beta\gamma$ integral at two-loop oder.

\section{ Divergence Transmission From Momentum Space to UVDP Parameter Space}

The Feynman diagram for the $\alpha\beta\gamma$ integral is shown in Fig. (\ref{abcoverall}). By using the internal momenta $k_j$ and making the particular choice of loops defined therein, the  $\alpha\beta\gamma$ integral can be expressed as follows
\begin{eqnarray}\label{abc in BJ}
I_{\alpha\beta\gamma} &=&
\int\frac{d^4k_1}{(2\pi)^4}\int\frac{d^4k_2}{(2\pi)^4}
\frac{1}{(k_1^2-m_1^2)^\alpha(k_2^2-m_2^2)^\beta(k_3^2-m_3^2)^\gamma}\nonumber\\
&=& \int\frac{d^4k_1}{(2\pi)^4}\int\frac{d^4k_2}{(2\pi)^4}
\frac{\Gamma(\alpha+\beta+\gamma)}{\Gamma(\alpha)\Gamma(\beta)\Gamma(\gamma)}
\int^\infty_0\prod^3_{i=1}\frac{dv_i}{(1+v_i)^{\alpha_i+1}} \nonumber \\
& & \cdot \  \delta(1-\sum^3_{j=1}\frac{1}{1+v_j}) \frac{1}{[\frac{k_1^2-m_1^2}{1+v_1}+
\frac{k_2^2-m_2^2}{1+v_2}+\frac{k_3^2-m_3^2}{1+v_3}]^{\alpha+\beta+\gamma}} \ , 
\end{eqnarray}
with the notations $\alpha_i$ (i=1,2,3) corresponding to $\alpha,\beta,\gamma$. The momentum conservation laws for overall diagram and both vertices read 
\begin{eqnarray}\label{mcl}
& & p_1=-p_2\equiv p  \  , \nonumber \\
&&p_1-k_1-k_2-k_3=0  \  , \\
&&p_2+k_1+k_2+k_3=0 \  . \nonumber 
\end{eqnarray}
According to Eq. (\ref{decomp}), the internal momenta $k_j$ are decomposed into the following forms
\begin{eqnarray}\label{dcabc}
k_1 &=& q_1 + l_1 \  , \nonumber \\
k_2 &=& q_2 + l_2 \  , \\
k_3 &=& q_3-l_1-l_2  \  .  \nonumber 
\end{eqnarray}
Replacing the $k_j$ with $q_j$ and $l_r$ in Eq. (\ref{abc in BJ}), changing the integral variables to $l_r$, we have
\begin{eqnarray}
I_{\alpha\beta\gamma} & = & \frac{\Gamma(\alpha+\beta+\gamma)}{\Gamma(\alpha)\Gamma(\beta)\Gamma(\gamma)}
\int\frac{d^4l_1}{(2\pi)^4}\int\frac{d^4l_2}{(2\pi)^4}
\int^\infty_0\prod^3_{i=1}\frac{dv_i}{(1+v_i)^{\alpha_i+1}} \nonumber \\
& & \cdot\ \delta(1-\sum^3_{j=1}\frac{1}{1+v_j})
\frac{1}{D^{\alpha+\beta+\gamma}} \  , 
\end{eqnarray}
where $D$ is defined as 
\begin{equation}\label{abc denom}
D = \sum^3_{j=1}\frac{q_j^2-m_j^2}{1+v_j}+
2(\frac{q_1}{1+v_1}-\frac{q_3}{1+v_3})l_1+
2(\frac{q_2}{1+v_2}-\frac{q_3}{1+v_3})l_2+ L^T M L  \  , 
\end{equation}
with
\begin{eqnarray}
L \equiv \left( \begin{array}{c} l_1\\ l_2
\end{array}
\right),\quad M \equiv \left(
\begin{array}{cc}
\frac{1}{1+v_1}+\frac{1}{1+v_3} & \frac{1}{1+v_3}\\
\frac{1}{1+v_3} & \frac{1}{1+v_2}+\frac{1}{1+v_3}
\end{array}
\right) \  . 
\end{eqnarray}
The elimination of the cross terms in the denominator $D$ requires that
\begin{eqnarray} \label{voltcons abc}
\frac{q_1}{1+v_1}-\frac{q_3}{1+v_3}=0, \nonumber \\
\frac{q_2}{1+v_2}-\frac{q_3}{1+v_3}=0 \  , 
\end{eqnarray}
which correspond to the Kirchhoff's law for two loops in the analogy of electrical circuit. From Eqs. (\ref{voltcons abc}), (\ref{dcabc}) and (\ref{mcl}), the momenta $q_i$ can completely be determined
\begin{eqnarray}
q_1 &=& \frac{1+v_1}{3+v_1+v_2+v_3}p \  , \nonumber \\
q_2 &=& \frac{1+v_2}{3+v_1+v_2+v_3}p\  , \\
q_3 &=& \frac{1+v_3}{3+v_1+v_2+v_3}p \  . \nonumber 
\end{eqnarray}

By diagonalizing the matrix $M$ with a $2$x$2$ orthogonal matrix transformation $O$
\begin{eqnarray}
L = O L^\prime, \qquad O^T M O = \left(
\begin{array}{cc}
\lambda_1 & 0\\
0 & \lambda_2
\end{array}
\right)
\end{eqnarray}
we explicitly obtain the eigenvalues  $\lambda_{1,2} = \lambda_{+,-}$ corresponding to two eigenvectors $L'= (l'_+,\ l'_{-})$
\begin{eqnarray} \lambda_{\pm} &=&
\frac{(1+\frac{1}{1+v_3})\pm
\sqrt{(1+\frac{1}{1+v_3})^2+ 4\Delta} }{2}\\
\Delta = \det |M| &=&
\frac{1}{(1+v_1)(1+v_2)}+\frac{1}{(1+v_2)(1+v_3)}+\frac{1}{(1+v_3)(1+v_1)}
\end{eqnarray}
which indicates that the matrix $M$ is not always invertible as the determinant of $M$ vanishes when any two of $v_i$s tend to $\infty$. For instance, taking $v_1, v_3\rightarrow \infty$,  the eigenvalue $\lambda_-$ vanishes. 

In the new basis, the $\alpha\beta\gamma$ integral can be rewritten as:
\begin{eqnarray}\label{abc med}
I_{\alpha\beta\gamma} 
&=&  \frac{\Gamma(\alpha+\beta+\gamma)}{\Gamma(\alpha)\Gamma(\beta)\Gamma(\gamma)}
\int\frac{d^4l^\prime_+}{(2\pi)^4}\int\frac{d^4l^\prime_-}{(2\pi)^4}
\int^\infty_0\prod^3_{i=1}\frac{dv_i}{(1+v_i)^{\alpha_i+1}}
\delta(1-\sum^3_{j=1}\frac{1}{1+v_j})\nonumber\\
&&\frac{1}{[\sum^3_{j=1}\frac{q_j^2-m_j^2}{1+v_j}+
\lambda_{+}l^{\prime2}_{+}+\lambda_{-}l_{-}^{\prime2}]^{\alpha+\beta+\gamma}} \  , 
\end{eqnarray}
which shows that the two combination quantities $\lambda_+ l^\prime_+$ and $\lambda_- l^\prime_-$  are on equal footing in the
denominator, namely  $\lambda_+ l^\prime_+$ and $\lambda_- l^\prime_-$ approach to infinity at the same speed when
both $l^\prime_{\pm}\rightarrow \infty$. It indicates that once $\lambda_-\rightarrow 0$ and $\lambda_+$ finite, the speed of $l^\prime_-$ going to infinity is faster than that of $l^\prime_+$.  

Thus the integration over $l^\prime_-$ represents the subintegrals, while the one over $l^\prime_+$ is an overall integral. In general, the integral over $l^\prime_-$ reflects to the asymptotic behavior of subintegrals when the corresponding UVDP parameters approach to infinity. Such an explicit demonstration shows why and how the divergences in the subdiagrams are transmitted to the corresponding divergences in the UVDP parameter space, which illustrates intuitively the electrical circuit analogy of Feynman diagrams.

To be more clear, we may explicitly integrate over $l^\prime_-$ and yield
\begin{eqnarray}
I_{\alpha\beta\gamma} &=& \frac{i}{16\pi^2}
\int\frac{d^4l^\prime_+}{(2\pi)^4}
\frac{\Gamma(\alpha+\beta+\gamma-2)}{\Gamma(\alpha)\Gamma(\beta)\Gamma(\gamma)}
\int^\infty_0\prod^3_{i=1}\frac{dv_i}{(1+v_i)^{\alpha_i+1}}
\delta(1-\sum^3_{j=1}\frac{1}{1+v_j})\frac{1}{\lambda^2_-}\nonumber\\
&&\frac{1}{[\sum^3_{j=1}\frac{q_j^2-m_j^2}{1+v_j}+
\lambda_{+}l^{\prime2}_{+}]^{\alpha+\beta+\gamma-2}} \  , 
\end{eqnarray}
which shows that when $\lambda_-$ goes to zero, that happens when
any two of the three UVDP parameters $v_i$ approach to infinity,
then the integrand becomes singular and the integrals over the UVDP
parameters yield some divergences.

By a scaling definition for a new momentum $l_+$
\begin{equation}
l_+\equiv  \sqrt{\lambda_+}l_+^{\prime},
\end{equation}
the $\alpha\beta\gamma$ integral can be rewritten in a more tractable form which is the same as Eq.(\ref{abc int_ed0}) achieved by the rescaling transformation given in Eq.(\ref{scaling})
\begin{eqnarray}\label{abc int_ed2}
I_{\alpha\beta\gamma}  &=& \frac{i}{16\pi^2}
\frac{\Gamma(\alpha+\beta+\gamma-2)}{\Gamma(\alpha)\Gamma(\beta)\Gamma(\gamma)}
\int^\infty_0\prod^3_{i=1}\frac{dv_i}{(1+v_i)^2}
\delta(1-\sum^3_{j=1}\frac{1}{1+v_j})F(v_k)\nonumber\\
&& \int\frac{d^4l_+}{(2\pi)^4} \frac{1}{[l_+^2 - {\cal
M}^2(p^2,m_k^2,v_k)]^{\alpha+\beta+\gamma-2}} \  , 
\end{eqnarray}
which shows the advantage when merging the UVDP parametrization and the evaluation of ILIs with the Bjorken-Drell's circuit analogy of Feynman diagrams. By applying the LORE method to the momentum integral over $l_+$  for possible overall divergence, we have\cite{HW1}
\begin{eqnarray}\label{abc int_ed3}
I_{\alpha\beta\gamma} &=& \frac{i}{16\pi^2}
\frac{\Gamma(\alpha+\beta+\gamma-2)}{\Gamma(\alpha)\Gamma(\beta)\Gamma(\gamma)}
\int^\infty_0\prod^3_{i=1}\frac{dv_i}{(1+v_i)^2}
\delta(1-\sum^3_{j=1}\frac{1}{1+v_j})F(v_k)\nonumber\\
&& \lim_{N,M_i^2}\sum^N_{l=0}c_l^N \int\frac{d^4l_+}{(2\pi)^4} \frac{i(-1)^{\alpha+\beta + \gamma} }{[l_+^2 + M_l^2 + {\cal
M}^2(p^2,m_k^2,v_k)]^{\alpha+\beta+\gamma-2}} \nonumber \\
& = & \frac{i}{16\pi^2}
\frac{\Gamma(\alpha+\beta+\gamma-2)}{\Gamma(\alpha)\Gamma(\beta)\Gamma(\gamma)}
\int^\infty_0\prod^3_{i=1}\frac{dv_i}{(1+v_i)^2}
\delta(1-\sum^3_{j=1}\frac{1}{1+v_j})\nonumber \\
& & \cdot\ F(v_k)\  I^R_{-2(\alpha +\beta +\gamma-4)}( {\cal M}^2 )  \  , 
\end{eqnarray}
from which one can directly read off the consequence with the case $\alpha=\gamma=1, \beta=2$ as given in Eq.(\ref{I121})
\begin{eqnarray}
I_{121} &=&
\frac{i}{16\pi^2}\int^\infty_0\prod^3_{i=1}\frac{dv_i}{(1+v_i)^2}
\delta(1-\sum^3_{j=1}\frac{1}{1+v_j}) \frac{(1+v_1)^2(1+v_2)(1+v_3)^2}{(3+v_1+v_2+v_3)^2}  \  I^R_0( {\cal M}^2 )
\nonumber\\
&\to & -\frac{1}{(16\pi^2)^2}\int^\infty_0\prod^3_{i=1}dv_i \delta(1-\sum^3_{j=1}\frac{1}{1+v_j}) \frac{1}{(3+v_1+v_2+v_3)^2(1+v_2)}\nonumber\\
&&(\ln\frac{M_c^2}{\mu_M^2}-\gamma_E+ \varepsilon(\frac{\mu_M^2}{M_c^2})) \  , 
\end{eqnarray}
where the singular behavior arising from the region $v_1,v_3\rightarrow \infty$ becomes manifest as $\det|M|=\Delta=0$ due
to the zero eigenvalue $\lambda_-\to 0$. In contrast, for the other two regions: $v_1, v_2\rightarrow\infty $ and $v_2, v_3\rightarrow\infty$, there is an additional factor $\frac{1}{(1+v_2)}$ which leads the integration to be finite.

\section{Divergence Correspondence in Circuit Analogy of Feynman Diagrams}

The LORE method merging with the Bjorken-Drell's circuit analogy enables us to treat a more complicated overlapping divergence structure of Feynman diagrams. For an explicit illustration, let us examine a typical case with $\alpha=\beta=\gamma=1$  as it causes both the quadratic divergence and the complicated overlapping divergence structure. 

From the general form of $\alpha\beta\gamma$ integral, one can directly read off the result for the case $\alpha=\beta=\gamma=1$
\begin{eqnarray}\label{I_111 def}
I_{111} &= & \frac{i}{16\pi^2}
\int^\infty_0\prod^3_{i=1}\frac{dv_i}{(1+v_i)^{2}}
\delta(1-\sum^3_{j=1}\frac{1}{1+v_j})\frac{1}{(\det|M|)^2}\nonumber\\
&&\lim_{N,M_l^2}\sum^N_{l=1}c_l\int\frac{d^4 l_+}{(2\pi)^4}
\frac{-i}{\sum^3_{j=1}\frac{q_j^2-m_j^2}{1+v_j}+ l_+^2+ M_l^2}
\nonumber\\
&\to &
\frac{1}{(16\pi^2)^2}\int^\infty_0\prod^3_{i=1}\frac{dv_i}{(1+v_i)^2}
\delta(1-\sum^3_{j=1}\frac{1}{1+v_j})\frac{\prod^3_{j=1}(1+v_j)^2}{(3+v_1+v_2+v_3)^2}\nonumber\\
&&[M_c^2-{\cal M}^{2}(\ln\frac{ M_c^2}{{\cal
M}^{2}}-\gamma_E+1)] \label{I111} \  ,  
\end{eqnarray}
where the overall quadratic divergence for the loop momentum integral $l_+$ has been regularized by the LORE method with the mass factor ${\cal M}$ is given in Eq.(\ref{MASS}). It is seen from the expression of integral $I_{111}$ that the three subintegrals $\alpha\gamma$, $\beta\gamma$, and $\alpha\beta$ are all divergent as the UV divergences due to the large internal loop momenta are transmitted to the asymptotic regions of UVDP parameter space. The corresponding divergent conductances are associated to the following asymptotic regions in the circuits:
\begin{eqnarray}
Circuit~1:~ \alpha\gamma ~divergence &\Leftrightarrow& v_1\rightarrow\infty,v_3\rightarrow\infty,v_2\rightarrow0, \nonumber \\
Circuit~2:~ \beta\gamma ~divergence &\Leftrightarrow& v_2\rightarrow\infty, v_3\rightarrow\infty,v_1\rightarrow0, \\
Circuit~3:~ \alpha\beta ~ divergence &\Leftrightarrow& v_1\rightarrow\infty,
v_2\rightarrow\infty, v_3\rightarrow0  \  , \nonumber
\end{eqnarray}
which have similar behaviors due to a permutation $Z_3$ symmetry among the three pairs of parameters $(v_1, m_1)$, $(v_2, m_2)$,$(v_3, m_3)$ as shown in Eq.(\ref{I111}), thus the treatment on three asymptotic regions in the circuits is the same.  Without losing generality, it only needs to examine one of the cases. 

Let us consider the region in Circuit 1: $v_1\rightarrow \infty$, $v_3\rightarrow\infty$ and $v_2\rightarrow0$. As it has been discussed in previous section how to treat the divergence in the UVDP parameter space,  the integral domain in this region can be written as $\int^\infty_{v_o}dv_1\int^\infty_{v_o} dv_3$ with ${\cal M}^2\rightarrow m_2^2$ and $F(v_j)\to \frac{(1+v_1)^2(1+v_3)^2}{(v_1+v_3)^2}$
\begin{eqnarray}
I_{111}^{(0)(\alpha\gamma)} &\simeq &
\frac{1}{(16\pi^2)^2}\int^\infty_{v_o}  \frac{dv_1}{(1+v_1)^2}
\int^\infty_{v_o}  \frac{dv_3}{(1+v_3)^2}
\frac{(1+v_1)^2(1+v_3)^2}{(v_1+v_3)^2} \nonumber\\
&&
[M_c^2-m_2^2(\ln\frac{M_c^2}{m^2_2}-\gamma_E+1)]\nonumber\\
&=&
\frac{1}{(16\pi^2)^2}[M_c^2-m_2^2(\ln\frac{M_c^2}{m_2^2}-\gamma_E+1)]\int^\infty_{v_o}
dv_1\frac{1}{v_1+ v_o}+É\nonumber\\
&=&
\frac{1}{(16\pi^2)^2}[M_c^2-m_2^2(\ln\frac{M_c^2}{m_2^2}-\gamma_E+1)]
(\ln\frac{M_c^2}{\mu_o^2}-\gamma_E)+... \  ,  
\end{eqnarray}
where we have adopted the LORE method in the treatment on the UVDP parameter as shown in Eq.(\ref{treatment}). The dots represent other terms including the single logarithmic divergent term and finite terms which are irrelevant to our purpose here for a check on the cancelation of harmful divergences. 

It is noticed that the divergences are factorizable. In order to make the comparison between the divergence structure in the UVDP parameter space and that in the subdiagram $(\alpha\gamma)$, it is helpful to calculate the counterterm diagram $I^{(c)(\alpha\gamma)}_{111}$:
\begin{eqnarray}
I^{(c)(\alpha\gamma)}_{111} &=& -
\int\frac{d^4k_2}{(2\pi)^4}\frac{1}{k_2^2-m_2^2}\textsc{DP}
\{\int\frac{d^4k_1}{(2\pi)^4}
\frac{1}{(k_1^2-m_1^2)[k_3^2-m_3^2]}\}\nonumber\\
&\to &
-\frac{1}{(16\pi^2)^2}
[M_c^2-m_2^2(\ln\frac{M_c^2}{m_2^2}-\gamma_E+1)]  (\ln\frac{M_c^2}{\mu^2}-\gamma_E) \label{I111_1} \  , 
\end{eqnarray}
where \textsc{DP}\{...\} denotes the divergence part of the integral in the bracket, and $\mu^2$ is the renormalization scale. It becomes manifest that by choosing $\mu^2=\mu_o^2$, the harmful divergence parts cancel exactly.

Based on the permutation $Z_3$ symmetry, it is easy to demonstrate that the harmful divergence parts in other two regions in Circuit~2 and Circuit~3 also cancel exactly. We then arrive at the conclusion that there is no harmful divergence for the case $\alpha=\beta=\gamma=1$ when adding the corresponding counterterm diagrams.

\section{Finite Renormalization Scheme in LORE}

The renormalization scheme was initiated to remove the divergences in QED\cite{Dyson}. The development of renormalization group analysis indicates that the renormalization is actually needed to define physics quantities at any interesting energy scale. In LORE, the finite quadratic and logarithmic forms corresponding to $(M_c^2- {\cal M}^2)$ and $\ln M_c^2/{\cal M}^2 $ are resulted intrinsically to avoid infinities as the CES $M_c$ can be taken to be finite, thus there are in principle no divergences in QFTs when applying for the LORE method. While the divergence structure of QFTs is maintained when taking the CES $M_c$ to be infinitely large $M_c^2 \to \infty$.  It is noticed that the presence of the quadratic term to the mass correction of scalar particles does not allow us to make a mass independent renormalization. To realize a consistent renormalization and make a renormalization group analysis, a well-defined subtraction scheme is necessary and proposed as follows\cite{HW1}

(i) For quadratic term $(M_c^2- {\cal M}^2) $, subtract $(M_c^2-\mu^2)$ and leave $(\mu^2- {\cal M}^2)$ in the finite expression; 

(ii) For logarithmic term $(\ln \frac{M_c^2}{{\cal M}^2}-\gamma_E)$, subtract $(\ln \frac{M_c^2}{\mu^2}-\gamma_E)$ and leave term $\ln \frac{\mu^2}{{\cal M}^2}$ in the finite expression.

It is seen that the subtraction for the quadratic term is analogous  to the usual momentum subtraction, and the one for the logarithmic term is similar to the $\bar{MS}$ scheme in the dimensional regularization, which may be called as the energy scale subtraction scheme at $\mu^2$.  In such an energy scale subtraction scheme,  the quadratic and logarithmic terms are set up in terms of the correlative forms $(M_c^2-\mu^2)$ and $\ln M_c^2/\mu^2$ with a single subtracted energy scale $\mu^2$.  It is useful to make a postulation that such a correlative form at one-loop level with a single subtracted energy scale $\mu^2$ is maintained at high loop level, which prevents us to make either the rescaling transformation $\mu^2\to e^{\alpha_0} \mu^2$ or the shifting operation $\mu^2 \to \mu^2 - \alpha_0 m^2$ for the subtracted energy scale $\mu^2$, and leads the mass renormalization to be well-defined at high loop level.

In general, we arrive at the following theorems to achieve the consistent regularization and renormalization in LORE:

{\em Factorization Theorem for Overlapping Divergences}: Overlapping divergences which contain divergences of subintegrals and overall divergences in the general Feynman loop integrals become factorizable in the corresponding asymptotic regions of circuit analogy of Feynman diagrams.

{\em Substraction Theorem for Overlapping Divergences}: For general scalar-type two-loop integral $I_{\alpha\beta\gamma}$, when including the corresponding subtraction integrals (which is composed of divergent subintegrals multiplied by an overall integral), the sum will only contain harmless divergence.

{\em Harmless Divergence Theorem}: If the general loop integral contains no divergent subintegrals, then it contains only a harmless single divergence arising from the overall divergence.

{\em Trivial Convergence Theorem}: If the general loop integral contains no overall divergence and also no divergent subintegrals,
then it is convergent.

These theorems together with the energy scale subtraction scheme enable us to carry out a consistent finite renormalization scheme in QFTs.

\section{Consistency and Advantages in Applications of LORE Method}

\subsection{Slavnov-Taylor-Ward-Takahaski Identities in LORE} 

As a consistent check and practical calculation for the LORE method, it is useful to apply to the Yang-Mills gauge theories and make a direct computing for all two-, three- and four-point Green functions. An explicit calculation was carried out to verify the Ward-Takahaski-Slavnov-Taylor identities among the renormalization constants\cite{Cui:2008uv}. To define the physics processes at any interesting scale, it is necessary to renormalize the theory by rescaling the fields and redefining the masses and coupling constant. This procedure is equivalent to the introduction of some counterterms to the Lagrangian of Yang-Mills gauge theory or QCD in Eq.(\ref{GTQCD})
\begin{eqnarray}
\delta\mathcal{L}&=&[(z_2-1)\bar{\psi}_ni\gamma^{\mu}\partial_{\mu}\psi_n-(z_2z_m-1)m\bar{\psi}_n\psi_n]
+(z_3-1)[-\frac{1}{4}(\partial_{\mu}A^a_{\nu}-\partial_{\nu}A^a_{\mu})^2]\nonumber\\
& &+(\tilde{z}_3-1)[\partial^{\mu}\bar{c}^a\delta^{ac}\partial_{\mu}c^c] +(\tilde{z_1}-1)gf^{abc}\partial^{\mu}\bar{c}^aA_{\mu}^bc^c +(z_{1F}-1)g\bar{\psi}_n\gamma_{\mu}A^{a\mu}T^a\psi_{n}\nonumber\\
& &-(z_1-1)\frac{1}{2}gf^{abc}(\partial_{\mu}A^a_{\nu}-\partial_{\nu}A^a_{\mu})A^{b\mu}A^{c\nu}
+(z_4-1)\frac{1}{4}g^2f^{abc}f^{ade}A^b_{\mu}A^c_{\nu}A^{d\mu}A^{e\nu} \  , \nonumber \\
\end{eqnarray}
with $z_1, \cdots, z_4$ being the so-called renormalization constants. They must satisfy the so-called Slavnov-Taylor identities\cite{st} which are the generalization of the usual Ward-Takahaski identities and also the consequence of gauge symmetry. These identities indicate that the renormalization constants should satisfy the following relations\cite{relationofz}:
\begin{eqnarray}
\frac{z_{1F}}{z_3^{1/2}z_2}=\frac{\tilde{z_1}}{z_3^{1/2}\tilde{z}_3}=\frac{z_1}{z_3^{3/2}}=\frac{z_4^{1/2}}{z_3} \  . 
\end{eqnarray}
In fact, the gauge independence and the unitarity of the renormalized S matrix require that the gauge symmetry must be maintained after the renormalization\cite{smatrix}, namely the renormalization constants of $g$ obtained from each vertex renormalization must be the same, which actually leads to the above relations. 

With the detailed calculations performed in ref. \cite{Cui:2008uv}  by applying for the LORE method, all the renormalization constants are found to be
\begin{eqnarray}
z_2=1-\frac{g^2}{8\pi^2}C_2\xi\frac{1}{2}(\ln\frac{M_c^2}{\mu_s^2}-\gamma_E)
\end{eqnarray}
for the fermion fields, and
\begin{eqnarray}
z_3=1+\frac{g^2}{16\pi^2}\left[(\frac{13}{3}-\xi)C_1-\frac{g^2}{6\pi^2}N_fT_2
\right] \frac{1}{2}(\ln\frac{M_c^2}{\mu_s^2}-\gamma_E)
\end{eqnarray}
for the gluon fields,  and
\begin{eqnarray}
\tilde{z_3}&=&1+\frac{g^2}{16\pi^2}C_1(\frac{3}{2}-\frac{\xi}{2})\frac{1}{2}(\ln\frac{M_c^2}{\mu_s^2}-\gamma_E)
\end{eqnarray}
for the ghost fields, and
\begin{eqnarray}
z_{1F}&=&1-\frac{g^2}{8\pi^2}\left[(\frac{3}{4}+\frac{\xi}{4})C_1+{\xi}C_2
\right] \frac{1}{2}(\ln\frac{M_c^2}{\mu_s^2}-\gamma_E)
\end{eqnarray}
from the Fermion-gluon vertex, and
\begin{eqnarray}
\tilde{z_1}=1-\frac{g^2}{16\pi^2}{\xi}C_1\frac{1}{2}(\ln\frac{M_c^2}{\mu_s^2}-\gamma_E)
\end{eqnarray}
from the Ghost-gluon vertex, and
\begin{eqnarray}
z_1=1+\left[\frac{g^2}{12\pi^2}[1 + \frac{9}{8} (1 -
\xi)]C_1-\frac{g^2}{6\pi^2}N_fT_2 \right]
\frac{1}{2}(\ln\frac{M_c^2}{\mu_s^2}-\gamma_E)
\end{eqnarray}
from the three-gluon vertex, and 
\begin{eqnarray}
z_4=1-\left[\frac{g^2}{24\pi^2}(1 + 3(\xi -1)
)C_1+\frac{g^2}{6\pi^2}N_fT_2\right]\frac{1}{2}(\ln\frac{M_c^2}{\mu_s^2}-\gamma_E)
\end{eqnarray}
from the four-gluon vertex.

It becomes manifest to verify the Ward-Takahaski-Slavnov-Taylor identities
\begin{eqnarray}
z_g =
\frac{z_{1F}}{z_3^{1/2}z_2}=\frac{\tilde{z_1}}{z_3^{1/2}\tilde{z}_3}=\frac{z_1}{z_3^{3/2}}=\frac{z_4^{1/2}}{z_3}\  ,
\end{eqnarray}
and obtain explicitly the gauge independent renormalization constant for
the gauge coupling constant $g = z_g^{-1} g_0$
\begin{eqnarray}
z_g&=&1-(\frac{11}{48\pi^2}C_1-\frac{1}{12\pi^2}N_fT_2)g^2\frac{1}{2}(\ln\frac{M_c^2}{\mu_s^2}-\gamma_E) \  , 
\end{eqnarray}
which leads to the well-known one-loop $\beta$ function via the definition
\begin{eqnarray}
\beta(g)&{\triangleq}&\lim_{M_c\to \infty} \mu_s\frac{\partial}{\partial\mu_s}g\mid_{g_0,m_0} = \lim_{M_c\to \infty} g\mu_s\frac{\partial}{\partial\mu_s}\ln{z_g}\mid_{g_0,m_0}\nonumber\\
&{\simeq}&g\mu_s\frac{\partial}{\partial\mu_s}[(\frac{11}{48\pi^2}C_1-\frac{1}{12\pi^2}N_fT_2)g^2
\frac{1}{2}(\ln\frac{M_c^2}{\mu_s^2}-\gamma_E)]\nonumber\\
&=&-\frac{g^3}{(4\pi)^2}(\frac{11}{3}C_1-\frac{4}{3}N_fT_2)  \  .
\end{eqnarray}

The LORE method has also been applied to verify several supersymmetric Ward identities in different supersymmetric models, it arrived at the conclusion that the LORE method can preserve both the supersymmetry and gauge symmetry as all the Ward identities hold~\cite{Cui:2008bk}. The explicit computation shows that in the supersymmetric theories the verification of Ward identities relies on the four-dimensional Dirac algebra and the shift of integration variable, which strongly indicates that the consistent regularization scheme for supersymmetric theories should be realized in the physical four dimension with translational invariance for the integration variable. By applying the LORE method to perform a complete one-loop renormalization for the massive Wess-Zumino model,  it was shown in ~\cite{Cui:2008bk} that the quadratic divergences vanish as expected and the relations among masses and coupling constants hold by renormalization, which agrees with the non-renormalization theorem.

\subsection{Quantum Chiral Anomaly in LORE }

The anomaly as quantum effects has been studied substantially in QFTs. In perturbation theory, the anomaly has been calculated by using different regularization schemes. In the dimensional regularization, it is well-known to have a difficulty of defining $\gamma_5$. In the Pauli-Villars regularization, it usually changes the field contents of original theory by the introduction of super massive regulator fields. In contrast, the LORE method realized without modifying original theory has advantages in these aspects.  

To show explicitly the advantage of the LORE method, let us begin with the massless QED
\begin{eqnarray}
{\cal L}=\bar{\psi}\gamma^\mu(i\partial_\mu-{\cal A}_\mu)\psi \ . 
\end{eqnarray}
The vector current $V_\mu(x)$ and axial-vector current $A_\mu(x)$ are defined as
\begin{eqnarray}
V_\mu(x)=\bar{\psi}(x)\gamma_\mu\psi(x),\qquad A_\mu(x)=\bar{\psi}(x)\gamma_\mu\gamma_5\psi(x) \ ,
\end{eqnarray}
which are conserved classically
\begin{eqnarray}
\partial^\mu V_\mu(x)=0,\qquad \partial^\mu A_\mu(x)=0\label{divcurrents} \ . 
\end{eqnarray}

To investigate the quantum corrections, one may consider three-point Green function
\begin{eqnarray}
T^{AVV}_{\mu\nu\lambda}(p,q;(p+q))&=&\int d^4x_1d^4x_2e^{ipx_1+iq
x_2}\langle0|T[V_\mu(x_1) V_\nu(x_2) A_\lambda(0)]|0\rangle  \  .
\end{eqnarray}
The corresponding classical Ward identity (\ref{divcurrents}) requires
\begin{eqnarray}
&&p^\mu T^{AVV}_{\mu\nu\lambda}(p,q;(p+q))=0, \nonumber \\
&&q^\nu T^{AVV}_{\mu\nu\lambda}(p,q;(p+q))=0,  \nonumber \\
&&(p+q)^\lambda T^{AVV}_{\mu\nu\lambda}(p,q;(p+q))=0 \ .
\end{eqnarray}
In perturbative calculation, one can simply compute the corresponding contributions from the triangle loop diagram to $T^{(1),AVV}_{\lambda\mu\nu}$ 
\begin{eqnarray}
T^{(1),AVV}_{\lambda\mu\nu}&=&(-1)\int\frac{d^4k}{(2\pi)^4}{\rm
tr}\{\gamma_\lambda\gamma_5\frac{i}{(k\hspace{-0.2cm}\slash+k\hspace{-0.2cm}\slash_2)}
\gamma_\nu\frac{i}{(k\hspace{-0.2cm}\slash+k\hspace{-0.2cm}\slash_1)}
\gamma_\mu\frac{i}{(k\hspace{-0.2cm}\slash+k\hspace{-0.2cm}\slash_3)}\}\nonumber\\
&=&-i\int\frac{d^4k}{(2\pi)^4}\frac{(k+k_2)_\alpha(k+k_1)_\beta(k+k_3)_\xi}{(k+k_2)^2(k+k_1)^2(k+k_3)^2}{\rm tr}\{\gamma_5\gamma_\lambda\gamma_\alpha\gamma_\nu\gamma_\beta\gamma_\mu\gamma_\xi\}\label{TAVV}  \  . 
\end{eqnarray}
Here the momentum associated with the axial-vector vertex is $(k_3-k_2)$.  For the trace of gamma matrices, there are several ways to treat it. It is interesting to notice that there is a unique solution when treating all three currents symmetrically by adopting the definition of $\gamma_5$
\begin{eqnarray}
\gamma_5={i\over 4!}\epsilon_{\mu\nu\alpha\beta}\gamma^\mu\gamma^\nu\gamma^\alpha\gamma^\beta,
\quad \epsilon_{0123}=1 \ .
\end{eqnarray}
With repeatedly using the relation $\gamma_{\rho}\gamma_{\sigma} = 2g_{\rho\sigma} - \gamma_{\sigma}\gamma_{\rho}$, one can obtain the following result\cite{Ma:2005md}
\begin{eqnarray}
& & Tr\{\gamma_5\gamma_\lambda\gamma_\alpha\gamma_\nu\gamma_\beta\gamma_\mu\gamma_\xi\}
= {i\over 4!}\epsilon_{\mu' \nu' \alpha' \beta' }Tr\{\gamma^{\mu'}\gamma^{\nu'}\gamma^{\alpha'}
\gamma^{\beta'} \gamma_\lambda\gamma_\alpha\gamma_\nu\gamma_\beta\gamma_\mu\gamma_\xi\}\nonumber\\
& & \qquad \quad =4i\{\epsilon_{\lambda\alpha\beta\xi}g_{\mu\nu}-\epsilon_{\lambda\alpha\nu\beta}
g_{\mu\xi}+\epsilon_{\lambda\alpha\nu\mu}g_{\beta\xi}-\epsilon_{\lambda\alpha\nu\xi}
g_{\mu\beta}-\epsilon_{\lambda\alpha\beta\mu}g_{\nu\xi}\nonumber\\
&&\qquad \quad -\epsilon_{\lambda\alpha\mu\xi}g_{\nu\beta}+\epsilon_{\lambda\nu\beta\mu}
g_{\alpha\xi}-\epsilon_{\lambda\nu\beta\xi}g_{\alpha\mu}+\epsilon_{\lambda\nu\mu\xi}
g_{\alpha\beta}-\epsilon_{\lambda\beta\mu\xi}g_{\alpha\nu}\nonumber\\
&&\qquad \quad -\epsilon_{\alpha\nu\beta\mu}g_{\lambda\xi}+\epsilon_{\alpha\nu\beta\xi}
g_{\lambda\mu}-\epsilon_{\alpha\nu\mu\xi}g_{\lambda\beta}+\epsilon_{\alpha\beta\mu\xi}
g_{\lambda\nu}+\epsilon_{\nu\mu\beta\xi}g_{\lambda\alpha}\}\label{trace10gamma}  \  , 
\end{eqnarray}
which gives the most general form respecting all the symmetries of the Lorentz indices and eliminates the ambiguities caused by the trace of gamma matrices with $\gamma_5$. With such a general form, the amplitude $T^{(1),AVV}_{\lambda\mu\nu}$ is given by 
\begin{eqnarray}
& & T^{(1),\{AVV\}}_{\lambda\mu\nu}= T^{(1),\{AVV\}}_{L,\lambda\mu\nu}
+T^{(1),\{AVV\}}_{C,\lambda\mu\nu}\nonumber\\
& & T^{(1),\{AVV\}}_{L,\lambda\mu\nu}= 4\int\frac{d^4k}{(2\pi)^4}
\bigg\{    \times\bigg[\frac{1}{(k+k_1)^2(k+k_2)^2(k+k_3)^2}\bigg]          \nonumber\\
&& \qquad \times\{-\epsilon_{\lambda\alpha\nu\beta}(k+k_2)_\alpha(k+k_1)_\beta(k+k_3)_\mu
-\epsilon_{\lambda\alpha\nu\rho}(k+k_2)_\alpha(k+k_1)_\mu(k+k_3)_\rho\nonumber\\
&&\qquad -\epsilon_{\lambda\alpha\beta\mu}(k+k_2)_\alpha(k+k_1)_\beta(k+k_3)_\nu
+\epsilon_{\lambda\alpha\beta\rho}g_{\mu\nu}(k+k_2)_\alpha(k+k_1)_\beta(k+k_3)_\rho\nonumber\\
&&\qquad -\epsilon_{\lambda\alpha\mu\rho}(k+k_2)_\alpha(k+k_1)_\nu(k+k_3)_\rho
-\epsilon_{\lambda\nu\beta\rho}(k+k_2)_{\mu}(k+k_1)_\beta(k+k_3)_\rho\nonumber\\
&&\qquad -\epsilon_{\lambda\beta\mu\rho}(k+k_2)_\nu(k+k_1)_\beta(k+k_3)_\rho
-\epsilon_{\alpha\nu\beta\mu}(k+k_2)_\alpha(k+k_1)_\beta(k+k_3)_\lambda\nonumber\\
&&\qquad +\epsilon_{\alpha\nu\beta\rho}g_{\lambda\mu}(k+k_2)_\alpha(k+k_1)_\beta(k+k_3)_\rho
-\epsilon_{\alpha\nu\mu\rho}(k+k_2)_\alpha(k+k_1)_\lambda(k+k_3)_\rho\nonumber\\
&&\qquad +\epsilon_{\alpha\beta\mu\rho}g_{\lambda\nu}(k+k_2)_\alpha(k+k_1)_\beta(k+k_3)_\rho
+\epsilon_{\nu\mu\beta\rho}(k+k_2)_\lambda(k+k_1)_\beta(k+k_3)_\rho\}\nonumber\\
&&\qquad +\frac{\epsilon_{\lambda\alpha\nu\mu}}{2}\bigg[\frac{(k+k_2)_\alpha}{(k+k_2)^2(k+k_3)^2}
+\frac{(k+k_2)_\alpha}{(k+k_2)^2(k+k_1)^2}\bigg]\nonumber\\
&&\qquad +\frac{\epsilon_{\lambda\nu\beta\mu}}{2}\bigg[\frac{(k+k_1)_\beta}{(k+k_1)^2(k+k_3)^2}
+\frac{(k+k_1)_\beta}{(k+k_2)^2(k+k_1)^2}\bigg]\nonumber\\
&&\qquad +\frac{\epsilon_{\lambda\nu\mu\rho}}{2}\bigg[\frac{(k+k_3)_\rho}{(k+k_1)^2(k+k_3)^2}
+\frac{(k+k_3)_\rho}{(k+k_2)^2(k+k_3)^2}\bigg]\bigg\}\\
& & T^{(1),\{AVV\}}_{C,\lambda\mu\nu} = -2\epsilon_{\lambda\alpha\nu\mu}\int\frac{d^4k}{(2\pi)^4}
\frac{(k_3-k_1)^2(k+k_2)_\alpha}{(k+k_1)^2(k+k_2)^2(k+k_3)^2}\nonumber\\
&&\qquad \qquad -2\epsilon_{\lambda\nu\beta\mu}\int\frac{d^4k}{(2\pi)^4}
\frac{(k_3-k_2)^2(k+k_1)_\beta}{(k+k_1)^2(k+k_2)^2(k+k_3)^2}\nonumber\\
&&\qquad \qquad -2\epsilon_{\lambda\nu\mu\rho}\int\frac{d^4k}{(2\pi)^4}
\frac{(k_1-k_2)^2(k+k_3)_\rho}{(k+k_1)^2(k+k_2)^2(k+k_3)^2} \  .
\end{eqnarray}

When applying the LORE method to the amplitude, one can safely shift the integration variable and make some algebra. The regularized amplitude in LORE gets the following form\cite{Ma:2005md}
\begin{eqnarray}
T^{R,(1),\{AVV\}}_{\lambda\mu\nu}&=&T^{R,(1),\{AVV\}}_{0,\lambda\mu\nu}
+T^{R,(1),\{AVV\}}_{-2,\lambda\mu\nu} \  ,
\end{eqnarray}
with the divergent part
\begin{eqnarray}\label{divermasslessTAVV}
T^{R,(1),\{AVV\}}_{0,\lambda\mu\nu}&=&2\int_0^1dx[\epsilon_{\lambda\alpha\nu\mu}(k_2-k_1)_\alpha
I_0^R(x,\mu_1)   +\epsilon_{\lambda\mu\alpha\nu}(k_3-k_1)_\alpha I_0^R(x,\mu_3) \nonumber\\
&&  +\epsilon_{\lambda\alpha\nu\mu}(2x-1)(k_3-k_2)_\alpha I_0^R(x,\mu_2) ]  \nonumber\\
&&-2\epsilon_{\lambda\mu\nu\alpha}\int_0^1dx_1\int_0^{x_1}dx_2(-2k_3-2k_2+4k_1)_\alpha
I_0^R(x_i,\mu) \  , 
\end{eqnarray}
and the convergent part
\begin{eqnarray}\label{convermasslessTAVV}
T^{R,(1),\{AVV\}}_{-2,\lambda\mu\nu}&=&-8\int_0^1dx_1\int_0^{x_1}dx_2
 \bigg\{\epsilon_{\lambda\alpha\nu\beta}(-\Delta+k_2)_\alpha(-\Delta+k_1)_\beta(-\Delta+k_3)_\mu\nonumber\\
&&+\epsilon_{\lambda\alpha\nu\rho}(-\Delta+k_2)_\alpha(-\Delta+k_1)_\mu(-\Delta+k_3)_\rho\nonumber\\
&&+\epsilon_{\lambda\alpha\beta\mu}(-\Delta+k_2)_\alpha(-\Delta+k_1)_\beta(-\Delta+k_3)_\nu\nonumber\\
&&+\epsilon_{\lambda\alpha\mu\rho}(-\Delta+k_2)_\alpha(-\Delta+k_1)_\nu(-\Delta+k_3)_\rho\nonumber\\
&&+\epsilon_{\lambda\nu\beta\rho}(-\Delta+k_2)_{\mu}(-\Delta+k_1)_\beta(-\Delta+k_3)_\rho\nonumber\\
&&+\epsilon_{\lambda\beta\mu\rho}(-\Delta+k_2)_\nu(-\Delta+k_1)_\beta(-\Delta+k_3)_\rho\nonumber\\
&&+\epsilon_{\alpha\nu\beta\mu}(-\Delta+k_2)_\alpha(-\Delta+k_1)_\beta(-\Delta+k_3)_\lambda\nonumber\\
&&+\epsilon_{\alpha\nu\mu\rho}(-\Delta+k_2)_\alpha(-\Delta+k_1)_\lambda(-\Delta+k_3)_\rho\nonumber\\
&&-\epsilon_{\nu\mu\beta\rho}(-\Delta+k_2)_\lambda(-\Delta+k_1)_\beta(-\Delta+k_3)_\rho\nonumber\\
&&+\frac{\epsilon_{\lambda\alpha\nu\mu}}{2}
(k_3-k_1)^2(-\Delta+k_2)_\alpha+\frac{\epsilon_{\lambda\nu\beta\mu}}{2}(k_3-k_2)^2(-\Delta+k_1)_\beta\nonumber\\
&&+\frac{\epsilon_{\lambda\nu\mu\rho}}{2}(k_2-k_1)^2(-\Delta+k_3)_\rho\bigg\}I_{-2}^R(x_i,\mu)  \  , 
\end{eqnarray}
with $x$ and $x_i$ the Feynman parameters and $\mu^2=\mu_s^2+M^2$ and $\mu_i^2=\mu_s^2+M_i^2$. Some definitions are made as follows
\begin{eqnarray}
& & \Delta = (1-x)k_1+(x_1-x_2)k_2+x_2k_3  \nonumber \\
& & M^2 = (x_2-x_1)(1-x_1)(k_1-k_2)^2  - x_2(1-x_1)(k_3-k_1)^2+x_2(x_2-x_1)(k_3-k_2)^2  \nonumber \\
& & M_1^2 = x(x-1)(k_1-k_2)^2,\quad M_2^2 = x(x-1)(k_3-k_2)^2,\quad  M_3^2 = x(x-1)(k_3-k_1)^2  \  . \nonumber 
\end{eqnarray}

The corresponding Ward identities can be written as follows\cite{Ma:2005md} 
\begin{eqnarray}
& & (k_1-k_2)_\nu
T^{R,(1),\{AVV\}}_{\lambda\mu\nu} = 4\epsilon_{\lambda\mu\nu\alpha}(k_1-k_2)_\nu(k_3-k_1)_\alpha
I_{0,(00)}\nonumber\\
&& \qquad \qquad -8\epsilon_{\lambda\mu\nu\alpha}(k_1-k_2)_\nu(k_3-k_1)_\alpha  \{(k_1-k_2)^2[2I_{-2,(01)}-2I_{-2,(02)}-{1\over2}I_{-2,(00)}]\nonumber\\
&& \qquad \qquad +(k_1-k_2)\cdot(k_3-k_1)[2I_{-2,(11)}-I_{-2,(10)}]\}\label{A-V-V-VWard1}\\
& & (k_3-k_1)_\mu
T^{R,(1),\{AVV\}}_{\lambda\mu\nu}= -4\epsilon_{\lambda\nu\alpha\mu}(k_1-k_2)_\alpha(k_3-k_1)_\mu
I_{0,(00)}^\prime\nonumber\\
&& \qquad \qquad +8\epsilon_{\lambda\nu\alpha\mu}(k_1-k_2)_\alpha(k_3-k_1)_\mu  \{(k_1-k_2)\cdot(k_3-k_1)[2I_{-2,(11)}-I_{-2,(01)}]\nonumber\\
&&\qquad \qquad +(k_3-k_1)^2[2I_{-2,(10)}-2I_{-2,(20)}-{1\over2}I_{-2,(00)}]\}\label{A-V-V-VWard2}\\
& & (k_3-k_2)_\lambda
T^{R,(1),\{AVV\}}_{\lambda\mu\nu} = -4\epsilon_{\mu\nu\lambda\alpha}(k_3-k_1)_\lambda(k_1-k_2)_\alpha
I_{0,(00)}^\prime\nonumber\\
&&\qquad \qquad -4\epsilon_{\mu\nu\lambda\alpha}(k_3-k_1)_\lambda(k_1-k_2)_\alpha
I_{0,(00)}\nonumber\\
&&\qquad \qquad -4\epsilon_{\mu\nu\lambda\alpha}(k_3-k_1)_\lambda(k_1-k_2)_\alpha \{(k_3-k_1)^2I_{-2,(00)}+(k_1-k_2)^2I_{-2,(00)}\nonumber\\
&&\qquad \qquad +2(k_1-k_2)\cdot(k_3-k_1)[I_{-2,(10)}+I_{-2,(01)}]\}\label{A-V-V-AWard}  \  , 
\end{eqnarray}
where we have introduced the following definitions for the integrals
\begin{eqnarray}
& & I_{-2,(ij)}(\mu_s^2)= {-i\over32\pi^2}\int_0^1dx_1
\int_0^{x_1}dx_2{x_2^i(x_1-x_2)^j\over M^2+\mu_s^2}
\{ 1- ({\mu_M^2\over M_c^2})[  \varepsilon' ({\mu_M^2\over M_c^2})- 1] \}\label{defineI(-2ij)} \nonumber \\
& & I_{0,(00)}(\mu_s^2)\equiv
{i\over16\pi^2}\{\int^1_0dx_1\int_0^{x_1}dx_2
 [\ln(\frac{M_c^2}{\mu_M^2})-\gamma_E+ \varepsilon ({\mu_M^2\over M_c^2})]\nonumber\\
&&\;\;\;\;\;\;\;\;\;\;\;-\int_0^1dxx[\ln(\frac{M_c^2}{\mu_{M_3}^2})-\gamma_E+ \varepsilon
({\mu_{M_3}^2\over M_c^2})]\}\label{defineI(00)} \nonumber \\
& & I_{0,(00)}^\prime(\mu_s^2)\equiv
{i\over16\pi^2}\{\int^1_0dx_1\int_0^{x_1}dx_2
 [\ln(\frac{M_c^2}{\mu_M^2})-\gamma_E+ \varepsilon ({\mu_M^2\over M_c^2})]\nonumber\\
&&\;\;\;\;\;\;\;\;\;\;\;-\int_0^1dxx[\ln(\frac{M_c^2}{\mu_{M_1}^2})-\gamma_E+ \varepsilon
({\mu_{M_1}^2\over M_c^2})]\}\label{defineI('00)} \   , 
\end{eqnarray}
with $\mu_{M_i}^2 = \mu_s^2 + M_i^2$. Where $I_{0,(00)}$ is given by the difference of two logarithemically divergent integrals, which actually leads to a finite result.  As a consequence, it is easy to check 
\begin{eqnarray}
& & (k_1-k_2)_\nu T^{R,(1),\{AVV\}}_{\lambda\mu\nu} = 0\label{A-V-Vvector1} \nonumber \\
& & (k_3-k_1)_\mu T^{R,(1),\{AVV\}}_{\lambda\mu\nu} = 0\label{A-V-Vvector2}  \  ,
\end{eqnarray}
due to the cancellation of all terms in the expressions, which means that the vector currents are conserved.

For the axial-vector current, it gets the following result
\begin{eqnarray}
& & (k_3-k_2)_\lambda
T^{R,(1),\{AVV\}}_{\lambda\mu\nu}=16\mu_s^2\epsilon_{\mu\nu\lambda\alpha}(k_1-k_2)_\lambda(k_3-k_1)_\alpha I_{-2,(00)}(0,\mu_s^2)\label{SymmaxialWard}  \nonumber \\
& & \qquad  + {i\over2\pi^2}\epsilon_{\mu\nu\lambda\alpha}(k_1-k_2)_\lambda(k_3-k_1)_\alpha
\int_0^1dx_1\int_0^{x_1}dx_2e^{-(\mu_s^2 + M^2)/M_c^2}  \  .
\end{eqnarray}
When taking $\mu_s=0$ and $M_c\to \infty$, one has 
\begin{eqnarray}
(k_3-k_2)_\lambda
T^{R,(1),\{AVV\}}_{\lambda\mu\nu}&=&{i\over4\pi^2}\epsilon_{\mu\nu\lambda\alpha}(k_1-k_2)_\lambda(k_3-k_1)_\alpha \  . 
\end{eqnarray}
By including the cross diagrams, we arrive at the Ward identities with anomaly of axial-vector current 
\begin{eqnarray}
(k_3-k_1)_\mu
T^{R,\{AVV\}}_{\lambda\mu\nu}&=&0,  \nonumber \\
(k_1-k_2)_\nu
T^{R,\{AVV\}}_{\lambda\mu\nu}&=&0 , \nonumber \\
(k_3-k_2)_\lambda T^{R,\{AVV\}}_{\lambda\mu\nu}&=&{i\over2\pi^2}
\epsilon_{\mu\nu\alpha\beta}(k_1-k_2)_\alpha(k_3-k_1)_\beta \  ,
\end{eqnarray}
which can be expressed in terms of the well-known operator form as follows
\begin{eqnarray}
\partial_\mu V_\mu(x)&=&0 \  , \nonumber \\
\partial_\mu
A_\mu(x)&=&{e^2\over4\pi^2}\epsilon_{\mu\nu\alpha\beta}\partial^\alpha
\mathcal{A}^\mu(x)\partial^\beta \mathcal{A}^\nu(x)
= {e^2\over8\pi^2}F^{\mu\nu}(x)\widetilde{F}_{\mu\nu}(x)\label{operatormasslessanomal}  \  .
\end{eqnarray}

In comparison with Pauli-Villars scheme in which the triangle anomaly arises from different sources,  in LORE method, the triangle anomaly can appear in the axial-vector Ward identity when the trace of gamma matrices is treated symmetrically for three currents by using the definition of $\gamma_5$, which reflects the intrinsic property of original theory as the LORE method is realized without modifying the original theory. In Pauli-Villars regularization, the vector Ward identity is always preserved as the anomalies arising from the original spinor and heavy regulator spinors cancel each other, thus the anomaly exists only in the axial-vector Ward identity, but the cancellation mechanism is different in different treatments. It then becomes unclear whether the anomaly arises directly from the axial-vector Ward identity of original theory or due to the introduction of heavy regular fields as the anomaly is likely caused by the heavy regulator spinors in Pauli-Villars scheme.

In the dimensional regularization in which the triangle anomaly receives contributions from both the $n-4$ dimensions and the original four dimensions, thus both the vector and axial-vector Ward identities appear anomaly. In particular, when acting the external momentum of the axial-vector current on the AVV diagram before evaluating the integrals, the resulting triangle anomaly depends only on the extended dimensions and appears only in the axial-vector Ward identity. It implies that the triangle anomaly of vector and axial-vector currents relies on the procedures of operation although the total anomaly when normalizing to the conserved vector current has the same standard form. Besides such an ambiguity, when the acting external momentum operates on the vector current momentum, a similar calculation leads the vector Ward identity to be anomalous. 

In conclusion, a unique solution for the Ward identity anomaly of axial-vector current can be obtained in LORE with the definition of $\gamma_5$ to eliminate the ambiguity caused from the trace of gamma matrices by treating all the three currents symmetrically. It has been shown in LORE that the quantum chiral anomaly reflects the infrared behavior of QFTs\cite{Ma:2005md}.

In addition, based on the LORE method, the well-known ambiguities in calculating the radiatively induced Lorentz and CPT violating Chern-Simons term in the extended QED can be clarified when relating to the calculations of chiral anomaly\cite{Ma:2006yc}. The main ambiguities arise from the finite term in the relation for a linear divergent integral due to momentum translation. It has been shown that one should apply the LORE method directly to the linearly divergent integrals. Furthermore, the QED trace anomaly was also calculated based on the LORE method\cite{cui:2011}, it can be shown that the dilation Ward identity which relates the three-point diagrams and the vacuum polarization diagrams gets the standard form of trace anomaly through quantum corrections, where the use of the consistency conditions are crucial for obtaining a consistent result.

\subsection{ Two Loop Renormalization of Scalar Interaction and Power-Law Running of Scalar Mass }

The discovery of Higgs boson arises a great interest to investigate the quantum contributions to the mass of scalar particles. The loop contributions to the mass of scalar boson is in general quadratic, thus the dimensional regularization is not suitable to make a reliable calculation for the mass renormalization of scalar boson. In contrast, the LORE method can maintain the quadratic structure of original theory, it is then applicable to study the mass renormalization of scalar bosons. 

The well-known scalar $\phi^4$ theory has explicitly been examined in ref.\cite{HW1} to demonstrate the mass and coupling constant renormalization at two loop level. The Lagrangian density for $\phi^4$ theory is given as follows:
\begin{equation}
{\cal L} = \frac{1}{2}\partial_\mu\phi\partial^\mu\phi
-\frac{1}{2}m^2\phi^2-\frac{\lambda}{4!}\phi^4  \  . 
\end{equation}
Here we focus on the mass renormalization of scalar boson in LORE. 

\begin{figure}[ht]
\begin{center}
  \includegraphics[scale=0.8]{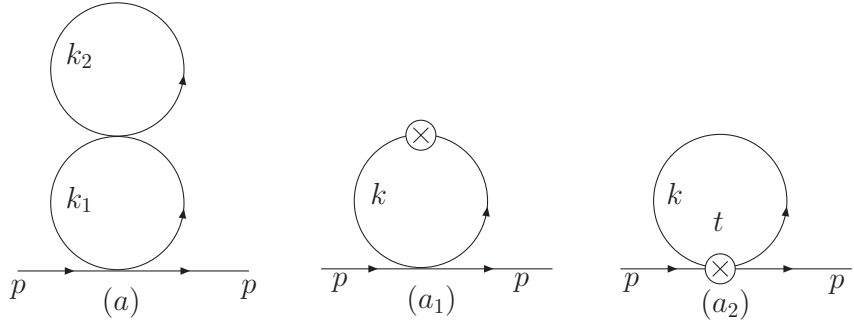}\\
  \caption{Two-loop tadpole and counterterm diagrams}\label{groupA}
\end{center}
\end{figure}

\begin{figure}[ht]
\begin{center}
  \includegraphics[scale=0.7]{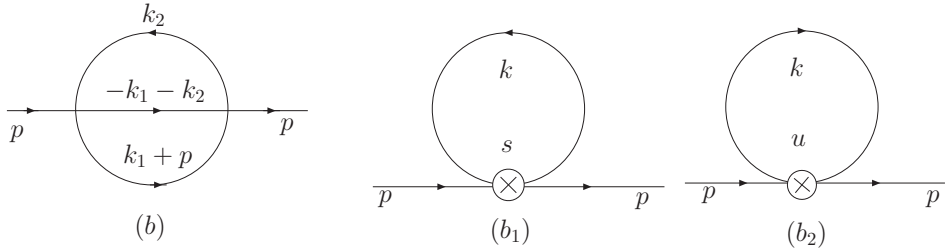}\\
  \caption{Two-loop sunrise and counterterm diagrams}\label{groupB}
\end{center}
\end{figure}

In general, there are two types of diagrams contributing to the two-loop self-energy
corrections, which are presented in Figs. (\ref{groupA}) and (\ref{groupB}).

\begin{figure}[ht]
\begin{center}
  \includegraphics[scale=0.8]{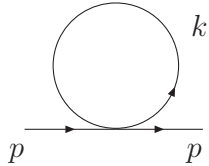}\\
  \caption{One-loop tadpole diagram }\label{1loop2}
\end{center}
\end{figure}

\begin{figure}[ht]
\begin{center}
  \includegraphics[scale=0.8]{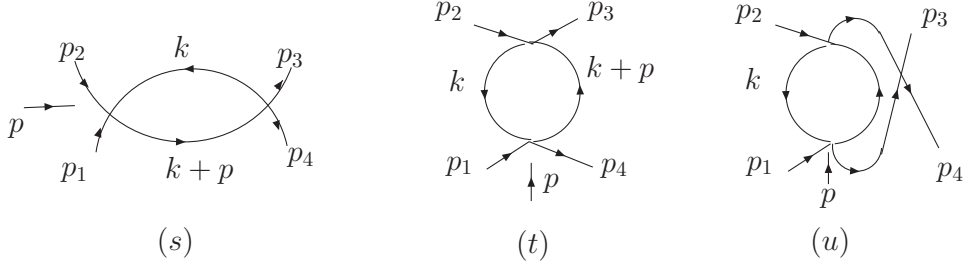}\\
  \caption{One-loop vertex diagarms}\label{1loop4}
\end{center}
\end{figure}

The Figs.\ref{groupA} $(a_1)$-$(a_2)$ and Figs.\ref{groupB} $(b_1)$-$(b_2)$ correspond to the counterterm diagrams with the insertion of one loop counterterms. Thus before making a detailed calculation at two-loop level, it is necessary to compute first the one-loop counterterms. In Fig.\ref{groupA} $(a_1)$ , it needs to insert the self-energy correction at one loop level shown in Fig. (\ref{1loop2}), and in Fig.\ref{groupA} $(a_2)$ and Figs.\ref{groupA} $(b_1)$-$(b_2)$, the vertex corrections at one loop level are required, the corresponding diagrams are shown in Fig.  (\ref{1loop4}) for the so-called $s$, $t$ and $u$ channels.


The calculation for the self-energy correction is straightforward 
\begin{eqnarray}\label{1_selfenergy}
-iM^2_{(1)} &=& -i\lambda\cdot\frac{1}{2}\int\frac{d^4k}{(2\pi)^4}
\frac{i}{k^2-m^2}\nonumber\\
&\to &
-\frac{i\lambda}{2(4\pi)^2}[(M_c^2- m^2) - m^2(\ln\frac{M_c^2}{m^2}-\gamma_E)]  \  , 
\end{eqnarray}
where the LORE method has been adopted to obtain the regularized result and the small terms suppressed by $m^2/M_c^2$ have been neglected. Such a result differs from the one yielded by using the dimensional regularization. The difference arises from the quadratic behavior in the renormalization counterterm
which may greatly change the renormalization group analysis. With the energy scale subtraction scheme described in the previous section, we have the following mass and wave function counterterms:
\begin{eqnarray}
-i\delta^{(1)}_{m^2} &=&
\frac{i\lambda}{2(4\pi)^2}[(M_c^2-\mu^2)-m^2(\ln\frac{M_c^2}{\mu^2}-\gamma_E)]\label{delta_m1}\\
i\delta^{(1)}_Z &=& 0\label{delta_z1} \  , 
\end{eqnarray}
and the renormalized mass of scalar boson is found to be
\begin{equation}
-iM^2_{(1)} =
-\frac{i\lambda}{2(4\pi)^2}[(\mu^2-m^2)-m^2\ln\frac{\mu^2}{m^2}] \ .
\end{equation}

The calculation for the vertex corrections from the one-loop four-point Green function is similar, and the correction for the s-channel is given by:
\begin{eqnarray}
-i\Lambda^{(s)}_{(1)} &=& \frac{(-i\lambda)^2}{2} \int\frac{d^4 k}{(2\pi)^4}
\frac{i}{k^2-m^2}\frac{i}{(k+p)^2-m^2}\nonumber\\
&=& \frac{\lambda^2}{2}\int\frac{d^4k}{(2\pi)^4}\int^1_0 dx
\frac{1}{[k^2+x(1-x)p^2-m^2]^2}\nonumber\\
&\to & \frac{i\lambda^2}{2(4\pi)^2}\int^1_0 dx
[\ln\frac{M_c^2}{m^2-x(1-x)p^2}-\gamma_E] \  , 
\end{eqnarray}
with $-p^2=-(p_1+p_2)^2\equiv s$. For the t- and u-channels, the same expression can be obtained with the replacement $p^2$: $-p^2=-(p_1-p_4)^2\equiv t$ for the t-channel and $-p^2=-(p_1-p_3)^2\equiv u$ for the u-channel.With the energy scale subtraction scheme in LORE,  the corresponding counterterm is given by:
\begin{equation}\label{delta_lmd1}
-i\delta^{(1)}_\lambda = -\frac{3i\lambda^2}{2(4\pi)^2}
[\ln\frac{M_c^2}{\mu^2}-\gamma_E] \  , 
\end{equation}
where the factor 3 is due to three contributions from the $s$, $t$, $u$-channels. The renormalized vertex has the following form:
\begin{eqnarray}
-i\Lambda_{(1)} = \frac{i\lambda^2}{2(4\pi)^2}\int^1_0 dx
[\ln\frac{\mu^2}{m^2+x(1-x)s} + (s\to t) + (s\to u) ]  \  .
\end{eqnarray}

Let us turn to the computation for the mass renormalization at two loop level. The calculation of the diagram in Fig.\ref{groupA} $(a)$ is straightforward:
\begin{eqnarray}
-iM^{2(a)}_{(2)} &=&
\frac{1}{4}(-i\lambda)^2\int\frac{d^4k_1}{(2\pi)^4}\frac{i}{k_1^2-m^2}\frac{i}{k_1^2-m^2}
\int\frac{d^4k_2}{(2\pi)^4}\frac{i}{k_2^2-m^2}\nonumber\\
& \to & \frac{1}{4}\frac{i\lambda^2}{(16\pi)^2}(\ln\frac{M_c^2}{m^2}-\gamma_E)\cdot
[M^2_c-m^2(\frac{M_c^2}{m^2}-\gamma_E+1)] \   .
\end{eqnarray}
The contributions from the counterterm diagrams in Fig.\ref{groupA} $(a_1)$-$(a_2)$ are found to be:
\begin{eqnarray}
-iM^{2(a1)+(a2)}_{(2)} &=&
\frac{1}{2}(-i\lambda)(-i\delta_{m^2})\int\frac{d^4k}{(2\pi)^4}\frac{i^2}{(k^2-m^2)^2}+
\frac{1}{2}(-i\delta^t_\lambda)\int\frac{d^4k}{(2\pi)^4}\frac{i}{k^2-m^2}\nonumber\\
&\to & -\frac{1}{4}\frac{i\lambda^2}{(16\pi^2)^2}\{[(M_c^2-\mu^2)-m^2(\ln\frac{M_c^2}{\mu^2}-\gamma_E)](\ln\frac{M_c^2}{m^2}-\gamma_E)\nonumber\\
&&+(\ln\frac{M_c^2}{\mu^2}-\gamma_E)\cdot
[M_c^2-m^2(\ln\frac{M_c^2}{m^2}-\gamma_E+1)]\} \  , 
\end{eqnarray}
where $\delta_{m^2}$ is the one-loop mass counterterm and $\delta_\lambda^t$ only the t-channel vertex
counterterm. The sum of three diagram contributions is given by:
\begin{eqnarray}\label{A_2}
-iM^{2(a)+(a1)+(a2)}_{(2)} &=&
\frac{1}{4}\frac{i\lambda^2}{(16\pi^2)^2}\{- [(M_c^2-\mu^2)-m^2(\ln\frac{M_c^2}{\mu^2}-\gamma_E)] 
(\ln\frac{M_c^2}{\mu^2}-\gamma_E)\nonumber\\
&&+[(\mu^2-m^2)-m^2\ln\frac{\mu^2}{m^2}]\ln\frac{\mu^2}{m^2} \}  \   .
\end{eqnarray}

With the energy scale subtraction scheme in LORE, the overall counterterm for two loop diagram Fig.\ref{groupA} $(a)$ is given by:
\begin{equation}\label{A_counter}
-i\delta^{(a)}_{m^2} =
\frac{1}{4}\frac{i\lambda^2}{(16\pi^2)^2}(\ln\frac{M_c^2}{\mu^2}-\gamma_E)
[(M_c^2-\mu^2)-m^2(\ln\frac{M_c^2}{\mu^2}-\gamma_E)] \  , 
\end{equation}
and the corresponding renormalized result  is:
\begin{equation}\label{A result}
-iM^{2(a)}_{(2)R}
=\frac{1}{4}\frac{i\lambda^2}{(16\pi^2)^2}[(\mu^2-m^2)-m^2\ln\frac{\mu^2}{m^2}]\ln\frac{\mu^2}{m^2} \  .
\end{equation}

For the diagram in Fig.\ref{groupB} $(b)$, its contribution is calculated as follows:
\begin{eqnarray}\label{M(b)}
-iM^{2(b)}_{(2)} &=&
\frac{1}{6}(-i\lambda)^2\int\frac{d^4k_1}{(2\pi)^4}\int\frac{d^4k_2}{(2\pi)^4}
\frac{i}{(k_1+p)^2-m^2} \frac{i}{k_2^2-m^2}\frac{i}{(k_1+k_2)^2-m^2}\nonumber\\
&\to & \frac{i\lambda^2}{6(16\pi^2)^2}
\int^\infty_0\prod^3_{j=1}\frac{dv_j}{(1+v_j)^2}
\delta(1-\sum^3_{j=1}\frac{1}{1+v_j})
\frac{\prod^3_{j=1}(1+v_j)^2}{(3+v_1+v_2+v_3)^2}\nonumber\\
&&[ M_c^2-{\cal M}^{2}(\ln\frac{M_c^2}{{\cal
M}^{2}}-\gamma_E+1)] \  ,
\end{eqnarray}
with $ {\cal M}^2 = m^2- p^2/(3+v_1+v_2+v_3) $. As the above integral is identified to the $\alpha\beta\gamma$ integral with $\alpha=\beta=\gamma=1$ and the same masses $m_i^2=m^2$, the general result in Eq. (\ref{I_111 def}) has straightforwardly been adopted to yield the regularized result. To compute the contribution to the two-point Green function, it is useful to change to a new set of UVDP parameters $u,v, w$ via
\begin{eqnarray}
\frac{1}{1+v_1} &\equiv& \frac{1}{(1+u)(1+w)};\quad \frac{1}{1+v_2} \equiv  \frac{u}{1+u};\quad
\frac{1}{1+v_3} \equiv \frac{1}{(1+u)(1+v)}  \  , \nonumber 
\end{eqnarray}
which leads to the corresponding change for Eq.(\ref{M(b)}):
\begin{eqnarray}\label{parts}
-iM^{2(b)}_{(2)} &=& \frac{i\lambda^2}{6(16\pi^2)^2} \int^\infty_0
du \int^\infty_0 \frac{dw}{(1+w)^2} \frac{dv}{(1+v)^2}
\delta(1-\frac{1}{1+w}-\frac{1}{1+v})  \nonumber\\
&&\frac{1+u}{[u+\frac{1}{(1+w)(1+v)}]^2} [M_c^2-{\cal M}^{2}(\ln\frac{M_c^2}{{\cal
M}^{2}}-\gamma_E+1)] \   , 
\end{eqnarray}
with
\begin{eqnarray}
{\cal M}^2 &=& m^2-\frac{u}{1+u}\frac{1}{u(1+w)(1+v)+1}p^2  \  .
\end{eqnarray}
For the quadratic term, it can easily be integrated out:
\begin{eqnarray}\label{B_quad}
-iM^{2(b)}_{(2)quad}&=& \frac{i\lambda^2}{6(16\pi^2)^2} M_c^2
\int^\infty_0 \frac{dw}{(1+w)^2}
\frac{dv}{(1+v)^2} \delta(1-\frac{1}{1+w}-\frac{1}{1+v})\nonumber\\
&&\int^\infty_0 du \frac{1+u}{[u+\frac{1}{(1+w)(1+v)}]^2}\nonumber\\
&\to& \frac{i\lambda^2}{6(16\pi^2)^2}
M_c^2[3(\ln\frac{M_c^2}{q_o^2}-\gamma_E)+1] \  , 
\end{eqnarray}
which is local by choosing $q_o^2=\mu^2$.

For the logarithmic term, the integral becomes complicated as there are three parameter regions which contain divergent contributions:
\begin{eqnarray}
-iM^{2(b)}_{(2)log} &=& -\frac{i\lambda^2}{6(16\pi^2)^2} \int^\infty_0
\frac{dw}{(1+w)^2} \frac{dv}{(1+v)^2}
\delta(1-\frac{1}{1+w}-\frac{1}{1+v})  \nonumber\\
&& \int^\infty_0 du \frac{1+u}{[u+\frac{1}{(1+w)(1+v)}]^2} {\cal M}^2 [\ln\frac{M_c^2}{{\cal M}^2}-\gamma_E+1] \  .
\end{eqnarray}
It is useful to consider four parameter regions corresponding to the following extreme
asymptotic behaviors in the UVDP parameter space:
\begin{eqnarray}
I:  &&\quad\quad u\rightarrow \infty\, \quad \quad \quad vw =1 ; \qquad v_1\to \infty\, \ v_3\to \infty\, \  v_2\to 0  \nonumber\\
II: &&\quad\quad v\rightarrow \infty\, \ u\to 0\, \  w\to 0; \qquad v_2\to \infty\, \ v_3\to \infty\, \  v_1\to 0 \nonumber \\
III: &&\quad\quad w\rightarrow \infty\,  \ u\to 0\, \  v\to 0; \qquad v_2\to \infty\, \ v_1\to \infty\, \  v_3\to 0  \nonumber \\
IV: &&\quad\quad  p^2\gg m^2  \nonumber
\end{eqnarray}
In two asymptotic regions II and III, they are symmetric under the exchange of parameters $v$ and $w$ or $v_1$ and $v_3$. 

{\bf In Region (I)}: $u\rightarrow \infty$, ${\cal M}^2 \simeq m^2 $, its contribution is approximately  given by:
\begin{eqnarray} \label{B_I}
-iM^{2(b)(I)}_{(2)log} &\simeq & -\frac{i\lambda^2}{6(16\pi^2)^2}
\int^\infty_0 \frac{dw}{(1+w)^2} \frac{dv}{(1+v)^2}
\delta(1-\frac{1}{1+w}-\frac{1}{1+v})\nonumber\\
&& \int^\infty_0\frac{du}{u+\frac{1}{(1+w)(1+v)}}m^2
(\ln\frac{M_c^2}{m^2}-\gamma_E+1)\nonumber\\
& \to &  -\frac{i\lambda^2}{6(16\pi^2)^2} m^2
[(\ln\frac{M^2_c}{m^2}-\gamma_E)^2
+3(\ln\frac{M_c^2}{m^2}-\gamma_E)]  \  . 
\end{eqnarray}

{\bf In Region (II+III)}: $v \to \infty$ $or$ $w \to \infty$, ${\cal M}^2 \simeq m^2$, the contribution is found to be
\begin{eqnarray} \label{B_II}
-iM^{2(b)(II+III)}_{(2)log} &\simeq & -\frac{i\lambda^2}{6(16\pi^2)^2} m^2
\int^\infty_0 \frac{dw}{(1+w)^2} \frac{dv}{(1+v)^2}
\delta(1-\frac{1}{1+w}-\frac{1}{1+v})\nonumber\\
&&[1-\frac{1}{(1+w)(1+v)}] \int^\infty_0
\frac{du}{[u+\frac{1}{(1+w)(1+v)}]^2}
[\ln\frac{M_c^2}{m^2}-\gamma_E+1]\nonumber\\
&=& -\frac{i\lambda^2}{6(16\pi^2)^2} m^2
[\int^\infty_0\frac{dw}{1+w}+\int^\infty_0
\frac{dv}{1+v}-1](\ln\frac{M_c^2}{m^2}-\gamma_E+1)\nonumber\\
& \to & -\frac{i\lambda^2}{6(16\pi^2)^2} m^2
[2(\ln\frac{M_c^2}{m^2}-\gamma_E)^2+(\ln\frac{M_c^2}{m^2}-\gamma_E)] \  .
\end{eqnarray}
Note that in obtaining the above result the choice $\mu_o^2 = m^2$ has been made as the only mass scale in the limit $u\to \infty$ or $v\to \infty$ or $w\to \infty$ is the mass of the particle $m^2$. It is seen that there are three logarithmic divergences in the UVDP parameter space, which concern the calculations of $-iM^{2(b)(I)}_{(2)log}$ and $-iM^{2(b)(II+III)}_{(2)log}$ in three regions and correspond to subdivergences in the subdiagrams of Fig.\ref{groupB} $(b)$. 

{\bf In Region (IV):  $-p^2\gg m^2$}, in this region there is no harmful divergence as all the integrals of UVDP parameters are
convergent. All the terms proportional to $m^2$ in ${\cal M}^2$ will be neglected, the integral is simply given by:
\begin{eqnarray}\label{B_III}
-iM^{2(b)(IV)}_{(2)log} &\simeq & -\frac{i\lambda^2}{6(16\pi^2)^2}(-p^2)
\int^\infty_0 \frac{dw}{(1+w)^3} \frac{dv}{(1+v)^3}
\delta(1-\frac{1}{1+w}-\frac{1}{1+v})\nonumber\\
&& \int^\infty_0 \frac{du u}{[u+\frac{1}{(1+w)(1+v)}]^3}
[\ln\frac{M_c^2}{\frac{u}{(1+u)[u(1+w)(1+v)+1]}(-p^2)}-\gamma_E+1]
\nonumber\\
&=& \frac{i\lambda^2}{6(16\pi^2)^2}p^2[\frac{1}{2}(\ln\frac{M_c^2}{-p^2}-\gamma_E+1) + c_0 ]  \  .
\end{eqnarray}
with $c_0 =\frac{1}{108}(-81-2~\psi^{(1)}(\frac{1}{6})-2~\psi^{(1)}(\frac{1}{3})+2~\psi^{(1)}(\frac{2}{3}) +2~\psi^{(1)}(\frac{5}{6}))$ and $\psi^{(1)}(z)\equiv \frac{d^{2}}{d z^{2}}\ln \Gamma(z)$ bing the polygamma function of order 1. This gives the first order correction to the wave function renormalization in the $\phi^4$ theory. 

The contributions of two loop sunrise diagram Fig.\ref{groupB} $(b)$ is found to be
\begin{eqnarray}\label{B_2}
-iM^{2(b)}_{(2)} &\simeq&
\frac{i\lambda^2}{6(16\pi^2)^2}\{[3(\ln\frac{M_c^2}{\mu^2}-\gamma_E)+1]M_c^2-
3m^2(\ln\frac{M_c^2}{m^2}-\gamma_E)^2 \nonumber\\
& & -4m^2(\ln\frac{M_c^2}{m^2}-\gamma_E) +\frac{1}{2}p^2(\ln\frac{M_c^2}{-p^2}-\gamma_E)\}  \  . 
\end{eqnarray}
The contribution of the counterterm diagrams Fig.\ref{groupB} $(b_1)$-$(b_2)$ is given by:
\begin{eqnarray}\label{B_2(1)}
-iM^{2(b_1)+(b_2)}_{(2)} &=&
\frac{1}{2}(-i\delta_\lambda^{s+u})\int\frac{d^4k}{(2\pi)^4}\frac{i}{k^2-m^2}\nonumber\\
&\to &-\frac{i\lambda^2}{2(16\pi^2)^2}(\ln\frac{M_c^2}{\mu^2}-\gamma_E)
[M_c^2-m^2(\ln\frac{M_c^2}{m^2}-\gamma_E+1)] \  . 
\end{eqnarray}
The sum of all contributions yields:
\begin{eqnarray}
-iM^{2(b)+(b_1)+(b_2)}_{(2)} &=& \frac{i\lambda^2}{(16\pi^2)^2}\{[
\frac{1}{6}(M_c^2-\mu^2)-\frac{1}{6}m^2(\ln\frac{M_c^2}{\mu^2}-\gamma_E) \nonumber \\
& & +\frac{1}{12}p^2(\ln\frac{M_c^2}{\mu^2}-\gamma_E)] +[\frac{1}{6}(\mu^2-m^2)-\frac{1}{2}m^2(\ln\frac{\mu^2}{m^2})^2 \nonumber \\
& & -\frac{2}{3}m^2\ln\frac{\mu^2}{m^2}+\frac{1}{12}p^2\ln\frac{\mu^2}{-p^2}]\}+... \  .
\end{eqnarray}
 The overall counterterm has the following form:
\begin{eqnarray}\label{B_counter}
i(p^2\delta^{(b)}_Z-\delta^{(b)}_{m^2}) &=&
-\frac{i\lambda^2}{(16\pi^2)^2}[\frac{1}{6}(M_c^2-\mu^2)
-\frac{1}{6}m^2(\ln\frac{M_c^2}{\mu^2}-\gamma_E) \nonumber \\
&& +\frac{1}{12}p^2(\ln\frac{M_c^2}{\mu^2}-\gamma_E)]  \   . 
\end{eqnarray}
The renormalized result for Fig.\ref{groupB} is given by:
\begin{eqnarray}\label{B_result}
-iM^{2(b)}_{(2)R} & = &
\frac{i\lambda^2}{(16\pi^2)^2}[\frac{1}{6}(\mu^2-m^2)
-\frac{1}{2}m^2(\ln\frac{\mu^2}{m^2})^2 -\frac{2}{3}m^2\ln\frac{\mu^2}{m^2} \nonumber \\
& & +\frac{1}{12}p^2\ln\frac{\mu^2}{-p^2}]+...  \  . 
\end{eqnarray}
 
The total contributions to the two-loop self-energy of scalar boson as shown in Fig.\ref{groupA} and Fig.\ref{groupB} are found to be
\begin{eqnarray}
-iM^2_{(2)R}&=&
\frac{i\lambda^2}{(16\pi^2)^2}[\frac{1}{4}(\mu^2-m^2)\ln\frac{\mu^2}{m^2}
+\frac{1}{6}(\mu^2-m^2) \nonumber\\
&& -\frac{3}{4}m^2(\ln\frac{\mu^2}{m^2})^2-\frac{11}{12}m^2\ln\frac{\mu^2}{m^2}
+\frac{1}{12}p^2\ln\frac{\mu^2}{-p^2}]+...  \   .
\end{eqnarray}
In the massless limit  $m^2\to0$ and without including the quadratic contribution $\mu^2$, one arrives at the result yielded by using the dimensional regularization\cite{Peskin:1995ev}
\begin{equation}
-iM^2_{(2)R}=\frac{i\lambda^2}{12(16\pi^2)^2}p^2\ln\frac{\mu^2}{-p^2} \  .
\end{equation}

From Eqs. (\ref{A_counter}) and (\ref{B_counter}), the two-loop mass and wave function counterterms are found to be as follows:
\begin{eqnarray}
-i\delta^{(2)}_{m^2} &=& \frac{i\lambda^2}{(16\pi^2)^2}[
\frac{1}{4}(M_c^2-\mu^2)(\ln\frac{M_c^2}{\mu^2}-\gamma_E)-\frac{1}{6}(M_c^2-\mu^2)\nonumber\\
&&-\frac{1}{4}m^2(\ln\frac{M_c^2}{\mu^2}-\gamma_E)^2
+\frac{1}{6}m^2(\ln\frac{M_c^2}{\mu^2}-\gamma_E)]\label{delta_m2}\\
i\delta^{(2)}_{Z}&=&
-\frac{i\lambda^2}{12(16\pi^2)^2}(\ln\frac{M_c^2}{\mu^2}-\gamma_E)\label{delta_z2} \  .
\end{eqnarray}
 The renormalized mass is defined as:
\begin{eqnarray}
m^2 & = & Z_{\phi}m_0^2-\delta_{m^2} = m_0^2 +m_0^2\delta_Z
-\delta_{m^2} \nonumber \\
 &=& m_0^2 +\frac{\lambda_0}{2(16\pi^2)}[(M_c^2-\mu^2)-m_0^2(\ln\frac{M_c^2}{\mu^2}-\gamma_E)]\nonumber\\
&& +\frac{\lambda_0^2}{(16\pi^2)^2}[
\frac{1}{4}(M_c^2-\mu^2)(\ln\frac{M_c^2}{\mu^2}-\gamma_E)-\frac{1}{6}(M_c^2-\mu^2)\nonumber\\
&& - \frac{1}{4}m_0^2(\ln\frac{M_c^2}{\mu^2}-\gamma_E)^2+\frac{1}{12}m_0^2(\ln\frac{M_c^2}{\mu^2}-\gamma_E)] \  ,
\end{eqnarray}
which differs from the result obtained by using the dimensional regularization due to the appearance of the quadratic terms.
The anomalous mass dimension by summing up all the leading quadratic and logarithmic terms (without considering the logarithmic-squared term and quadratic-logarithmic cross term) is found to be:
\begin{eqnarray}
\gamma_{\phi^2} &=& \frac{\mu^2}{m^2}\frac{d m^2}{d\mu^2} =
-\frac{\lambda_0}{(16\pi^2)m^2}(\frac{1}{2}\mu^2-\frac{1}{2}m_0^2)
+\frac{\lambda^2_0}{(16\pi^2)^2m^2}(\frac{1}{6}\mu^2-\frac{1}{12}m_0^2)\nonumber\\
&\approx&
-\frac{\lambda}{16\pi^2}(\frac{1}{2}\frac{\mu^2}{m^2}-\frac{1}{2})+
\frac{\lambda^2}{(16\pi^2)^2}(\frac{1}{6}\frac{\mu^2}{m^2}-\frac{1}{12}) \nonumber \\
& = &  \frac{1}{2} \frac{\lambda}{16\pi^2}   - \frac{1}{12} \left(\frac{\lambda}{16\pi^2}\right)^2
- \frac{\mu^2}{m^2} \left[\frac{1}{2} \frac{\lambda}{16\pi^2}   - \frac{1}{6} \left(\frac{\lambda}{16\pi^2}\right)^2 \right]  \  .
\end{eqnarray}
In the second line the bare mass and coupling constant have been replaced with the renormalized ones. 

Note that the result differs from the one obtained in ref. \cite{Kazakov}  by using the dimensional regularization with the $\bar{MS}$ subtraction scheme. The difference appears in both the power-law running terms and the logarithmic running terms. The power-law running terms with the form $\mu^2/m^2$ reflects the fact that the LORE method maintains the quadratic structure of the original theory. For the logarithmic terms, it is known that the two-loop anomalous mass dimension in $\phi^4$ theory is subtraction scheme dependent. It may be seen by rescaling the energy scale $\mu^2 \to e^{\alpha_0} \mu^2$, the leading logarithmic term at two-loop level will be changed by an additional contribution caused from the logarithmic-squared term, and the resulting $\gamma_{\phi^2}$ for the logarithmic running becomes
\begin{eqnarray}
\gamma_{\phi^2}|_{log} &=&  \frac{1}{2} \frac{\lambda}{16\pi^2}   - \frac{1}{12}(1+6\alpha_0) \left(\frac{\lambda}{16\pi^2}\right)^2 \  . \nonumber
\end{eqnarray}
Meanwhile both the $\mu^2$-independent term and the quadratic $\mu^2$-dependent terms will be changed correspondingly. Alternatively, by shifting the energy scale $\mu^2\to \hat{\mu}^2\equiv \mu^2 - \alpha_0 m^2$, the leading logarithmic term will receive an extra contribution from the quadratic-logarithmic cross term, and the resulting $\gamma_{\phi^2}$ for the logarithmic running is modified to be
\begin{eqnarray}
\gamma_{\phi^2}|_{log} & = &  \frac{1}{2} \frac{\lambda}{16\pi^2}   - \frac{1}{12}(1+3\alpha_0) \left(\frac{\lambda}{16\pi^2}\right)^2\  , \nonumber 
\end{eqnarray}
and the quadratic-logarithmic cross term is given in terms of two energy scales $\mu^2$ and $\hat{\mu}^2$ instead of a single energy scale, $(M_c^2-\mu^2)(\ln M_c^2/\hat{\mu}^2 -\gamma_w)$. Thus either the rescaling or the shifting of the subtracted energy scale $\mu^2$ will modify the initial correlative form $(M_c^2-\mu^2)$ and $\ln M_c^2/\mu^2$. It is interesting to note that the arbitrariness caused by the subtraction scheme for the scalar mass renormalization at high loop order may be removed by the requirement of keeping the correlative form $(M_c^2-\mu^2)$ and $\ln M_c^2/\mu^2$ with a single subtracted energy scale.

For the two-loop vertex contributions, a detailed calculation in LORE is referred to Ref.\cite{HW1}. In
the perturbative calculation, the renormalized $\lambda$ at two-loop level is given by
\begin{eqnarray}
\lambda &\approx&
\lambda_0-\frac{3\lambda_0^2}{2\cdot(16\pi^2)}(\ln\frac{M_c^2}{\mu^2}-\gamma_E) \nonumber \\
& & -\frac{\lambda_0^3}{(16\pi^2)^2}[\frac{3}{4}(\ln\frac{M_c^2}{\mu^2}-\gamma_E)^2
-\frac{17}{6}(\ln\frac{M_c^2}{\mu^2}-\gamma_E)]  \  .
\end{eqnarray}
From the definition of $\beta$-function for the renormalized coupling constant $\lambda$, which is considered to sum up all the leading logarithmic terms (not including the logarithmic-squared term), we have 
\begin{eqnarray}
\beta_{\lambda} &=& \mu\frac{d\lambda}{d\mu}= \frac{3\lambda^2_0}{16\pi^2}-\frac{2\lambda_0^3}{(16\pi^2)^2}\frac{17}{6}\nonumber\\
&\approx& \frac{3\lambda^2}{16\pi^2}-\frac{17}{3}\frac{\lambda^3}{(16\pi^2)^2} \  ,
\end{eqnarray}
which agrees with the standard result $\beta_\lambda$\cite{Brezin,Chetyrkin,Dittes:1977aq}. Where we has replaced in the last line the bare constant $\lambda_0$ by its renormalized one.

\subsection{Quantum Gravitational Effects and Asymptotic Free Power-Law Running of Gauge Couplings}

The discovery of Higgs boson has completed the test on the standard model, so far all particles predicted in the standard model have been founded. Thus the graviton becomes the only undetected particle which is predicted from the gravitational force$-$one of the four basic forces in nature. The quantization of general relativity becomes the most interesting and frustrating question in theoretical physics. It is known that the general relativity is not regarded as a renormalizable QFT as the mass dimension of its coupling constant $\kappa=\sqrt{32\pi G}$ is negative. Whereas it is undoubtable that the gravity has effects on matter fields as all matter has gravitational interaction.

Robinson and Wilczek \cite{RW} carried out a calculation, in the framework of traditional background-field method, for gravitational corrections to gauge theories by adopting the naive cut-off regularization, and found that these corrections lead all gauge theories to be asymptotically free through changing the gauge couplings to power-law running in a specific gauge condition. While it was showed\cite{Pietrykowski} that the result obtained in \cite{RW}  depends gauge condition, and in the harmonic gauge the gravitational correction to $\beta$ function is absent at one-loop order. Late on, it was demonstrated in \cite{Toms}  by using gauge-condition independent formalism\cite{Vilkovisky,DeWitt} that the gravitational corrections to the $\beta$ function vanished in Dimensional Regularization\cite{DR}. Then a diagrammatic calculation was performed in\cite{Ebert} by using both cut-off and dimensional regularization schemes, it yielded the same conclusion that the quadratic corrections are absent in the harmonic gauge. In ref. \cite{Tang:2008ah} , all the calculations were checked in the framework of diagrammatic and traditional background field methods, it was found that the results are not only gauge condition dependent but also regularization scheme dependent. By applying for the consistent LORE method, it was concluded that there was asymptotic freedom with power-law running in the harmonic gauge condition. To make a gauge-condition independent calculation in LORE,  a further calculation has been carried out in refs.\cite{Tang:2010cr,Tang2011} by using the gauge-condition independent Vilkovisky-DeWitt formalism in the framework of background field method. 

For simplicity, consider the $U(1)$ electromagnetic theory with gravitational interaction. The classical action functional of Einstein-Maxwell theory is
\begin{equation} \label{eq:GravityL}
S =\int d^4x |g(x)|^{1/2}\left(\frac{1}{4} F_{\mu\nu}F^{\mu\nu} -\frac{2}{\kappa^2}(R-2\Lambda)\right) \  ,
\end{equation}
with $F_{\mu\nu}=\partial_{\mu}A_{\nu}-\partial_{\nu}A_{\mu}$ and $\kappa^2=32\pi G$.  

Before proceeding, it is useful to make a brief introduction for the Vilkovisy-DeWitt formalism, a general description is referred to refs. \cite{DJToms,Tomsbook}. For a comparison, we shall use the DeWitt's condensed index notation \cite{DeWittbook} and Riemannian. Let $S[\varphi]$ represent the classical action  which is gauge invariant under the transformation
\begin{equation}
\delta \varphi^i = K^i_{\alpha}[\varphi]\delta \epsilon^\alpha \  , 
\end{equation}
where $K^{i}_{\alpha}[\varphi]$ is regarded as the generator of gauge transformations. A gauge condition is needed to quantize gauge theory, it is supposed to satisfy $\chi ^\alpha [\varphi]=0$ and $\chi^\alpha[\varphi+\delta\varphi]=\chi^\alpha[\varphi]$, which leads to
\begin{equation}
\chi^{\alpha}{}_{,i}[\varphi] K^{i}_{\beta}[\varphi]\delta \epsilon^{\beta} \equiv Q^{\alpha}{}_{\beta}[\varphi] \delta \epsilon^{\beta} =0  \  .
\end{equation}
Thus $\det Q^{\alpha}{}_{\beta}$ defines the Faddeev-Popov factor \cite{FaddeevPopov}. In the background field approach, the fields $\varphi^i$ are usually written as the sum of background-fields $\bar{\varphi}^i$ and quantum fields $\eta^i$
\begin{equation}
\varphi ^i = \bar{\varphi}^i +\eta ^i.
\end{equation}
For a practical calculation, it is convenient to choose the Landau-DeWitt gauge condition \cite{FradkinTseytlin} to simplify the calculation
\begin{equation}
\chi_\alpha =K_{\alpha i}[\bar{\varphi}]\eta^i=0.
\end{equation}
Taking $g_{ij}[\varphi]$ as the metric of the field space and Landau-DeWitt gauge,  the effective action at one-loop order is given by
\begin{eqnarray}\label{Eq.VDaction}
& & \Gamma[\bar{\varphi}]  =  S[\bar{\varphi}]-\ln\det Q_{\alpha\beta}[\bar{\varphi}] \nonumber \\
& & +  \frac{1}{2}\lim_{\Omega\rightarrow0}\ln\det\left(\nabla^i\nabla_j S[\bar{\varphi}] + \frac{1}{2\Omega}K^{i}_{\alpha}[\bar{\varphi}] K^{\alpha}_{j}[\bar{\varphi}]\right)
\end{eqnarray}
with 
\[ \nabla_i\nabla_j S[\bar{\varphi}]=S_{,ij}[\bar{\varphi}] - \Gamma^{k}_{ij} S_{,k}[\bar{\varphi}]\]
where the Christoffel connection $\Gamma^{k}_{ij}$ is determined by $g_{ij}[\varphi]$. It is the connection term $\Gamma^{k}_{ij} S_{,k}[\bar{\varphi}]$ which distinguishes the Vilkovisky-DeWitt's method from the traditional background-field method. It is convenient to rewrite the determinant by an integral
\begin{eqnarray}\label{Eq.ordertwo}
& & \Gamma_G = \frac{1}{2} \ln\det \left\{ \nabla^i\nabla_j S[\bar{\varphi}] +\frac{1}{2\Omega}K^{i}_{\alpha}[\bar{\varphi}] K^{\alpha}_{j}[\bar{\varphi}]\right\} = -\ln\int\left[ d\eta\right]\,e^{-S_q}\;, \nonumber \\
& & \Gamma_{GH}  =  -\ln\det Q_{\alpha\beta}=-\ln\int\left[ d\bar{\eta}d\eta\right] e^{-S_{GH}}, \nonumber \\
& & S_q   =  \frac{1}{2}\eta^i\eta^j(\nabla_i\nabla_j S  +\frac{1}{2\Omega}K_{\alpha\,i}K^{\alpha}_{j}),\quad S_{GH} = \bar{\eta}_\alpha Q^{\alpha}{}_{\beta}\eta^\beta
\end{eqnarray}
where $\Gamma_{GH}$ is the ghost contribution with $\bar{\eta}_\alpha$ and $\eta^\beta$ are anti-commuting ghost fields. The limit $\Omega\rightarrow 0$ ensures the Landau-DeWitt gauge condition.

Let us now apply the above formalism to the Einstein-Maxwell theory. The cosmological constant term will lead a logarithmic contribution to change $\beta$ function of gauge coupling\cite{DJToms}. It can be checked in LORE that the resulting logarithmic term agrees with the one obtained in \cite{DJToms}. We shall focus on the quadratic contributions and expand the fields $\varphi^i=(g_{\mu\nu}, A_{\mu})$ with the corresponding background fields as follows
\begin{eqnarray}
g_{\mu\nu}&=&\delta_{\mu\nu}+\kappa h_{\mu\nu}; \quad A_{\mu}=\bar{A}_\mu + a_{\mu}
\end{eqnarray}
which leads to the effective action
\begin{eqnarray}\label{Eq:vConnection}
& & S_q=\frac{1}{2}\eta^i\eta^j\left(S_{,ij}-v\Gamma^{k}_{ij}S_{,k} +\frac{1}{2\Omega}K_{\alpha\,i}K^{\alpha}_{j}\right), \nonumber \\
& & S_{GF}=\frac{1}{4\Omega}\eta^i\eta^jK_{\alpha\,i}K^{\alpha}_{j}=\frac{1}{4\xi}(\chi_{\lambda})^2+\frac{1}{4\zeta}(\chi)^2
\end{eqnarray}
with $\xi$ and $\zeta$ corresponding to the gauge fixing parameters of gravity and gauge fields. 
$\chi_{\lambda}$ and $\chi$ are given by the Landau-DeWitt gauge conditions
\begin{eqnarray}
\chi _{\lambda} &=&\frac{2}{\kappa}(\partial^\mu h_{\mu\lambda}-\frac{1}{2}\partial_\lambda h)+\omega(\bar{A}_\lambda \partial^\mu a_\mu+a^\mu\bar{F}_{\mu\lambda}), \label{Eq:GravityGaugefix} \nonumber \\
\chi & = & {}-\partial^\mu a_\mu. \label{Eq:GaugeGaugefix}
\end{eqnarray}
Note that both $\omega$ and $v$ are not real gauge condition parameters. They are introduced just for an advantage of comparing with traditional background field method in harmonic gauge. Namely,  taking the Landau-DeWitt gauge condition  with $\omega =1$, $v =1$, $\xi \rightarrow 0$ and $\zeta \rightarrow 0$ to obtain a gauge condition independent result,  and the harmonic gauge condition with $\omega =0$, $v =0$, $\xi = 1/\kappa^2$ and $\zeta = 1/2$  to read off straightforwardly the results in the traditional background field method.

In principle, the Vilkovisky-DeWitt formalism is applicable in any gauge condition as it has been verified to be gauge condition independent\cite{FradkinTseytlin,BV,Huggins}. In a practical calculation, such a formalism becomes much simple in Landau-DeWitt gauge. It is useful to separate $S_q$ into several terms 
\begin{equation}
S_q = S_0 + S_1 + S_2
\end{equation}
with the subscript denoting the order in the background gauge field $\bar{A}_\mu$ after expansion of action \cite{DJToms}.The contributions from gravity-gauge coupling are given by
\begin{equation}
\Gamma_{G}=\langle S_2\rangle-\frac{1}{2}\langle S_1^2\rangle,\quad \langle S_2\rangle=\langle S_{21}\rangle+\langle S_{22}\rangle \  .
\end{equation}
All the contributions involve the quadratically divergent tensor- and scalar-type ILIs $I_{2\mu\nu}(m=0)$ and $I_2(m=0)$.  For a comparison with difference regularization schemes,  we take the general relation
\begin{equation}
I^R_{2\mu\nu} = \frac{1}{4}a_2 \delta_{\mu\nu} I^{R}_2\label{Eq.TensorScalar} \  ,
\end{equation}
with $a_2=2$ in the LORE method which satisfies the gauge invariance consistency condition, and $a_2=1$ in the naive cut-off regularization which violates gauge invariance. In the dimensional regularization, 
$I^R_{2\mu\nu} = I^{R}_2(m=0) =0$. 

The effective action from the ghost's contribution is easily obtained
\begin{equation}
\Gamma_{GH}=-\left[\langle S_{GH2}\rangle-\frac{1}{2}\langle S_{GH1}^2\rangle\right]  \  , 
\end{equation} 
with the quadratic contributions 
\begin{eqnarray}
\langle S_{GH2}\rangle &=& -\kappa^2 \omega I^{R}_2 \frac{1}{4}\int d^4x \bar{F}^2\ ; \quad
\left\langle S_{GH1}^2 \right\rangle =  0 \   ,
\end{eqnarray}
which is independent of $a_2$.

The quadratic parts from the gravity contributions are found to be
\begin{eqnarray}
\langle S_{2}\rangle &= & \kappa^2 (C_{21} + C_{22})I^{R}_2 \frac{1}{4} \int d^4x \bar{F}^2 \  ,  \nonumber \\
\langle S_{1}^2\rangle & = & \kappa^2 C_{11}I^{R}_2 \frac{1}{4} \int d^4x \bar{F}^2 \  , \nonumber \\
C_{21} & = & \frac{1}{2}\Big( [v(1-a_2)+3a_2/2](\kappa^2\xi-1)+3 \Big) \  ,  \\
C_{22} & = & \frac{v}{8}(a_2-1)(2\zeta-1)+\frac{\omega^2}{\kappa^2\xi} [( 2\zeta-1)a_2/4+1] \  , \nonumber \\
C_{11} & = & \frac{2\omega^2}{\kappa^2\xi}[(2\zeta-1)a_2/4+1]+2\kappa^2\xi(1-a_2/4) + 3a_2/2-4\omega(1-a_2/4)  \  , \nonumber 
\end{eqnarray}
which leads to the effective action
\begin{eqnarray}
\Gamma_{G} & = & \langle S_2 \rangle - \frac{1}{2}\langle S_1^2\rangle = \kappa^2 C_G I^{R}_2 \frac{1}{4} \int d^4x \bar{F}^2 \  , 
\end{eqnarray}
with
\begin{eqnarray}
C_G & = &\frac{(a_2-1)}{8}\Big(v\left[(2\zeta -1)-4(\kappa^2\xi-1)\right] \nonumber \\
& + & 8(\kappa^2\xi-1)- 16\omega - 4\Big)+3\omega a_2/2 \  .
\end{eqnarray}

The total effective action at one loop order  from all quadratic contributions has the following form\cite{Tang:2010cr,Tang2011}
\begin{eqnarray}\label{Eq.EAfinal}
\Gamma &=& \frac{1}{4}\int d^4x \bar{F}^2 +  \kappa^2 C I^{R}_2\frac{1}{4}\int d^4x \bar{F}^2  \   , 
\end{eqnarray}
with
\begin{eqnarray}
C & = & C_G-\omega = \frac{a_2-1}{8}\Big(v\left[(2\zeta -1)-4(\kappa^2\xi-1)\right]\nonumber \\
& + &8(\kappa^2\xi-1) - 16\omega - 4\Big) + \omega(-1+3a_2/2) \  .
\end{eqnarray}

The renormalized gauge action is given by
\begin{eqnarray}
S_M &=& \frac{1}{4}(1 + \delta_A) \int d^4x \bar{F}_{\mu\nu}\bar{F}^{\mu\nu}
\end{eqnarray}
with
\begin{equation}
 \delta_A \simeq - \kappa^2 C I^{R}_2  \  .
\end{equation}
Here only the quadratic contributions are considered and the logarithmic contributions have been ignored. The charge renormalization constant $Z_e$ is defined as
\begin{equation}
Z_A=1+\delta_A ,\qquad Z_eZ^{1/2}_{A}=1  \  , 
\end{equation}
which leads the gravitational correction to the $\beta$ function as follows
\begin{eqnarray}
\beta^{\kappa}_{e}= \mu \frac{\partial}{\partial\mu}e=\mu \frac{\partial}{\partial\mu} Z^{-1}_e e^{0}
         =\frac{1}{2}e^0\mu \frac{\partial}{\partial\mu} \delta_A  \  .
\end{eqnarray}
In the LORE method, we have
\begin{equation}\label{Eq:QuadInt}
I^{R}_{2}(m=0) \simeq \frac{1}{16\pi^2}(M_c ^2-\mu ^2) \  , 
\end{equation}
which results the extra $\beta$ function from gravitational contributions as follows
\begin{equation}\label{Eq.Betafunc}
\beta^{\kappa}_{e} = \frac{\mu ^2}{16\pi^2} e \kappa^2 C  \  .
\end{equation}
With including the cosmological constant $\Lambda$ which has a logarithmic contribution to the $\beta$ function, the above result is extended to be
\begin{equation}\label{Eq.Betaall}
\beta^{\kappa}_{e} = \frac{\mu ^2}{16\pi^2} e \kappa^2 C -\frac{3\Lambda}{64\pi^2}e \kappa^2  \   .
\end{equation}
To obtain an explicit result, taking the Landau-DeWitt gauge condition $v=1,\ \omega =1,\ \zeta = 0,\  \xi = 0$, and the gauge invariance consistency condition $a_2=2$, we have
\begin{eqnarray}
& & C = -1 - \frac{25}{8}(a_2 - 1) + 3a_2/2 = -\frac{9}{8},\nonumber \\
& & \beta^{\kappa}_{e} = - \frac{9\mu ^2}{128\pi^2} e \kappa^2   \  , 
\end{eqnarray}
which also holds for non-Abelian gauge theories\cite{Tang2011}. Such a result is not only gauge condition independent guaranteed by the Vilkovisky-DeWitt formalism, but also regularization scheme independent for any consistent regularization satisfying the consistency condition and preserving the divergence structure.  

Before ending, let us make comments on the regularization scheme dependence. In the dimension regularization, as $I^{R}_{2} (m=0)= 0$, so that $\delta_A = 0$, there is no quadratic gravitational contribution due to the disadvantage of dimensional regularization. In the cut-off regularization, $a_2=1$ and $C=1/2$. Thus one has
\begin{eqnarray}
& & \beta_e^{\kappa}=0, \qquad \qquad \qquad \qquad \mbox{Dim.~regularization} \  , \nonumber \\
& & \beta^{\kappa}_{e} = \mu ^2/(32\pi^2) e \kappa^2 ,\qquad \quad \mbox{Cut-off~regularization}  \  , \nonumber
\end{eqnarray}
which leads to no asymptotic freedom in cut-off regularization.

To make an independent check, it is useful to revisit the traditional background field method in the harmonic gauge. This can be done by simply taking $v=0,\ \omega =0,\ \zeta = 1/2,\  \xi = 1/\kappa^2$ in the above Vilkovisky-DeWitt formalism, which gives
\begin{equation}
C = 1/2 - a_2/2 \  .
\end{equation}
It then becomes manifest that in the cut-off regularization, $a_2=1$ and $C=0$, which confirms the results obtained in\cite{Pietrykowski,Ebert,Tang:2008ah}. In the LORE method $a_2 =2$ and $C=-1/2$. Thus we have 
\begin{eqnarray}
& & \beta^{\kappa}_{e} = 0, \qquad \qquad \qquad \qquad  \mbox{Cut-off~\&~Dim.~ regularization}  \ , \nonumber \\
& & \beta^{\kappa}_{e} = -\mu ^2/(32\pi^2) e \kappa^2, \qquad \mbox{in~LORE} \  .
\end{eqnarray}
which reproduces the result given in ref.\cite{Tang:2008ah}. It comes to the conclusion that the quadratic gravitational contributions to the gauge coupling constant are asymptotic free in the traditional background field or diagrammatic method with the harmonic gauge.

In conclusion, the quadratic gravitational contributions to the gauge coupling constant are non zero and lead to the asymptotic free power-law running of gauge couplings.

\section{ Conclusions and Remarks}

Based on the symmetry-preserving and infinity-free LORE method, it has been shown explicitly that QFTs can be defined fundamentally with physically meaningful energy scale $M_c$ to avoid infinities.  It is then not difficult for us to understand why the laws of nature can well be described by a quantum theory of fields when the energy scale in the considered phenomena is sufficiently lower than the characterizing energy scale $M_c$, and also the laws of nature at an interesting energy scale can well be dealt with by a renormalization group analysis at the sliding energy scale $\mu_s$ which can be chosen to be the order of magnitude of energy scale concerned in the considered process. 

It has been seen that the key concept of the LORE method is the introduction of the irreducible loop integrals(ILIs), which are generally evaluated from the Feynman diagrams by using the ultraviolet-divergence-preserving(UVDP) parametrization. It is interesting to notice that the evaluation of ILIs with the UVDP parametrization naturally merges with the Bjorken-Drell's circuit analogy of Feynman diagrams, where the UVDP parameters can be regarded as the conductance or resistance in the electric circuit analogy, the sets of conditions required for evaluating the ILIs and the momentum conservations are associated with the conservations of electric voltages in the loop and the
conservations of electric currents at each vertex respectively. Consequently, the divergences in Feynman
diagrams correspond to infinite conductances or zero resistances in electric circuits. The LORE method merging with the Bjorken-Drell's circuit analogy has its advantage to analyze the overlapping divergence structure of Feynman diagrams, which enables us to clarify the origin of UV divergences in the UVDP parameter space and identify the correspondence of the divergences between subdiagrams and UVDP parameters. From the explicit demonstration for the general $\alpha\beta\gamma$ integral, especially for the case with
$\alpha=\beta=\gamma=1$, the divergences arising from the subintegrals become manifest themselves in the integration in some asymptotic regions of UVDP parameter space. The calculations of the corresponding
counterterm diagrams show that all the harmful divergences cancel exactly in the final result. 

It has been demonstrated that the LORE method has advantages to treat the quadratic behavior of QFTs, such as the scalar interaction and gravitational interaction. The discovery of Higgs boson brings one's attention to investigate the features of scalar boson, such as  the mass renormalization of Higgs mass with power-law running, the self-interaction of Higgs boson, the stability of the electroweak vacuum state.  Also the graviton becomes the only undetected particle which is predicted from the quantum gravity theory. It is expected that the LORE method shows its merit in the applications to the electroweak symmetry breaking and the quantization of gravitational interaction.


\vspace{1 cm}

\centerline{{\bf Acknowledgement}}

I am grateful to K.K. Phua for his kind invitation and hospitality during the Conference in Honour of the 90th Birthday of Freeman Dyson.  This work was supported in part by the National Science Foundation of China (NSFC) under Grant \#No.10821504, 10975170 and the key project of the Chinese Academy of Science.


\end{document}